\input amstex
\vcorrection{1.8cm}
\hcorrection{-0.35cm}
\hyphenation{pa-ra-me-tri-za-tion pa-ra-me-ters dia-go-na-li-ze
re-fe-ren-ces re-fe-ren-ce}
\def\Ha{{\Cal H}}
\def\al{\alpha}
\def\sg{\sigma}
\def\lm{\lambda}
\def\om{\omega}
\def\gm{\gamma}
\def\vphi{\varphi}
\def\ep{\varepsilon}
\def\pr{P$\ $r$\ $o$\ $o$\ $f$\ $.\enspace}
\def\tr{\Delta}
\def\kp{\kappa}
\def\dd#1#2{\dfrac{d#1}{d#2}}
\def\sh{\;\text{sh}\;}
\def\ch{\;\text{ch}\;}

\magnification=1200
%\scaledocument{\magstep1}
\TagsOnLeft

\font\trm=cmr10            %cmr10
\font\bf=cmbx10            %cmbx10
\font\moj=cmcsc10          %csc Caps and small caps
\font\dbf=cmbx12 scaled1440

\pagewidth{14cm}
\pageheight{18cm}
\baselineskip=18pt
\parindent=10pt

\nopagenumbers
\trm
\catcode`@=11

\def\NoBlackBoxes{\global\overfullrule\z@}
\NoBlackBoxes
\def\binrel@#1{\setbox\z@\hbox{\thinmuskip0mu
\medmuskip\m@ne mu\thickmuskip\@ne mu$#1\m@th$}%
\setbox\@ne\hbox{\thinmuskip0mu\medmuskip\m@ne mu\thickmuskip
 \@ne mu${}#1{}\m@th$}%
 \setbox\tw@\hbox{\hskip\wd\@ne\hskip-\wd\z@}}
\def\headline{{\trm\hfil}\vskip18pt}

\def\makeheadline{\vbox to\z@{\vskip-22.5\p@
  \vbox {\headline}\vss}\nointerlineskip}

\outer\def\proclaim #1. #2\par{\medbreak
 {\moj #1.\enspace}{\sl#2}\par
  \ifdim\lastskip<\medskipamount \removelastskip\penalty55\medskip\fi}

\def\overset#1\to#2{\binrel@{#2}\ifdim\wd\tw@<\z@
 \mathbin{\mathop{\kern\z@#2}\limits^{#1}}\else\ifdim\wd\tw@>\z@
 \mathrel{\mathop{\kern\z@#2}\limits^{#1}}\else
 {\mathop{\kern\z@#2}\limits^{#1}}{}\fi\fi}

\def\underset#1\to#2{\binrel@{#2}\ifdim\wd\tw@<\z@
 \mathbin{\mathop{\kern\z@#2}\limits_{#1}}\else\ifdim\wd\tw@>\z@
 \mathrel{\mathop{\kern\z@#2}\limits_{#1}}\else
 {\mathop{\kern\z@#2}\limits_{#1}}{}\fi\fi}

 \def\ov#1#2{\overset{#1}\to{#2}}
 \def\ovj#1{\ov{(1)}{#1}}
 \def\ovd#1{\ov{(2)}{#1}}

\centerline{M. W. KALINOWSKI}
\vskip2cm
%\centerline{Department of Solid State Physics}
%\centerline{University of {\L}{\'o}d{\'z}, {\L}{\'o}d{\'z}, Poland}
\vskip3cm

\dbf
\centerline{Nonlinear Waves}
%\centerline{obtained by an improved Riemann method}
%\centerline{and their physical applications}

\vfill
%{\sevenrm Partially supported
%by Badania W{\l}asne U{\L} Nr. 505/485.}
\trm

\eject
\def\headline{{\trm\hfil  {--\ {\folio}\ --\quad}\hfill }\vskip18pt}
\pageno=1
%\newpage
\vglue20pt
\centerline{\bf Contents}
\vskip\baselineskip
\item{}
Abstract \dotfill\ 2
\item{1.}
Introduction \dotfill\ 3
\item{2.} The algebraization procedure \dotfill\ 5
\item{3.}
Equation for potential stationary
flow of a compressible perfect gas\hfill\break
(the first example of an application of the method) \dotfill\ 18
\item{4.} Simple integral elements (for the first
example of an application  of  the \ \ \ \
\linebreak method) \dotfill\ 20
\item{5.} Simple waves (the first example of an application of the method)
\dotfill\ 24
\item{6.} Gauge and B\"acklund transformation \dotfill\ 53
\item{7.} Equation for potential nonstationary
flow of a compressible perfect gas\hfill\break
(the second example of an application of the method) \dotfill\ 59
\item{8.} Simple integral elements (the second example of an application
of the \ \ \ \
\linebreak method) \dotfill\ 61
\item{9.} Simple waves (the second example of an application
of the method) \dotfill\ 65

Conclusions \dotfill\ 82

Appendix A -- simple elements (the first example) \dotfill\ 83

Appendix B -- simple elements (the second example) \dotfill\ 86

Acknowledgements \dotfill\ 88

References \dotfill\ 89

\vfill\eject
\vglue20pt

\centerline{\bf Abstract}
\vskip\baselineskip
A method for solving a quasilinear nonelliptical equation of the second order
is developed and we give classification and parametrization of simple elements
of the equation. The equation for potential stationary flow of a compressible gas
in a supersonic region is considered as
the first example. A new exact solution
is obtained which may be treated as a nonlinear analogue of a stationary wave.
A gauge structure for the equation and an analogue of the B\"acklund transformation are introduced.
Certain classes of the exact solution for equation  of nonstationary potential
flow of a compressible gas are found.
This is the second example. Finally a physical analysis of
the results is carried out.

\vskip\baselineskip


\vfill\eject

\vglue20pt

\centerline{\bf 1. Introduction}
\vskip\baselineskip

In this work we present a method for solving nonlinear and nonelliptical partial
differential equations. The paper is devoted to the theory of a differential equation
of second order with coefficients depending on first derivatives. The method
is old, for we employ Riemann invariants which have in fact been known for
a long time [1, 2, 3]. Nevertheless, we have found a new feature of this old theory
related to the problem of the algebraization  of a differential equation.

We
calculate exactly a characteristic cone for a Riemann wave and classify all simple elements for such
an equation. All these simple elements depend on parameters and we proved that these parameters
are parameters of some orthogonal or pseudo--orthogonal groups (which is very important
in a later analysis (see  section 6)).

Having classified the simple elements we may
classify simple waves. Every simple element belongs to some submanifold $F_i\subset
\Cal E^*$. The dimension of this submanifold equals the number of parameters
and these parameters are
arbitrary functions of dependent variables. Since we are dealing with
simple wares, the parameters become functions of $R$ (Riemann invariant). Due to
these arbitrary functions we may integrate the equations for simple waves,
and this indicates
that it is possible to introduce a gauge structure for every class of simple elements.

The paper is divided into 10 sections. In the second section we describe a parametrization
and a classification of simple elements and we write down all possible elements.
In the third section we deal with the first  example from
gas--dynamics. It is
the equation of potential stationary flow for a perfect gas in a suppersonic region. In the
fourth section we calculate all possible simple elements for this equation and
we collect them in the Appendix A (Table 1, 2, 3). In section 5 we calculate the simple
wave corresponding to one of the simple elements, find a new exact solution
of the equation and examine its properties. The solution depends on three arbitrary functions.
We shift the arbitrariness from the parameters (depending on $R$ -- Riemann
invariant) to more convenient functions. This results in some restrictions on the
arbitrary functions and on the range of the Riemann invariant $R$. The new freedom
associated with the parameters of the characteristic cone, and the restriction imposed
on them are new points in the Riemann invariants method. Simultaneously we
obtain a restriction on the range of the dependent variable, $R$. This seems
to be a new feature also. We derive the solution for the equation of potential
stationary flow
of the perfect gas (in three dimensions). This solution depends on three arbitrary functions
and restrictions which we have to impose on the functions lead to physical effects.
The solution may be considered as an analogue of a nonlinear stationary wave
and we obtain planes of ``density and pressure nodes'' and planes of
``magnitude of velocity antinodes''. In section 6 we deal with a ``gauge''
structure for the equation and its simple waves and we derive a transformation of gauge
type connecting two simple waves (from the same class). The transformation may be treated
as a nonlinear representation of a gauge group originating from the orthogonal
or pseudo--orthogonal group. The transformation is similar to the classical
B\"acklund transformation. In~section~7 we deal with the second example
from gas dynamics i.e. with an equation for a  nonstationary flow~of
compressible gas described by a potential of velocity and density. In the
eight section we find simple integral elements for this equation, which are
presented in Appendix~ B. Section 9 is devoted to simple waves for
this equation with some physical
interpretation. We find several classes of exact solutions.

\newpage
%############################################
%                SEC 2
%############################################
\vskip20pt
\centerline{\bf 2. The algebraization procedure}
\vskip\baselineskip

In this section we consider the algebraization procedure.

Let us consider a system of partial differential equations
$$\aligned
a_j^{s\nu}\bigl(u^1,u^2,\dots,u^l\bigr)\;\frac\partial{\partial x^\nu}\;u^j\bigl(x^1,
x^2,\dots,x^n\bigr)=0\endaligned
\qquad
\aligned&\nu\text{=}1,2,\dots,n.\\
&s\text{=}1,2,\dots,m.\\
&j\text{=}1,2,\dots,l.\endaligned
\qquad
\aligned m\ge l \endaligned
\tag 2.1 $$
$$x=(x^1,x^2,\dots,x^n)\in\Cal E\ ,\ \ u(x)=\bigl(u^1(x), u^2(x),\dots,u^l(x)\bigr)
\in \Ha $$
which is a quasilinear homogeneous system of first order with coefitients depending
only on the unknown functions. We suppose that $a^{s\nu}_j(u)$
are smooth functions of $u$ (at least of $C^1$ class) in a
certain open set $\vartheta \subset \Cal H$.
This system may be over--determined i.e. $m\ge l$.
Let us suppose that this system is a nonelliptical one. This means that there
exist some nontrivial solutions of the ``algebraic system'' of equations:
$$
a_j^{s\nu}\gamma^j(u_0)\lambda_\nu(u_0)=0\qquad\text{where}\quad
\text{rank}\;\|a_j^{s\nu}(u_0)
\lambda_\nu(u_0)\|<l,\quad u_0\in \Cal H=R^l\tag 2.2
$$
for vectors $\gamma\in \Cal H = R^l$ and $\lm\in \Cal E^*=R^n$.

The above algebraic system of equations defines respectivelly
so--called knotted
characteristic vectors in a {\it hodograph}\/ space $\Ha=R^l$ (i.e., the space of the values
of the functions $u^j$) and in a physical space
$\Cal E=R^n$ (i.e., the space of independent variables). The pair $\gamma$ and
$\lambda$ will be called a knotted pair for $u_0 \in \Cal H$
iff it obeys the equation (2.2). This
fact will be denoted by $\gamma (u_0)\sim \lambda(u_0)$. The matrix
$L^j_\nu(u_0)=\gamma^j(u_0)\lambda_\nu(u_0)$
created by a pair of knotted vectors will be a simple integral element, because
 $\text{rank}\;\|L^j_\nu(u_0)\|=1$, where $u_0\in\Ha$.

In this case we can consider several notions of hyperbolicity. We say
that the system (2.1) is hyperbolic at a~point $u_0\in \Cal H$
in a direction $\Cal E^*\ni v \ne
0$ if for every $\Cal E^*\ni w \ne 0$ the equation
$$
 \det\; \| a_j^{s\nu} (u_0) (v_\nu+\zeta w_\nu) \| = 0\quad\
(\text{iff}\ l=m)
$$
has only real solutions for $\zeta$ or more generally if
$$
\text{rank}\;\| a_j^{s\nu} (u_0) (v_\nu + \zeta w_\nu) \| < l
$$
has only real solutions. The system is hyperbolic at a point $u_0$ if the
system is hyperbolic at $u_0$ for any nonzero $v\in \Cal E^*$.
The last means that (2.2) has real and only real solutions.
The system is
hyperbolic in an open set $\vartheta \subset \Cal H =R^l$ if it is hyperbolic
at any point of the set $\vartheta$. This means also that the system is
hyperbolic in a usual sens i.e.\ it has only real characteristics.

Strong hyperbolicity means that any
integral element $L_\nu^j (u_0)$ annihilated by the matrix $a_j^{s\nu}$, i.e. $a_j^{s\nu} (u_0)
L_\nu^j (u_0) =0$ must be of the form
$$ L_\nu^j (u_0) = \sum \xi^s \gm^j(u_0) \lm_\nu(u_0) $$
and a system is hyperbolic in an open set $\vartheta \subset \Cal H=R^l$.
We call the system of equations (2.1) hyperbolic in a strong sense if
all its integral elements are generated by simple elements
$$
du = \sum\limits_{s=1}^n \xi^s(x) \gm_s (u) \otimes \lm^s (u)
=L_\nu^j\, dx^\nu.
$$
If the matrix in (2.2) is square, i.e. $m=l$ we get a solvability
condition in a simpler form, i.e.
$$ \det\; \| a_j^{s\nu} \lm_\nu \| = 0. $$

It is  convenient to consider $\lambda$ as an element of the space $\Cal E^*$ which is
the space of linear forms, $\Cal E^*\ni\lambda:\/\Cal E\to R^1$. On the other hand,
in the terminology of tensor calculus if we consider $x\in\Cal E$ as a contravariant
vector then $\lambda\in\Cal E^*$ is a covariant vector. In these terms the element
$L$ is an element of the tensor space $T_u\Ha\otimes\Cal E^*$ of the form $L=\gamma\otimes\lambda$.
Now we introduce a simple wave, which suggests a separation of simple integral
elements from a set of integral elements. In order to do this we formulate the
theorem (which allows us to define a simple wave).

\proclaim Theorem--Definition.
Let the mapping $u:\/D\to\Ha$, \ $D\subset\Cal E$ be any solution of the system
(2.1). We call $u$ a simple wave for a homogeneous system if the tangent mapping
$du$ is a simple element at any point $x_0\in D$. Let us consider the smooth
curve $\Gamma:\/R\to f(R)$ in the hodograph space $\Ha$ parametrized by $R$, so
the tangent vector such that
$$\frac{df(R)}{dR}=\gamma\bigl(f(R)\bigr) \tag 2.3 $$
is a characteristic vector. Then there exists a field of characteristic covectors $\lambda(u)$
connected with $\gamma\bigl(f(R)\bigr)$ specified on the curve $\Gamma:\/\lambda=\lambda\bigl(f(R)\bigr)$. Hence
we may state the following. If the curve $\Gamma\subset\Ha$ obeys the condition (2.3)
and if $\varphi(\cdot)$ is any differentiable function of one variable, then the function
 $u=u(x)$ specified in an implicit way by relations (equations)
$$
\left\{\aligned &u=f(R)\\
&R=\varphi\bigl(\lambda_\nu(f(R))x^\nu\bigr)\quad \text{where $a^{s\nu}_j
\gamma^j\lambda_\nu=0$} \endaligned \right.\tag 2.4 $$
is the solution of the basic system (2.1). The solution is called a simple wave or
Riemann wave. (The second equation of (2.4) is not always solvable so one
should impose sertain conditions on $\varphi$ and~$f$. Moreover
it is always solvable locally i.e.\ in a certain open set $\Cal
O = (R_0,R_1)\times \Cal O'$, $R_0,R_1\in R^1$, $\Cal O'\subset
\Cal E$ under given assumptions by an implicit function theorem in
$R^{n+1}$.)

\vskip10pt
A proof may be obtained by direct differentiation of the implicit relations (2.4).
The covector $\lambda$ in (2.4) specifies the velocity and direction of the wave
propagation. The curve $\Gamma$ fulfilling the condition (2.3) is called a characteristic
curve in the hodograph space, $\Ha\;$. If the mapping $u:\/\Cal E\to\Ha$ is a
simple wave, then a characteristic curve in space $\Ha$ is the image of the
map~$u$.
The parameter $R$, specified on this curve, is called a Riemann invariant.

It is interesting to mention that on the hypersurface $S$ in the space
$\Cal E$ given by two equations:
$$\aligned
& R = \vphi(\lm_\mu(f (R)) x^\mu), \\
& 1 - {\dot\vphi}\frac{d\lm_\mu (f(R))}{dR} x^\mu =0, \endaligned $$
the gradient of the function $R(x)$ becomes infinite (that is the so--called
gradient catastrophe), i.e.\ the solution does not make sense
on the hypersurface $S$. In this case some discontinuities can take place
on the hypersurface $S$ --- e.g.\ shock waves. The covector
$\lm$ is orthogonal to a level surface of the function $R(x)$. We can
define a one--parameter family of hyperplanes.
by an implicit equation $r =\vphi(\lm_\mu (r) x^\mu)$, where
$r$ is a parameter of a family. In this way
$$
r = \vphi (\lm_\mu (r) x^\mu),\quad \text{where}\quad \lambda_\mu(r)=\lambda_\mu (f(r))
$$
$$ 1 -{\dot\vphi} \lm_{\mu,r}(r) \cdot x^\mu =0,\quad \hbox{where} \quad
\lm_{\mu,r}(r)= \frac{d\lm_\mu (f(r))}{dr}
$$
are equations of an envelope of this family. So the
hypersurface $S$ is an envelope of the family $R(x) = \text{const}$.
This hypersurface is the place for a gradient catastrophe.

Now let us consider a nonelliptical equation of the second order
$$\sum_{i,j=1}^na_{ij}\Bigl(\frac{\partial \Phi}{\partial x^1}\;,\;\frac{\partial\Phi}{\partial x^2}
\;,\dots,\;\frac{\partial\Phi}{\partial x^n}\Bigr)\frac{\partial^2\Phi}{\partial
x^i\partial x^j}=0, \tag 2.5 $$
where $\Phi:\ D\subset R^n\to R'$ is a real function of $n$ variables of
$C^{(2)}$ clars, and $D$ a domain of $\Phi$ is an open set. We suppose
that $a_{ij}$ are smooth (at least of $C^1$ class) functions of their variables.
Equation (2.5) may be transformed, by the introduction of new unknown
functions, to the system
of equations of the first order:
$$\left\{\aligned
&\sum_{i,j=1}^na_{ij}(u^1,\dots,u^n)\frac{\partial u^i}{\partial x^j}=0\\
&\frac{\partial u^i}{\partial x^j}-\frac{\partial u^j}{\partial x^i}=0\\
\endaligned\right.\tag 2.6
$$
where
$$
u^i=\frac{\partial\Phi}{\partial x^i}\quad i=1,2,\dots,n.
$$
Rewriting the system of equations (2.6) in terms of simple integral elements we have:
$$\align
&\sum_{ij=1}^na_{ij}\gamma^i\lambda_j=0\tag 2.7a
\\\\
&\gamma^i\lambda_j-\gamma^j\lambda_i=0
\qquad i,j=1,2,\dots,n\tag 2.7b
\endalign $$
From equation (2.7b) we find that the vector $\lambda$ is proportional to vector $\gamma$.
Thus from equation (2.7a) we get a quadratic form with respect to the variables $\lambda_1,\lambda_2,\dots,\lambda_n$:
$$Q(\lambda_1,\lambda_2,\dots,\lambda_n)=\sum_{i,j=1}^na_{ij}\lambda_i\lambda_j=0.
\tag 2.8 $$
Equation (2.8) is the equation of a cone of the characteristic covector $\lambda$, specifying
the velocity and direction of propagation of the simple wave.

Our aim will be to find the parametrical equations of this cone and at the same time
to parametrize the covector $\lambda$.

To do this we transform the quadratic form $Q$ to a canonical form, i.e., we
diagonalize the symmetrized matrix $A=(\widetilde a_{ij})$ of the form $Q$. Hence we look for eigenvalues
of this matrix and write the secular equation
$$\det\;(A-\omega I)=0\tag 2.9$$
We also search for a matrix $B$ which diagonalizes the matrix $A$

\vskip10pt
$$
B^TAB=\left(\matrix\omega_1& & &0\\
                           &\omega_2\\
                    &&\ddots\\
                     0&&&\omega_n\endmatrix
		     \right)\quad ,
\quad B^T=B^{-1}
\tag 2.10 $$
\vskip10pt
\flushpar
where $\omega_1, \omega_2,\dots,\omega_n$ are the eigenvalues of the matrix $A$. The
matrix $A$ is a symmetric real matrix and it has real eigenvalues. There always
exists at least one such orthogonal matrix $B$ responsible for a rotation
in $n$--dimensional Euclidean space from variables $V=(V^1,\dots,V^n)$ to variables
$\lambda=
(\lambda^1,\dots,\lambda^n)$

$$
\lambda=B\left(\matrix V^1\\ V^2\\ \vdots\\ V^n\endmatrix\right)=BV\tag 2.11
$$
and
$$
Q(\lambda_1,\dots,\lambda_n)=\omega_1(V_1)^2+\omega_2(V_2)^2+\cdots+\omega_n(V_n)^2=
(B^TAB)_{ij}V_iV_j
$$
We assume that $\omega_i\ne0$, $i=1,2,\dots,n$. The case with $\omega_i=0$
for some $i$ is
examined separately.

\vskip10pt
Now we aim to parametrize matrix $B$, i.e. we find all $B$'s obeying (2.10).

Let a matrix $A$ posses $K$ different eigenvalues with orders $l_j$, $j=1,2,\dots,K$
respectively so that $\sum\limits_{j=1}^Kl_j=n$. Thus we obtain the following
formula: (after suitable carranging of $V_i$)

$$
Q(\lambda_1,\lambda_2,\dots,\lambda_n)=\sum_{i=1}^K\omega_i\Bigl(\sum_
{\mu=1}^{l_i}
V^2_{p_i+\mu}\Bigr)\ ,\quad\text{where} \ p_i=\sum_{\nu=1}^{i-1}l_\nu\ ,
$$
$$B^TAB=\left(\matrix
\omega_1& & & & & & & & & & \\
&\omega_1& & & & & & & & 0\\
&&\ddots\\
&l_1&&\omega_1\\
&&&&\ddots\\
&&&&&\ddots\\
&&&&&&\omega_K\\
&&&&&&&\omega_K\\
&0&&&&&&&\ddots\\
&&&&&&&l_K&&\omega_K\endmatrix\right)
\tag 2.12 $$
%.................
Let $B_0$ be any matrix  diagonalizing the matrix $A$.
Observe that any orthogonal transformation of variables $V_1,V_2,\dots,V_{l_1}$
does not destroy the dia\-gonali\-zation, i.e. any orthogonal $l_1\times l_1$ matrix $C_1$,
acting on the first $l_1$ variables does not change the diagonal form.

So a matrix $B$ is defined modulo the following matrix
$$
\align
&
B_1=\left.\left(\matrix
\underset{\tsize l_1}\to
{\underbrace{\underline{\;\ C_1\ \;\Bigl|}}}\Bigr\}l_1
&&&&&0\\
&1\\
&&1\\
&&&1\\
&&&&1\\
&&&&&1\\
0&&&&&&1\endmatrix\right)\right\}n \tag 2.13
\\ \\
\noalign{\text{where }}
&C^T_1=C_1^{-1}\ .\endalign
$$
The same may be said about the remaining subsets of variables $V_i$, i.e. sequences
of $l_i$ elements.

The action of the $l_i\times l_i$ matrix $C_i^T=C_i^{-1}$ (the orthogonal one)
cannot destroy the diagonalization. It is easy to see that the matrix $C_i$ corresponds
to the following $n\times n$ matrix $B_i$,
$$
\align&
B_i=
\underset {\tsize n}\to
{\left.
\underbrace{\left(
\matrix 1\\
&1&&&&&&0\\
&&\ddots\\
&q_1&&1\\
&&&&l_i\Bigl\{
     \overset{\tsize l_i}\to
     {\overbrace{\overline{\underline{\Bigl|\ \ C_i\ \ \Bigr|}}}}\\
&&&&&1\\
&&&&&&1\\
&0&&&&q_2&&\ddots\\
&&&&&&&&1\endmatrix
\right) }
 \right\}}n \tag 2.14
\\ \\
\noalign{\text{where}}
& q_1=\sum_{j=1}^{i-1}l_i\; ,\qquad q_2=\sum_{j=i+1}^kl_i .
\endalign $$

Hence the matrix $B$ is defined {\it modulo\/} the product of $K$ matrices $B_i$ of the form
(2.14). From that we obtain
$$B=B_0\prod_{i=1}^k B_i \tag 2.15 $$
where $B_0$ is an arbitrary, but established, matrix obeying (2.12). The order
of multiplication of matrices $B_i$ is irrelevant, because  all matrices
commute with each another. So we see that the problem of the parametrization
of $B$ is reduced to the parametrization of each of the $K$ orthogonal matrices
 $C_i$.
Each of these matrices is responsible for an arbitrary rotation or reflection
in $l_i$--dimensional space, thus it depends on $\frac12\;l_i(l_i-1)$
parameters which are generalized Euler angles and some     		
set of discrete parameters.
So we have $C_i=C_i(\al_{j_i}^i)$, where $j_i=1,2,\dots,\frac{l_i(l_i-1)}2$,
and where $i=1,2,\dots, K$.

By enumerating them in the order of their occurence, i.e.\ first parameters of
matrix $C_1$, then $C_2$ etc., we can write:
$$\align
&B=B(\al_1,\al_2,\dots,\al_m,K_1,K_2,\dots,K_l)
\tag 2.16\\
&\text{where}\\
&m=\tfrac12\;\sum_{i=1}^Kl_i(l_i-1)
\endalign $$
and $K_1,K_2,\dots,K_l$ correspond to reflections.

Finally, we have
$$\gm\sim\lm=B(\al_1,\al_2,\dots,\al_m,K_1,K_2,\dots,K_l)\;V \tag 2.17 $$
After having matrix $B$ parametrized we deal with the form $Q$ in coordinates $V$,
and consequently with the
equation of the characteristic cone in these coordinates. So we have:
$$\sum_{i=1}^K\om_i\Bigl(\sum_{\mu=1}^{l_i}V^2_{p_ i+\mu}\Bigr)=0.
\tag 2.18 $$
Compare (2.12).

We find a parametric equation of this cone. In order to do this we write (2.18)
in the following form
$$
\sum_{j=1}^n\zeta_jV^2_j=0\quad\text{where \ $\zeta_j$ \ is one of \
$\om_i$}.
\tag 2.18' $$
The equation (2.18') is the equation of an $(n-1)$--dimensional quadric in
projective coordinates in  the canonical form [4].

Let us suppose that $V_{j_0}\ne0$ and divide both sides of (2.18') by $V_{j_0}^2$ and introduce
new coordinates
$$\ovj\sg_j=\frac{V_j}{V_{j_0}}\ ,\qquad j=1,2,\dots,n\ ,\quad j\ne j_0. $$
One gets
$$\sum\Sb j=1\\j\ne i\endSb ^n\zeta_j{\ovj\sg}{}^2_j+\zeta_{j_0}=0 \tag 2.19 $$
This equation can be written in parametric form
$$\aligned
&\ovj\sg_j=\ovj\sg_j\bigl(\ovj\tau_1,\ovj\tau_2,\dots,\ovj\tau_{n-2}\bigr)\\
&j=1,2,\dots,n,\quad j\ne j_0 \endaligned \tag 2.20 $$
$\ovj\tau_1,\ovj\tau_2,\dots,\ovj\tau_{n-2}$ are internal parameters of the
quadric whereas the functions
\newline
$\ovj\sg_j\bigl(\ovj\tau_1,\ovj\tau_2,\dots,\ovj\tau
_{n-2}\bigr)$ are expressed by either trigonometric or hyperbolic functions with
respect to the type of quadric (2.19) according to F.Klein's classification
[4]. We parametrize the vector $V$:
$$\aligned
&V_j=\ovj\sg_j\bigl(\ovj\tau_1,\ovj\tau_2,\dots,\ovj\tau_{n-2}\bigr)\ovj\tau_{n-1}
\quad, \qquad  j\ne j_0\\
\\
&V_{j_0}=\ovj\tau_{n-1}\quad,\qquad \tau_{n-1}\ne0 \endaligned \tag 2.21 $$
where $\ovj\tau_{n-1}$ is a new parameter.

In the case when $V_{j_0}=0$ we have
$$\sum\Sb j=1\\j\ne j_0\endSb ^n\zeta_jV^2_j=0\tag 2.22 $$
so in this case the dimension of quadric (2.22) is smaller than that of (2.18').

Similarly we choose $v_{j_1}\ne0$, $j_1\ne j_0$, $j_1=1,2,\dots,n$ and introduce new
variables
$$\ovd\sg_j=\frac{V_j}{V_{j_1}}\ ,\qquad j\ne j_1\ ,\quad j\ne j_0. \tag 2.23 $$
We obtain the equation (2.22) in the form
$$\sum\Sb j=1\\j\ne j_0\\ j\ne j_1\endSb ^n
\zeta_j{\ovd\sg}{}_j^2+\zeta_{j_1}=0. \tag 2.24 $$
Now we write its parametric form
$$\ovd\sg_j=\ovd\sg_j\bigl(\ovd\tau_1,\ovd\tau_2,\dots,\ovd\tau_{n-3}\bigr) $$
(one parameter less than above) and get:
$$\aligned
&V_j=\ovd\sg_j\bigl(\ovd\tau_1,\ovd\tau_2,\dots,\ovd\tau_{n-3}\bigr)
\ovd\tau_{n-2}
\ ,\qquad j\ne j_1\ ,\quad j\ne j_0\\
&V_{j_1}=\ovd\tau_{n-2}\ ,\quad \ovd\tau_{n-2}\ne0\\
&V_{j_0}=0 \endaligned \tag 2.25 $$
and we proceed likewise i.e. we consider the case $V_{j_0}=0=V_{j_1}$ and choose
$V_{j_2}\ne0$ etc. After $k$ steps of this procedure we get the following
equations for $V$:
$$\aligned
V_j=\ov{(k)}\sg_j&\bigl(\ov{(k)}\tau_1,\ov{(k)}\tau_2,\dots,\ov{(k)}\tau_
{n-(k+1)}
\bigr)\;\ov{(k)}\tau_{n-k}\ ,\quad \ov{(k)}\tau_{n-k}\ne0
\\
&\bigwedge_{\nu=0,1,\dots,k-2} V_{j_\nu}=0
\\\\
&V_{j_{k-1}}=\ov{(k)}\tau_{n-k}
\\
&\bigwedge_{\nu=0,1,2,\dots,k-1} j\ne j_\nu
\endaligned \tag 2.26 $$
It is easy to see that we exhaust all possibilities if $n=k_{\max}+1$.

Then we have
$$\aligned
&V_r=\ov{(n-1)}\sg\cdot \ov{(n-1)}\tau_1\quad ,\qquad \ov{(n-1)}\tau_1\ne0
\\
&V_{j_{n-1}}=\ov{(n-1)}\tau_1
\\
&\bigwedge _{\nu=0,1,\dots,n-2}r\ne j_\nu
\endaligned \tag 2.27 $$
and the rest of $V_i$ equals to zero.
The last case where $V=0$ is not interesting.

At this stage we can present a full parametrization of the covector $\lm$, which divides
the set of $\lm$'s into $(n-1)$ connected components. Since we have
$$\lm=B(\al_1,\al_2,\dots,\al_m,K_1,K_2,\dots,K_l)\;V\ ,$$
where $\al_1,\al_2,\dots,\al_m$ are generalized Euler angles, and $K_1,K_2,\dots,K_l$
are equal to $1$ or $(-1)$ and correspond  to reflections we also have
$$
\qquad\aligned
\ovj\gm\sim\ovj\lm=BV=B
\left(\matrix
\ovj\sg_1(\ovj\tau_1,\ovj\tau_2,\dots,\ovj\tau_{n-2})\\
\ovj\sg_2(\ovj\tau_1,\ovj\tau_2,\dots,\ovj\tau_{n-2})\\
\vdots\\
\ovj\sg_{j_0-1}(\ovj\tau_1,\ovj\tau_2,\dots,\ovj\tau_{n-2})\\
1\\
\ovj\sg_{j_0+1}(\ovj\tau_1,\ovj\tau_2,\dots,\ovj\tau_{n-2})\\
\vdots\\
\ovj\sg_n(\ovj\tau_1,\ovj\tau_2,\dots,\ovj\tau_{n-2})
\endmatrix\right)
\tau_{n-1}\endaligned\qquad\qquad
\aligned &\ \\ &j_0\endaligned
$$

(2.28)\hfill

$$
\aligned
\sim B(\al_1,\dots,\al_m,K_1,\dots,K_l)\cdot
\left(\matrix
\ovj\sg_1(\ovj\tau_1,\ovj\tau_2,\dots,\ovj\tau_{n-2})\\
\ovj\sg_2(\ovj\tau_1,\ovj\tau_2,\dots,\ovj\tau_{n-2})\\
\vdots\\
\ovj\sg_{j_0-1}(\ovj\tau_1,\ovj\tau_2,\dots,\ovj\tau_{n-2})\\
1\\
\ovj\sg_{j_0+1}(\ovj\tau_1,\ovj\tau_2,\dots,\ovj\tau_{n-2})\\
\ovj\sg_n(\ovj\tau_1,\ovj\tau_2,\dots,\ovj\tau_{n-2})
\endmatrix \right)\endaligned
 \qquad\qquad
 \aligned &\ \\&j_0\endaligned
$$
\vskip5pt
\flushpar
where $j_0$ is the number of the row in which $1$ appears.

\vskip10pt
It is easy to see that $\ovj\lm$ depends on $(m+n-2)$ arbitrary parameters and on
$l$
integers equal to $(\pm1)$. For $\ovd\gm$ we have:

%\newpage

$$
\ovd\gm\sim\ovd\lm\sim B(\al_1,\dots,\al_m,K_1,\dots,K_l)\cdot
\left(\matrix
\ovd\sg_1(\ovd\tau_1,\ovd\tau_2,\dots,\ovd\tau_{n-3})\\
\ovd\sg_2(\ovd\tau_1,\ovd\tau_2,\dots,\ovd\tau_{n-3})\\
\vdots\\
\ovd\sg_{j_0-1}(\ovd\tau_1,\ovd\tau_2,\dots,\ovd\tau_{n-3})\\
0\\
\ovd\sg_{j_0+1}(\ovd\tau_1,\ovd\tau_2,\dots,\ovd\tau_{n-3})\\
\vdots\\
\ovd\sg_{j_1-1}(\ovd\tau_1,\ovd\tau_2,\dots,\ovd\tau_{n-3})\\
1\\
\ovd\sg_{j_1+1}(\ovd\tau_1,\ovd\tau_2,\dots,\ovd\tau_{n-3})\\
\vdots\\
\ovd\sg_n(\ovd\tau_1,\ovd\tau_2,\dots,\ovd\tau_{n-3})
\endmatrix\right)
\tag 2.29 $$
which depends on $(m+n-3)$ arbitrary parameters and $l$ integers equal to $\pm1$.
Note that the $0$ appears in the $j_0{}^{\text{th}}$ row and that the $1$ appears in the
$j_1{}^{\text{th}}$ row.

In this way we find that $\ov{(n-k)}\gm$ depends on $(m+n-(k+1))$ free
parameters and $l$ integers equal to $(\pm1)$.

Then in the column on which $B(\al_1,\dots,\al_m,K_1,\dots,K_1)$ acts, zeroes
appear at the places of numbers $j_0,j_1,\dots,j_{k-2}$ and an integer 1 at
the place of number $j_{k-1}$. The last element
$$
\aligned
\ov{(n-1)}\gm\sim\ov{(n-1)}\lm\sim B(\al_1,\al_2,\dots,\al_m, K_1,\dots,K_l)\cdot
\left(\matrix
0\\ \vdots\\0\\ \ov{(n-1)}\sg\\ 0\\ \vdots\\ 0\\ 1\\ 0\\ \vdots\\0
\endmatrix \right)
\endaligned
\qquad
\aligned &r\\ &\ \\&j_{n-2}\endaligned
\tag 2.30
$$
(where $\ov{(n-1)}\sg\ne0$)
has only zeros in column $V$ except the places corresponding to the number $r$
and number $j_{n-2}$ and it depends on $m$ arbitrary parameters and $l$ integers
equal to $(\pm1)$.

The division and parametrization of simple elements is not unique, it depends on
the choice of sequence $j_0,j_1,\dots,j_{n-2}$. For any other choice it will
in general be different.

Now, let us consider the case where matrix $A$ has a zero eigenvalue. Let
$\om_1=0$ and have an order $l_1$. In such a case we may proceed as in the $(n-l_1)$--dimensional
case starting with a parametrization of the quadric (in projective coordinates)
$$\sum_{j=l_1+1}^n\zeta_jV^2_j=0 \tag 2.31 $$
assuming that $V_1,V_2,\dots,V_{l_1}$ are arbitrary, i.e. $V_j=\mu_j$,
$j=1,2,\dots,l_1$.

Thus we get a classification of simple elements of eq.(2.5). Let $F_i\subset\Cal E^*$
be a set of all simple elements belonging to one of these classes. If $\lm\in
F_i$ then $\lm$ is given by one of the formulas (2.28), (2.29) etc.
Thus in this way we get the theorem:

\proclaim The classification theorem.
Consider a partial differential equation of a nonelliptical type of the
second order with coefficients depending on first derivatives of $\Phi$
(i.e.\ quasilinear) and being smooth functions (at least of $C^1$ class)
of their variables,
$$ \sum\limits_{i,j=1}^n a_{ij} \Big( \frac{\partial \Phi}{\partial x^1},
\frac{\partial \Phi}{\partial x^2}, \ldots, \frac{\partial \Phi}{\partial
x^n} \Big) \frac{\partial^2 \Phi}{\partial x^i \partial x^j} =0 $$
where $ \Phi \colon D \subset R^n \to R^1$ is a real function of $n$
variables of $C^{(2)}$ class, and $D$ a domain of $\Phi$ is an open set.
All solutions of the above equation of rank~1 can be described by simple
elements obtained according to the procedure mentioned above. Rank~1
solution means that
$$ \text{rank}\; \Big| \frac{\partial^2 \Phi}{\partial x^i \partial x^j} \Big| =1 $$
These solutions are called simple waves (Riemann waves).

An outline of the proof has been given above.
Thus we have
found many types of simple elements depending on various number of free parameters.
Each of them has its own type of simple wave. Hence, according to
relations (2.4) we have
$$\aligned
&u=\ov{(i)}f(R^i) \\
&R^i=\vphi(\lm_\nu^ix^\nu),
\endaligned\tag 2.32
$$
where
$$\aligned
&\ov{(i)}{\frac{df}{dR^i}}=\ov{(i)}\lm\bigl(f^i(R^i)\bigr)\\
& \ov{(i)}\lm\in F_i,\endaligned
$$
and where $i$ enumerates the types of simple waves admissible by the (2.5).
Because of the free parameters, which are functions of $u$, and consequently of $R^i$ we may integrate
(2.32) and get its exact solutions. It is easy to see that $F_i$ is a
smooth submanifold
of $\Cal E^*$ whose dimension is equal to the number of parameters.
Thus we get the classification theorem for simple waves of the equation (2.5).
The solutions of (2.3) (simple waves) may have a gradient catastrophe.

%\newpage
%#########################################
%           SEC 3
%#########################################	
\vskip20pt
\centerline{\bf 3. Equation for potential  stationary flow}
\centerline{\bf of a compresible perfect gas}
\centerline{\bf (the first example of an application of the
method)}
\vskip\baselineskip

This section is devoted to the equation of potential stationary flow of
a compressible perfect gas.
We consider  the equation of potential of velocity for
a stationary flow of perfect gas
$$\aligned
&(c^2-\Phi_x^2)\Phi_{xx}+(c^2-\Phi_y^2)\Phi_{yy}+(c^2-\Phi_z^2)\Phi_{zz}+\\
&-2(\Phi_x\Phi_y\Phi_{xy}+\Phi_x\Phi_z\Phi_{xz}+\Phi_y\Phi_z\Phi_{yz})=0
\endaligned \tag 3.1 $$
where
$$\aligned
& c^2=c_0^2-\frac{\kappa-1}2\bigl(\Phi^2_x+\Phi_y^2+\Phi_z^2\bigr)
\quad\text{and}\quad \vec u=\vec\nabla\Phi, \\
& \Phi \colon  D \subset R^3 \to R \quad\text{and}\quad \Phi \in
C^{(2)}_D, \quad D \hskip5pt\text{open} \endaligned  \tag 3.2 $$
($\Phi$ is the velocity potential and $c$ the velocity of sound).
\newline The equation (3.1) is interesting for us in the supersonic region
$$0\le c^2\le \Phi_x^2+\Phi_y^2+\Phi_z^2 \ .\tag 3.3 $$
In this region the equation is nonelliptical and we have [5]
$$
c_0^2\Bigl(\frac2{\kappa+1}\Bigr)\le\Phi_x^2+\Phi_y^2+\Phi_z^2\le c_0^2\Bigl(
\frac2{\kappa-1}\Bigr). \tag 3.4 $$
Now we look for the solutions of (3.1) provided that (3.4) is satisfied. Using
the fact that for a perfect gas $c^2=\dfrac{\kappa p}{\rho_0}$
(the so called adiabatic sound), from
the adiabatic equation $p=a\rho_0{}^\kappa$~, \ $a=\text{const.}$ we can
calculate $\rho_0$ and $p$ from (3.2)
$$\aligned
&\rho_0=\Bigl[\frac1{a\kappa}\;\Bigl(c_0^2-\frac{\kappa-1}2\bigl(\Phi_x^2+\Phi_y^2+\Phi_z^2\bigr)\Bigr)\Bigr]
^{\tfrac1{(\kappa-1)}}\\
\\
&p=\Bigl[\frac1{a\kappa}\;\Bigl(c_0^2-\frac{\kappa-1}2\bigl(\Phi_x^2+\Phi_y^2+\Phi_z^2\bigr)
\Bigr)\Bigr]^{\tfrac\kappa{(\kappa-1)}}
\endaligned \tag 3.5$$
We are interested in  solutions of the equation (3.1) that are given in terms
of the Riemann invariants. Following the method we transform (3.1), by introducing
some new variables, into a quasilinear system of equations of the first
order:
$$\aligned
&(c^2-\vphi_1^2)\vphi_{1,x}+(c^2-\vphi_2^2)\vphi_{2,y}+(c^2-\vphi_3^2)\vphi_{3,z}+
\\
&-2(\vphi_1\vphi_2\vphi_{1,y}+\vphi_1\vphi_3\vphi_{1,z}+\vphi_2\vphi_3\vphi_{2,z})=0
\\
&\vphi_{1,y}-\vphi_{2,x}=0\\
&\vphi_{1,z}-\vphi_{3,x}=0\\
&\vphi_{2,z}-\vphi_{3,y}=0
\endaligned \tag 3.6 $$
where
$$\Phi_{x}=\vphi_1\ ,\quad\Phi_{y}=\vphi_2\ ,\quad\Phi_{z}=\vphi_3. \tag 3.7 $$
Thus equation (3.1) is transformed into the over--determined system of 4
equations for 3 functions.

Reclall that the equation (3.1) can be derived in the following way from a mass conservation
law
$$\text{div}\;(\rho_0\vec v)=\rho_0\;\text{div}\;\vec v+\vec v\cdot\vec\nabla
\rho_0=0\ .
\tag 3.8$$
By means of the Euler equation
$$(\vec v\cdot\vec\nabla)\vec v=-\frac{\vec\nabla p}{\rho_0}=
-\frac{c^2\vec\nabla\rho_0}{\rho_0}\tag 3.9$$
we obtain
$$c^2\;\text{div}\;\vec v-\vec v(\vec v\cdot\vec\nabla)\vec v=0\ .\tag 3.10$$
By introducing a potential according to equation (3.2) and substituting it to
(3.10) we find equation (3.1).

%####################################
%      SEC 4
%####################################
%\newpage
\vskip20pt
\centerline{\bf 4. Simple integral elements }
\centerline{\bf (for the first example of an application of
the method)}
\vskip\baselineskip

In this section we find simple integral elements for the equation of potential
stationary flow of a compressible perfect gas.

Now we write the characteristic equation of the system (3.6) in the form (2.8) introducing the covector $\lm$:
$$\aligned
\\
Q(\lm_1,\lm_2,\lm_3)&=(c^2-\vphi_1^2)\lm_1^2+(c^2-\vphi_2^2)\lm_2^2+(c^2-\vphi_3^2)\lm_3^2+
\\
&-2(\vphi_1\vphi_2\lm_1\lm_2+\vphi_1\vphi_3\lm_1\lm_3+\vphi_2\vphi_3\lm_2\lm_3)=0.
\\
\endaligned \tag 4.1 $$
Following the procedure described  in section 2 we parametrize the covector $\lm$. To do this
we are going to diagonalize the form (4.1) and write the secular equation
$$
\det(A-\mu I)=0,\quad\text{where}\quad
A=\left(\matrix
(c^2-\vphi_1^2)&-\vphi_1\vphi_2&-\vphi_1 \vphi_3\\\\
-\vphi_1\vphi_2&(c^2-\vphi_2^2)&-\vphi_2\vphi_3\\\\
-\vphi_1\vphi_3&-\vphi_2\vphi_3&(c^2-\vphi_3^2)\endmatrix\right).
\tag 4.2 $$
From (4.2) we get the third order equation for a value $\mu$
$$(c^2-\mu)^2\bigl(c^2-(\vphi_1^2+\vphi_2^2+\vphi_3^2)-\mu\bigr)=0 \tag 4.3 $$
Hence one obtains two different eigenvalues
$$\mu_1=c^2\ ,\quad  \mu_2=c^2-(\vphi_1^2+\vphi_2^2+\vphi_3^2) \tag 4.4 $$
from which the first  has order 2. Thus the quadratic form $Q$ reduced to canonical
form is:
$$Q(\lm,\lm)=c^2(y^2_1+y^2_2)+\bigl(c^2-(\vphi_1^2+\vphi_2^2+\vphi_3^2)\bigr)
y^2_3=0\ . \tag 4.5 $$
Now we find parametric equations for (4.5). Supposing that $y_1\ne0$, we simply get
$$
\frac{X^2}{\bigl(\tfrac ac\bigr)^2}-Y^2=1\qquad\text{where}\quad
X=\frac{y_2}{y_1}\quad ,\quad Y=\frac{y_3}{y_1}
\tag 4.6 $$
and where $a^2=\vphi_1^2+\vphi_2^2+\vphi_3^2-c^2>0$ (supersonic flow).

Then we have
$$Y=\text{sh}\;\rho\quad,\quad X=\frac ac\;\text{ch}\;\rho \tag 4.7 $$
i.e. the parametric equation of a hyperbola.

If $y_1=0$ then one gets
$$\align
&c^2y_2^2-a^2y_3^2=0\tag 4.8\\
&y_2=\ep \;\frac ac\; y_3\ ,\quad \ep=\pm1. \tag 4.9 \endalign $$
In our case there is only one eigenvalue of order larger than 1, i.e. 2. Thus
the diagonalizing matrix $B$ will depend on only one parameter $\al$, $B=B(\al)$.
The matrix $B$ may easily be built from eigen vectors of the matrix $A$ and one obtains
$$
\align
\\
B(\al,l)&\text{=}\left(\matrix
\dfrac{\text{--}1}{\chi_2}\bigl(\vphi_2\sin\al\text{+}
        \dfrac{\vphi_1\vphi_3}{\chi_1}\cos\al\bigr)
&\dfrac1{\chi_2}\bigl((\text{--}1)^K\vphi_2\cos\al\text{+}(\text{--}1)^{K\text{+}1}\dfrac{\vphi_1\vphi_3}{\chi_1}\sin\al\bigr)
\!\!&\dfrac{\vphi_1}{\chi_1}
\\\\ \\
\dfrac1{\chi_2}\bigl(\vphi_1\sin\al\text{--}\dfrac{\vphi_2\vphi_3}{\chi_1}\cos\al\bigr)
&\dfrac{(\text{--}1)^{K\text{+}1}}{\chi_2}\bigl(\vphi_1\cos\al\text{+}\dfrac{\vphi_2\vphi_3}{\chi_1}\sin\al\bigr)
\!\!&\dfrac{\vphi_2}{\chi_1}
\\\\ \\
\dfrac{\chi_2}{\chi_1}\cos\al
&\dfrac{(\text{--}1)^K\chi_2}{\chi_1}\sin\al
\!\!&\dfrac{\vphi_3}{\chi_1}\endmatrix \right)
\\\\
K&=0,1\ ,\quad l=(-1)^K \tag 4.10\endalign $$
where $\al$ is a parameter depending on $\vphi_1,\vphi_2,\vphi_3$ and
$$
\chi_1^2=\vphi_1^2+\vphi_2^2+\vphi_3^2\quad,\quad \chi_2^2=\vphi_1^2+\vphi_2^2
.$$
The integer $l$ equals $(\pm1)$ and is associated with a reflection in two dimensional space,
whereas $\al$ is associated with a rotation. According to section 1 we have
$$
\ovj\gm\sim\ovj\lm=B(\al,l)\left(\matrix 1\\X\\Y\endmatrix\right)\tag 4.11 $$
and $$
\ovd\gm\sim\ovd\lm=B(\al,l)\left(\matrix 0\\1\\\sg\endmatrix\right)\quad
,\quad\sg=\frac{\ep a}c \tag 4.12 $$
Then by inserting (4.10) into (4.11) and (4.12) we get explicitly the form of
the simple
elements. These elements are presented in Table 1 ($F_1$ and $F_2$ of
Appendix~A).

According to the results given in section 2 \ $\ovj\lm$ depends on two parameters $\al$ and $\rho$,
whereas $\ovd\lm$ depends only upon $\al$.

Let us consider the form (4.5) once more, now assuming that $y_3\ne0$ and let us introduce
$$X=\frac{y_2}{y_3}\quad,\quad Y=\frac{y_1}{y_3}. \tag 4.13 $$
Then we have
$$X^2+Y^2=\Bigl(\frac ac\Bigr)^2 \tag 4.14 $$
so
$$X=\Bigl(\frac ac \Bigr)\sin\rho\quad,\quad Y=\Bigl( \frac ac \Bigr)\cos\rho
\tag 4.15 $$
i.e. a parametric equation of a circle, where $\rho$ is a function of $\vphi_i$,
\ $i=1,2,3$. If $y_3=0$ we obtain $y_1=y_2=0$, thus a zero case.
\newline We have
$$\gm_1\sim\lm^{(1')}=B(\al,l)\;\left(\matrix Y\\X\\1\endmatrix\right). \tag 4.16$$
By inserting (4.10) and (4.15) into (4.16) we obtain the explicit form of
the simple elements. They are presented in Table~2 ($F_{1'}$ of
Appendix~A). $\lm^{(1')}$ depends on two parameters: $\al$ and $\rho$.

However, in one particular case a degeneration occurs and for  $K=0$ we get only
one parameter $\beta=\rho+\al$, whereas for $K=1$, $\rho-\al=\om$. In this
case $\lm^{(1')}$ depends, really, on only one parameter.

Let us consider (4.5) again, assuming that $y_2\ne0$. We introduce new coordinates
$$
Y=\frac{y_3}{y_2}\quad,\quad X=\frac{y_1}{y_2} \tag 4.17 $$
and obtain from (4.5)
$$\frac{Y^2}{\bigl(\tfrac ac\bigr)^2}-X^2=1 \tag 4.18 $$
and in a parametric form
$$\aligned
&Y=\frac ac\;\text{ch}\;\rho\\
&X=\text{sh}\;\rho. \endaligned \tag 4.19 $$
Thus we have
$$ \gm_{1''}\sim\lm^{(1'')}=B(\al,l)\left(\matrix
X\\1\\Y\endmatrix\right). \tag 4.20 $$
If $y_2=0$ we obtain from (4.5)
$$\align
&c^2y_1-a^2y_3=0\tag 4.21 \\
&y_1=\ep\frac ac\;y_3\quad,\quad \ep^2=1 \tag 4.22
\endalign $$
and
$$\gm_{2''}\sim\lm^{(2'')}=B(\al,l)\left(\matrix\sg\\0\\1\endmatrix\right)\quad
\text{where}\quad \sg=\frac{\ep a}c\, . \tag 4.23 $$
Inserting (4.10) into (4.20) and (4.23) we derive the explicit form of the simple
elements $\lm^{(1'')}$ and $\lm^{(2'')}$ which are presented in Table~3
($F_{1''}$ and $F_{2''}$ of Appendix~A).

Thus we have found here several types of simple elements which are used to construct
solutions, i.e. simple waves and their interactions, the so--called double and multiple waves.

\proclaim Theorem 1.
All simple elements of the equation for a potential stationary flow of a
perfect gas in a supersonic region fall into the classes: $F_1$, $F_2$,
$F_3$, $F_4$, $F_{1'}$, $F_{2'}$, $F_{1''}$, $F_{2''}$, $F_{3''}$.

In Appendix A we present all classes of simple elements -- $F_1$, $F_2$, $F_3$,
$ F_4$, $F_{1'}$, $F_{2'},$
$F_{1''}$, $F_{2''}$, $F_{3''}$ for (3.1).

%\vskip20pt
%########################
%     SEC 5
%########################
\vskip2\baselineskip
\centerline{\bf 5. Simple waves }
\centerline{\bf (the first example of an application of the
method)}
\vskip\baselineskip

This section is devoted to simple waves for
the equation of potential stationary flow of a compressible perfect gas.

Now we present the simplest solutions of the system (3.1) namely those that have been
constructed on the basis of homogeneous simple integral elements. The method of finding these
solutions is presented in papers [1], [2], [3]. In this section we deal with a method of solving
equation (2.3), using the results of section 2 about the parametrization of
simple elements.

The crucial point of our method is the freedom  of choice
of parameters occuring in simple elements. According to the terminology presented in [1],
[2] the elementary solution of the homogeneous system has been called a simple
wave. Those solutions may be interpreted as waves since they are moving disturbances,
the profile of which changes in a course of propagation (a sign of this is the implicit
form of the relation (2.4) for the $R(x)$ function). The form of the solution of
(2.4) suggests that the covector $\lm$ may be regarded as equivalent to the wave vector
$(\om,\vec k)$ which specifies the velocity and direction of propagation of the wave.
The specific profile of a simple wave is explicity determined by its initial data
but there is a certain amount of freedom of choice of one free function, which is a function of one variable.
The above remarks concern all simple waves which have been  found.

There
also exists another freedom, associated with parameters in simple elements in our method. This freedom
is of another kind and has a different origin. Using it we may integrate equation (2.3) and obtain
solutions with $(q+1)$ arbitrary function of one variable, where $q$ is
the number of independent parameters in the simple elements. For all functions we obtain
a certain restriction and it seems to be an interesting feature of the method.

The simple
wave obeys conditions (2.3) and (2.4). By substitution of the simple integral element (4.11)
into equation (2.3), we obtain for $K=0$
(i.e. $\underset{(K=0)
}\to{\overset{(1)}\to\lm}-F_1$ see Table 1 of Appendix A)

%\newpage
$$\aligned
&\frac{d\vphi_1}{dR}\text{=}\frac{(\chi_1^2\text{--}c^2)^{1/2}}{\chi_2}
  \Bigl[\text{--}\Bigl(\vphi_2\sin\al\text{+}\frac{\vphi_1\vphi_3}{\chi_1}\cos\al\Bigr)\text{+}
  \Bigl(\vphi_2\cos\al\text{--}\frac{\vphi_1\vphi_3}{\chi_1}\sin\al\Bigr)\text{sh}\;\rho\Bigr]
  \text{+}\frac{c\vphi_1}{\chi_1}\text{ch}\;\rho
\\\\
&\frac{d\vphi_2}{dR}\text{=}\frac{(\chi_1^2\text{--}c^2)^{1/2}}{\chi_2}
  \Bigl[\Bigl(\vphi_1\sin\al\text{--}\frac{\vphi_2\vphi_3}{\chi_1}\cos\al\Bigr)\text{--}
  \Bigl(\vphi_1\cos\al\text{+}\frac{\vphi_2\vphi_3}{\chi_1}\sin\al\Bigr)\text{sh}\;\rho\Bigr]
  \text{+}\frac{c\vphi_2}{\chi_1}\text{ch}\;\rho
\\\\
&\frac{d\vphi_3}{dR}\text{=}\frac{\chi_2(\chi_1\text{--}c^2)^{1/2}}{\chi_1}\Bigl[\cos\al\text{+}
   \sin\al\;\text{sh}\;\rho\Bigr]\text{+}\frac{c\vphi_3}{\chi_1} \text{ch}\;\rho
. \endaligned \tag 5.1
$$
Now we introduce the new dependent variables $\chi_1,\chi_2$ and $\mu_0=
\dfrac{\vphi_1}{\vphi_2}$. We get from the system (5.1):

$$
\aligned
&\frac{d\chi_1}{dR}=c\;\text{ch}\;\rho\ge c\\\\
&\frac{d\chi_2}{dR}=\frac1{\chi_1}\bigl[-\ep_1(\chi_1^2-\chi_2^2)^{1/2}
(\chi_1^2-c^2)^{1/2}(\cos\al+\sin\al\;\text{sh}\;\rho)+c\chi_2\;
\text{ch}\;\rho\bigr]\\\\
&\frac d{dR}\;\text{arc}\;\text{tg}\;\mu_0=\frac{(\chi_1^2-c^2)^{1/2}}{\chi_2}\;(\cos\al
-\sin\al) .
\endaligned \tag 5.2 $$
Let us observe that the quantities $\al$ and $\rho$, entering in the equations
(5.2), are arbitrary functions of $R$. Hence, it is convenient to shift the arbitrariness
from $\al$ and $\rho$ to $\chi_1$ and $\chi_2$. The right--hand sides of the equations
for $\chi_1$ and $\chi_2$ depend on $\al$ and $\rho$ but they do not contain derivatives
of these functions. Thus assuming that $\chi_1$ and $\chi_2$ are arbitrary we obtain equations
for $\al$ and $\rho$. These equations will express $\al$ and $\rho$ in terms of
the functions $\chi_1$
and $\chi_2$ and their derivatives with respect to $R$. The condition of
solvability of algebraic (or transcendental) equations for $\al$ and $\rho$
provides us with restrictions on $\chi_1$ and $\chi_2$.

In this way the problem
of solving algebraic equations for $\chi_1$ and $\chi_2$ is reduced to the problem
of solving algebraic equations and searching for restrictions for the arbitrary functions
$\chi_1$ and $\chi_2$. We will derive a set of such restrictions. After expressing
$\al$ and $\rho$ in terms of new arbitrary functions with restrictions we insert
these relations into the third equation of (5.2) and in a covector (4.11). In this way,
on the right--hand side of this equation we have the functions of $R$ and we can integrate the equation
$$\align
\mu_0=\text{tg}\Bigl(\int_{R_0}^R\frac{(\chi_1^2-c^2)^{1/2}}{\chi_2}\;(\cos\al-\sin\al)
&dR'+d_0\Bigr)
\tag 5.3 \\
&d_0=\text{const.}\endalign $$
Solving equations
$$
\mu_0=\frac{\vphi_1}{\vphi_2}\quad,\quad
\chi_1^2=\vphi_1^2+\vphi_2^2+\vphi_3^2\quad,\quad
\chi_2^2=\vphi_1^2+\vphi_2^2 $$
we find the solution which is a simple wave
$$\left\{\aligned
&\vphi_1=\vphi_1(R)\\
&\vphi_2=\vphi_2(R)\\
&\vphi_3=\vphi_3(R)\\
&R=\Psi\bigl(\ovj\lm_1x+\ovj\lm_2y+\ovj\lm_3z\bigr).
\endaligned\right. \tag 5.4 $$
It is easy to see that the conditions for $\chi_1$ and $\chi_2$ imply some restrictions
on the range of parameter $R$ and consequently (see (5.4) ) a restriction
on the function $\Psi$.
This restriction is responsible for the fact that the function $\Psi$ has its range
in a certain subset of the real axis.

Now let us realize this programme. For the convenience of calculations we assume that
$$\chi_1^2=e^{2H}\quad,\quad \chi_2^2=e^{2G}\qquad\text{where} \ \ e^{2G}\le
e^{2H}
\tag 5.5 $$
and we find restrictions for $H$ and $G$ ($H$ and $G$ are functions of $R$).

The first restriction for $e^{2H}$ is of course (3.4) and we have:
$$c_0^2\Bigl(\frac2{\kappa+1}\Bigr)\le e^{2H}\le
c_0^2\Bigl(\frac2{\kappa-1}\Bigr). \tag 5.6 $$
By substituting (5.5) into (5.2)${}_1$ and (5.2)${}_2$ and then using the relations between trigonometric
and hyperbolic functions we get:
$$\aligned
&\text{ch}\;\rho=\frac{e^H}c\;\frac{dH}{dR}\\\\
&t^2-\frac{2AB}{B^2+1}\;t+\frac{A^2-1}{B^2+1}=0 \endaligned \tag 5.7 $$
where
$$\aligned
&A=\frac{e^{(G+H)}\frac d{dR}\;(G-H)}{(e^{2H}-e^{2G})^{1/2}(e^{2H}-c^2)^{1/2}}
\ ,\\\\
&B=\text{sh}\;\rho=\ep_2\Bigl(\frac{e^{2H}}{c^2}\Bigl(\frac{dH}{dR}\Bigr)^2-1\Bigr)^
{1/2}\\\\
&t=\sin\al\quad,\quad \cos\al=\ep(1-t^2)^{1/2}\\\\
&\ep_2^2=\ep_3^2=1.
\endaligned \tag 5.8 $$
Thus, we get the first restrictions:
$$\aligned
&1^0.\quad 1\le \text{ch}\;\rho=\frac{e^H}c\;\frac{dH}{dR}\\
&2^0.\quad |t|\le1 . \endaligned \tag 5.9 $$
At the same time the quadratic equation (5.7)${}_2$ must have real roots, so its
discriminant cannot be negative. The latter condition yields
$$B^2-A^2+1\ge 0. \tag 5.10$$
By inserting $A$ and $B$ into (5.10) we obtain the following condition
$$\aligned
&\al_1z_1^2-\al_2z_2^2+\al_3z_1z_2\ge0\\
&z_1=\frac{dH}{dR}\quad,\quad z_2=\frac{dG}{dR} \endaligned \tag 5.11 $$
where
$$
\align
&\al_1=e^{4H}-e^{2(H+G)}-c^2e^{2H}\\
&\al_2=c^2e^{2G}\ge 0\quad,\quad \al_3=2c^2e^{2G}\ge 0. \endalign $$
Supposing that $G\ne\text{const}$, we introduce the new variable $z_3=\frac{z_1}{z_2}$
and we obtain:
$$\al_1z_3^2+\al_3z_3-\al_2\ge0. \tag 5.12 $$
Now we present the conditions for which (5.12) is always satisfied regardless
of the value of $z_3$. By doing this we avoid differential inequalities that are hard
to satisfy.
These conditions are the following: the discriminant $\Delta_1$ of the quadratic equation (5.12)
must be non--positive, whereas the coefficient of $z^2_3$ must be positive. So, we have:
$$\align
&0\ge\Delta_1=-e^{4H}+c^2e^{2G}+e^{2(G+H)}-c^2e^{2H}\tag 5.13\\
&0<\al_1=e^{4H}-e^{2(H+G)}-c^2e^{2H}. \tag 5.14 \endalign $$
The condition (5.13) is stronger than (5.14) and it is sufficient to
fulfill only
(5.13). By inserting (5.5) and (3.2) into (5.13) we obtain
$$0<(\kappa+1)e^{4H}+(\kappa-3)e^{2(H+G)}-2c_0^2(e^{2G}+e^{2H}). \tag 5.15 $$
Using (5.6) we easily conclude that $2c_0^2-(\kappa-3)e^{2H}>0$, so finally we have:
$$
e^{2G}\ge\Biggl[\frac{(\kappa-1)e^{2H}-2c_0^2}{2c_0^2-(\kappa-3)e^{2H}}\Biggr]\;e^{2H}
. \tag 5.16 $$
Let us turn now to condition $2^0$. We solve  the quadratic equation with respect to $t$
and we get
$$
t_{1,2}=\frac{AB\pm(B^2-A^2+1)^{1/2}}{B^2+1}. \tag 5.17 $$
We choose the root with the smaller modulus
$$
t=\ep_2\frac{|A|\cdot|B|-(B^2-A^2+1)^{1/2}}{B^2+1}\ ,\quad \ep_2^2=1. \tag 5.17'
$$
Now we prove:

\proclaim  Lemma.
$$\frac{\bigl||A|\cdot|B|-(B^2-A^2+1)^{1/2}\bigr|}{B^2+1}\le 1 \tag 5.18 $$
when $B^2-A^2+1\ge 0$.

\pr
Let us consider the two cases
$$
\aligned
&\text{a)}\quad |A|\cdot|B|-(B^2-A^2+1)^{1/2}\le0\\
&\text{b)}\quad |A|\cdot|B|-(B^2-A^2+1)^{1/2}\ge 0.\endaligned \tag 5.19 $$

\vskip8pt
\flushpar a)\quad We start from an obvious inequality $(|A|+|B|)^2\ge0$ and we have
$$(A^2+B^2+2|A|\cdot|B|)(B^2+1)\ge0 \tag 5.20 $$
i.e.
$$A^2B^2+A^2+B^4+B^2+2|A|\cdot|B|+2|A|\cdot|B|B^2\ge0 \tag 5.21 $$
or
$$
B^2\text{--}A^2+1\le B^4+1+A^2B^2+2|A|\cdot|B|+2B^2+2|A|\cdot|B|^3\text{=}
(B^2+1+|A|+|B|)^2 \tag 5.22 $$
and from this, $\sqrt{B^2-A^2+1}-|A|\cdot|B|\le(B^2+1)$ or
\vskip5pt
$$
\frac{\bigl|(B^2-A^2+1)^{1/2}-|A|\cdot|B|\bigr|}{B^2+1}\le1. \tag 5.23 $$
\vskip10pt
\flushpar b)\quad We have $B^2-A^2+1\ge0$ that is
$$A^2\le B^2+1 \ .\tag 5.24 $$
So
$$\align
&A^2B^2\le B^2(B^2+1)\le (B^2+1)^2 \tag 5.25\\
&|A\cdot B|\le(B^2+1) \tag 5.25' \endalign $$
Since $|AB|\ge(B^2-A^2+1)^{1/2}$ we have
$$\Biggl|\frac{|AB|-(B^2-A^2+1)^{1/2}}{B^2+1}\Biggr|\le1. \tag 5.26 $$
So the proof has been established.
\vskip10pt

Summing up, we see that $|t|\le 1$, so $2^0$ is satisfied.

Now let us turn to condition $1^0$. This is a differential inequality which may
be solved by applying well known results [6]. We have
$$
\frac{e^H}{c(e^H)}\;\frac{dH}{dR}\ge1\quad,\quad H(R_0)=H_0 \tag 5.27 $$
and the function $P$ obeys the equation
$$
\frac1{c(P)}\;\frac{dP}{dR}=1\qquad\text{and}\quad P(R_0)=e^{H_0}.
\tag 5.27a $$
Using the definition of $c$ we can write (5.27) and (5.27a) in the following form
$$
\frac d{dR}\Bigl[\sqrt{\tfrac2{\kappa-1}}\arcsin\Bigl(\frac
{\sqrt{\frac{\kappa-1}2}\;e^H}{c_0}
\Bigr)\Bigr]\ge1 \tag 5.28 $$
and $H(R_0)=H_0$
$$\aligned
&\frac d{dR}\Bigl[\sqrt{\tfrac2{\kappa-1}}\arcsin\Bigl(
\frac{\sqrt{\frac{\kappa-1}2}\;P}{c_0}
\Bigr)\Bigr]=1\\\\
&P(R_0)=e^{H_0}. \endaligned \tag 5.28a
$$
From (5.28a) we get
$$
P(R)=c_0\sqrt{\tfrac{2}{\kappa-1}}\;\sin\Bigl[\sqrt{\tfrac2{\kappa-1}}\;(R-R_0)+
\arcsin\Bigl(\sqrt{\tfrac{\kappa-1}2}\;\frac{e^{H_0}}{c_0}\Bigr)\Bigr]
 \tag 5.29 $$
We have
$$\sqrt{\tfrac2{\kappa-1}}\arcsin\Bigl(\frac{\sqrt{\frac{\kappa-1}2}\;e^{H(R)}}{c_0}\Bigr)\ge
\sqrt{\tfrac2{\kappa-1}}\;\arcsin\Bigl(\frac{\sqrt{\frac{\kappa-1}2}\;P(R)}{c_0}\Bigr)
\tag 5.30 $$
and
$$e^{H(R_0)}=P(R_0)=e^{H_0} \tag 5.31 $$
(see e.g. [6]) hence
$$\align
&e^H\ge c_0\sqrt{\tfrac 2{\kappa-1}}\;\sin\Bigl[\sqrt{\tfrac2{\kappa-1}}\;(R-R_0)+\arcsin\Bigl(
\frac{\sqrt{\frac{\kappa-1}2}e^{H_0}}{c_0}\Bigr)\Bigr] \tag 5.31a\\\\
&e^{H(R_0)}=e^{H_0}. \endalign $$
Simultaneously we have
$$\align
&c_0\sqrt{\tfrac2{\kappa+1}}\;\le \;e^{H(R)}\;\le\; c_0\sqrt{\tfrac2{\kappa-1}}
 \tag 5.32\\\\
&c_0\sqrt{\tfrac2{\kappa+1}}\;\le\; e^{H_0}\; \le\; c_0\sqrt{\tfrac2{\kappa-1}} \tag 5.33
\endalign $$
Since $P(R_k)=c_0\sqrt{\frac 2{\kappa-1}}$ , for
%.....................
$$\align
&R_k=R_0+\sqrt{\tfrac{\kappa-1}2}\Bigl(\frac\pi2+2k\pi-\arcsin\Bigl(\frac1{c_0}\;
\sqrt{\tfrac{\kappa-1}2}\;e^{H_0}\Bigr)\Bigr) \tag 5.34
\\\\
&k=0,\pm1,\pm2,\dots  \endalign $$
we have
$$e^{H(R_k)}=c_0\sqrt{\tfrac2{\kappa-1}}. \tag 5.35 $$
We have of course
$$|R_{k+1}-R_k|=\sqrt{2(\kappa-1)}\;\pi \tag 5.36 $$
Hence, the function $e^{H(R)}$ is defined in the interval $[R_0,+\infty)$ for an
established but arbitrary~$R_0$.

At points $R_k$~, \ $c^2=0$, the magnitude of the vector of velocity of flow
is maximal, and pressure and density simultaneously become equal to zero. Thus, the solution of the system (5.1) is
$$\aligned
&\vphi_1=\eta_2e^G\sin K(R)\\&\vphi_2=\eta_2e^G\cos K(R)\\&\vphi_3=\eta_1\sqrt{e^{2H}
-e^{2G}}\endaligned
\qquad
\aligned \eta_1^2=\eta_2^2=1\endaligned \tag 5.37 $$
where
$$\align
&K(R)=\int_{R_0}^R(\cos\al-\sin\al)\;\frac{(e^{2H}-c^2)^{1/2}}{e^G}\;dR'+c'
\\\\
&c'=\text{const}\quad,\quad R=\Psi\bigl(\ovj\lm_1x+\ovj\lm_2y+\ovj\lm_3z\bigr)\quad,\quad
R\in [R_0,+\infty)=L. \endalign $$
The smooth function $\Psi$ takes values only from the interval $L$. The conditions (5.16),
(5.31a), (5.32) and (5.33) are implied on functions $G$ and $H$.

For $\cos\al$ and $\sin\al$ we have the following expressions:
$$\align
&\sin\al=
\ep_1\Biggl\{\frac
{e^G\dfrac{d}{dR}(H-G)\Bigl(\dfrac{e^{2H}}{c^2}\Bigl(\dfrac{dH}{dR}\Bigr)^2
-1\Bigr)^{1/2}-\Bigl[\Bigl(\dfrac{dH}{dR}\Bigr)^2(e^{2H}-e^{2G})\dfrac{(e^{2H}-c^2)}
{c^2}+}
%.........
{\dfrac1{c^2}\;e^H\Bigl(\dfrac{dH}{dR}\Bigr)^2(e^{2H}-e^{2G})^{1/2}\cdot(e^{2H}-c^2)
^{1/2}}
\\\\\\
&\qquad -\frac{e^{2G}\Bigl(\dfrac d{dR}(H-G)\Bigr)^2\Bigr]^{1/2}}{\ }\Biggr\}
\endalign$$

$$\align
&\cos\al=\ep_3 (1-\sin\al)^{1/2}
 \tag 5.38
\endalign$$

$$\aligned
&\text{ch}\;\rho=\frac{e^H}c\;\frac{dH}{dR}\\\\
&\text{sh}\;\rho=\ep_2\Bigl(\frac{e^{2H}}{c^2}\Bigl(\frac
{dH}{dR}\Bigr)^2-1\Bigr)^{1/2}  \endaligned \tag 5.39 $$
where $\ep_1^2=\ep_2^2=\ep_3^2=1$.
Of course, expressions (5.38) and (5.39) should be inserted in the covector $\ovj\lm$,
as in expressions (5.37).

Now let us consider the following equation
$$\align
&R_k=\Psi\bigl(\ovj\lm_1(R_k)x+\ovj\lm_2(R_k)y+\ovj\lm_3(R_k)z\bigr)\tag 5.40\\
&R_k\in L \endalign $$
where $R_k$ is given by (5.34).
For a given $k$ equation (5.40) describes one of several planes with a normal vector
$\vec\lm(R_k) = \bigl(\ovj\lm_1(R_k), \ovj\lm_2(R_k), \ovj\lm_3(R_k)\bigr)$. If
the equation $R_k=\Psi(r)$ possesses $n$ roots
$r_i$, $i=1,2,\dots,n$, $\Psi(r_i)=R_k$,
then there are planes:
$$\align
&\lm_1(R_k)x+\lm_2(R_k)y+\lm_3(R_k)z=r_i\tag 5.41\\
&i=1,2,\dots,n .\endalign $$
Thus, they are parallel planes. When the function $\Psi$ is single valued we have
only one plane. For different $k_1\ne k_2$ the planes belonging to $k_1$ do not
have to be parallel and in general they cross each other. The sections of planes
may also cross.

Note that (5.41) is a place at which both density and pressure disappear and
the magnitude of the velocity reaches its maximal value. So in fact they are planes of
nodes of density and pressure and of antinodes of magnitude of the velocity
vector. Because planes may cross each other we also have straight lines and points of nodes
and antinodes. Thus, the above solution may be treated as a nonlinear analogue
of a standing wave.

Now let us turn to the case $G=G_0=\text{const.}$,
$\dfrac{dG}{dR}=0$. We must specify the condition for $\Delta
\ge0$ (compare (5.11) ). Then we have $z_2=0$ and (5.11) is reduced to:
$$\al_1z_1^2\ge0. \tag 5.42$$
Supposing that $z_1\ne0$~, \ $H\ne\text{const.}$ we obtain $\al_1\ge0$, which by using (5.14)
is reduced to:
$$e^{2H}\ge\Bigl(\frac2{\kappa-1}\Bigr)\;(c_0^2+e^{2G_0}) \tag 5.43 $$
and now we may repeat all the considerations concerning the conditions $1^0$ and $2^0$
and derive the same results as before with the substitution $G=G_0=\text{const.}$
The only difference will be another restriction for the lowest value of $e^H$
and $e^{H_0}$ and the absence of the restriction (5.16).

The case $H=\text{const}$ leads to a nonphysical solution. From the equation
\linebreak
$e^H\dfrac{dH}{dR}=
c\;\text{ch}\rho\ge c$ we have that $c=0$, which leads to a vanishing of density and pressure
everywhere, thus to the absence of gas.

\proclaim Theorem.
The expressions (5.37--5.39) together with conditions (5.16), (5.31a),
(5.32) and (5.33) describe an exact solution of equation (3.1). The
solution has remarkable properties being a nonlinear analogue of a
standing wave (see equations (5.40) and (5.41)).

Proof has been given above.

The gradient catastrophe for the above solution takes place on a
hypersurface $S$ defined by two relations
$$\align
& R = \Psi (\ovj{\lm}_1 x + \ovj{\lm}_2 y + \ovj{\lm}_3 z) \\
\noalign{\text{and}}
& 1 = {\dot\Psi} (\ovj{\lm}_1 x + \ovj{\lm}_2 y + \ovj{\lm}_3 z) \cdot
\Big( \frac{d\ovj{\lm}_1}{dR}x + \frac{d\ovj{\lm}_2}{dR}y +
\frac{d\ovj{\lm}_3}{dR}z \Big) . \endalign $$
For all simple elements from Appendix A it is possible to repeat these
considerations (see References [7, 8]).

Now let us consider simple waves generated by the simple element
$\ov{(2)}{\lm}$ for $K=0$, i.e. $F_2$ (see Table 1 of Appendix A).
In this case we have the following system of equations:
$$\aligned
\\
&\dd{\vphi_1}R=\frac{\ep(\chi^2_1-c^2)^{1/2}}{\chi_2}\Bigl(\vphi_2\cos\al-\frac{\vphi_1\vphi_3}{\chi_1}\sin\al\Bigr)
+\frac{c\vphi_1}{\chi_1}
\\\\
&\dd{\vphi_2}R=\frac{-\ep(\chi_1^2-c^2)^{1/2}}{\chi_2}\Bigl(\vphi_1\cos\al+\frac
{\vphi_2\vphi_3}{\chi_1}\sin\al\Bigr)+\frac{c\vphi_2}{\chi_1}
\\\\
&\dd{\vphi_3}R=\frac{\ep(\chi_1^2-c^2)^{1/2}}{\chi_1}\cdot\chi_2\sin\al+\frac{c\vphi_3}
{\chi_1}
\endaligned\tag 5.44$$
\vskip6pt
\flushpar
where $\chi_1^2=\vphi_1^2+\vphi_2^2+\vphi_3^2$, $\chi_2^2=\vphi_1^2+\vphi_2^2$, $\ep^2=1$.

By introducing  new variables $\chi_1$, $\chi_2$, $\nu_0=\dfrac{\vphi_1}{\vphi_2}$ we can
transform (5.44) to the following system:

\goodbreak

$$\align
&\dd{\chi_1}R=c=\Bigl(c_0^2-\frac{\kp-1}2\;\chi_1^2\Bigr)^{1/2}\tag 5.45a
\\\\
&\dd{\chi_2}R=\frac\ep{\chi_1}\bigl\{(\chi_1^2-c^2)^{1/2}
(\chi_1^2-\chi_2^2)^{1/2}\sin\al+c\chi_2\bigr\}\tag 5.45b
\\\\
&\dd{}R\arctan\nu_0=\frac{-\ep(\chi_1^2-c^2)^{1/2}}{\chi_2}\cos\al. \tag 5.45c
\endalign$$
We integrate the first equation of (5.45) and have:
$$\chi_1=c_0\sqrt{\frac2{\kp-1}}\sin\Bigl(
\sqrt{\frac{\kp-1}2}\;(R+c_1)\Bigr)\tag 5.46$$
where $c_1=$const.

In the remaining two equations of (5.45) there is an arbitrary function
$\al(R)$. According to the accepted procedure we shift the
arbitrariness from $\al$ to $\chi_2=e^G$ and we find the restriction
for the function $G(R)$. By introducing $\chi_2=e^G$ to the last two equations
of (5.45) we can calculate from b) $\sin\al$ and then $\cos\al$. After simple
calculation we have the following formulae
$$\align
&\sin\al=
\frac{\dsize\chi_1 e^G\dd GR-ce^G}
{(\chi_1^2-c^2)^{1/2}(\chi_1^2-e^{2G})^{1/2}}=
\tag 5.47
\\\\\\
&\frac
{e^G\Bigl(\sqrt{\frac2{\kp-1}}
\sin\bigl(\sqrt{\frac{\kp-1}2}(R+c_1)\bigr)\dfrac{dG}{dR}-
\cos\bigl(\sqrt{\frac{\kp-1}2}(R+c_1)\bigr)\Bigr)}
{\Bigl(\bigl(\frac{\kp+1}{\kp-1}\bigr)\sin^2\bigl(
\sqrt{\frac{\kp-1}2}(R+c_1)\bigr)-1\Bigr)
^{1/2}
\Bigl(c_0^2\bigl(\frac2{\kp-1}\bigr)\sin^2\bigl(\sqrt{\frac{\kp-1}2}(R+c_1)
\bigr)-
e^{2G}\Bigr)^{1/2}}
\endalign$$

\vskip12pt
$$\align
&\cos\al=\eta
\frac
{\Biggl\{\Biggr.
c_0^2\bigl(\frac2{\kp-1}\bigr)^2\sin^4\bigl(\sqrt{\frac{\kp-1}2}(R+c_1)\bigr)-
\Bigl[c_0^2\cos^2\bigl(\sqrt{\frac{\kp-1}2}(R+c_1)\bigr)+e^{2G}
\bigl(1+\dfrac{dG}{dR}\bigr)^2\bigr)\Bigr]\cdot}
{\Bigl(\bigl(\frac{\kp+1}{\kp-1}\bigr)\sin^2\bigl(\sqrt{\frac{\kp-1}2}(R+c_1)\bigr)
-1\Bigr)^{1/2}
}
\\\\\\
&\frac
{\cdot\bigl(\frac2{\kp-1}\bigr)\sin^2\bigl(\sqrt{\frac{\kp-1}2}(R+c_1)\bigr)-e^{2G}
\dfrac{dG}{dR}\;\sqrt{\frac2{\kp-1}}\sin\bigl(\sqrt{2(\kp-1)}(R+c_1)\bigr)
\Bigl.\Bigr\}^{1/2}}
{\Bigl(\dfrac{2c_0^2}{\kp-1}\sin^2\bigl(\sqrt{\frac{\kp-1}2}(R+c_1)\bigr)-
e^{2G}\Bigr)^{1/2}}
\tag 5.48
\endalign$$
For the function $\chi_1$ we have the following restriction
$$c_0\sqrt{\frac2{\kp+1}}\le\chi_1\le c_0\sqrt{\frac2{\kp-1}}. \tag 5.49 $$
By inserting (5.46) into (5.49) we obtain the following inequality
$$\sqrt{\frac{\kp-1}{\kp+1}}\le\sin\Bigl(\sqrt{\frac{\kp-1}2}(R+c_1)\Bigr)\le1
\tag 5.50$$
and this introduces the following restriction for $R$
$$\align
&2k\pi+\arcsin\sqrt{\frac{\kp-1}{\kp+1}}\le\sqrt{\frac{\kp-1}2}(R+c_1)\le\frac\pi2
+2k\pi-\arcsin\sqrt{\frac{\kp-1}{\kp+1}}
\tag 5.51
\\
&k=0,\pm1,\pm2,\ldots\endalign$$
Simultaneously we achieve a certain restriction for the function $G$ caused by the necessity
that the modulus of the expression (5.47) is less than $1$, i.e. a certain
differential inequality

$$\frac
{\Bigl|e^G\Bigl(\sqrt{\frac2{\kp-1}}\sin\bigl(\sqrt{\frac{\kp-1}2}(R+c_1)\bigr)\dfrac{dG}{dR}-
\cos\bigl(\sqrt{\frac{\kp-1}2}(R+c_1)\bigr)\Bigr)\Bigr|}
{\Bigl(\bigl(\frac{\kp+1}{\kp-1}\bigr)\sin^2\bigl(\sqrt{\frac{\kp-1}2}(R+c_1)\bigr)-1\Bigr)^{1/2}
\Bigl(\dfrac{2c_0^2}{\kp-1}\sin^2\bigl(\sqrt{\frac{\kp-1}2}(R+c_1)\bigr)
-e^{2G}\Bigr)^{1/2}}\le1
\tag 5.52$$
By inserting the following notations into (5.52)
$$\sqrt{\frac2{\kp-1}}\;c_0e^G=e^{G_1}\text{ and }
\beta=\sqrt{\frac{\kp-1}2}\;(R+c_1)\tag 5.53$$
we get
$$
\Biggl|\frac
{\dfrac {d}{d\beta}(e^{-G_1}\sin\beta)}
{\Bigl(\bigl(\frac{\kp+1}{\kp-1}\bigr)\sin^2\beta-1\Bigr)^{1/2}\bigl(\sin^2\beta-
e^{2G_1}\bigr)^{1/2}}
\Biggr|\le1.
\tag 5.54$$
We are able to solve this inequality by the substitutions
$$X=\frac{\sin\beta}{e^{G_1}},\qquad
\xi=\sin\beta.\tag 5.55$$
We transform (5.54) to the following expression
$$\Biggl|\dd {}\xi
\bigl(\arctan(X+\sqrt{X^2-1})\bigr)\frac
{2\xi(1-\xi^2)^{1/2}}
{\Bigl(\frac{\kp+1}{\kp-1}\;\xi^2-1\Bigr)^{1/2}}\Biggr|\le1.
\tag 5.56$$
By substituting for convenience $\eta=\xi^2$ and $\dfrac{\kp+1}{\kp-1}=\rho_1$
we get:
$$\Biggl|\dd{}\eta\bigl(\arctan(X+\sqrt{X^2-1}\bigr)\frac{\eta(1-\eta)^{1/2}}{(\rho_1\eta-1)^{1/2}}
\Biggr|\le\frac14\tag 5.57$$
and by making a final change of dependent variable
$$r=2\Bigl[(\sqrt{\rho_1}\arctan\Bigl(\sqrt{\frac{\rho_1-\eta}{\rho_1-\rho_1\eta}}
\Bigr)-\arctan\Bigl(\sqrt{\frac{\rho_1\eta-1}{1-\eta}}\Bigr)\Bigr]\tag 5.58$$
i.e. such that
$$\dd r\eta=\frac{(\rho_1\eta-1)^{1/2}}{\eta(1-\eta)^{1/2}}\tag 5.59$$
we derive the folowing inequality
$$\Bigl|\dd {}r\bigl(\arctan(X+\sqrt{X^2-1}\bigr)\Bigr|\le\frac14\tag 5.60$$
Let
$$\bigl(X(0)+\sqrt{X^2(0)-1}\bigr)=f_0\tag 5.61$$
and let
$$\dd{f_1}r=-\frac14\quad,\quad
\dd{f_2}r=\frac14\quad,\quad
f_1(0)=f_0=f_2(0).\tag 5.62$$
Then
$$f_1=-\frac14\;r+f_0\quad,\quad
f_2=\frac14\;r+f_0\tag 5.63$$
and
$$-\frac14\;r+f_0\le\arctan(X+\sqrt{X^2-1})\le\frac14\;r+f_0
\tag 5.64$$
for $r\ge0$.

By introducing $\eta=\sin^2\beta$ into (5.58) we get:
$$0=\sqrt\rho_1
\arctan\Bigl(\sqrt{\frac{\rho_1\sin^2\beta_0-1}{\rho_1-\rho_1\sin^2\beta_0}}
\Bigr)-\arctan\Bigl(\sqrt{\frac{\rho_1\sin^2\beta_0-1}{1-\sin^2\beta_0}}
\Bigr)\tag 5.65$$
from which we get
$$\sin^2\beta_0=\frac1{\rho_1}\tag 5.66$$
From (5.61) we obtain
$$f_0=\arctan\Bigl(\;\frac1{\rho_1e^{2G_1(\beta_0)}}+\sqrt{
\frac1{\rho_1e^{2G_1(\beta_0)}}-1}\;\Bigr)\tag 5.67$$
from which we have
$$0\le e^{G_1(\beta_0)}\le\frac1{\sqrt{\rho_1}}\tag 5.68$$
and
$$\arctan\Bigl(\frac1{\rho_1^2}+\sqrt{\frac1{\rho_1^2}-1}\Bigr)\le f_0\le\frac
\pi2.\tag 5.69$$
Examining the function (5.58) for $\dfrac1{\rho_1}\le\eta\le1$ we get
$$0\le r\le 2\Bigl[\sqrt{\rho_1}\;\frac\pi2-\frac\pi2\Bigr]=\pi
(\sqrt{\rho_1}-1).\tag 5.70$$
Now let us examine the expression $-\frac14\;r+f_0$:
$$\align
&\frac\pi2\ge-\frac14\;r+f_0\ge-\frac\pi4(\sqrt{\rho_1}-1)+
\arctan\Bigl(\frac1{\rho_1^2}+\sqrt{\frac1{\rho_1^2}-1}\Bigr)=
g(\rho_1)\tag 5.71
\\\\
&\rho_1=\frac{\kp+1}{\kp-1}\tag 5.72\endalign$$
and thus $1<\rho_1<\infty$, $\kp>1$.

The function $g(\rho_1)$ is  decreasing $\dfrac{dg}{d\rho_1}<0$, so
$$g(\rho_1)\le\max(g(\rho_1))=g(1)=\arctan 1=\frac\pi4.\tag 5.73$$
From this we have
$$\frac\pi2\ge-\frac14\;r+f_0\ge g(\rho_1)\ge\frac\pi4.\tag 5.74$$
On the other hand we have, that
$$\frac\pi4(1+\sqrt{\rho_1})\ge\frac14\;r+f_0\ge\arctan\Bigl(\frac1{\rho_1^2}
+\sqrt{\frac1{\rho_1^2}-1}\Bigr)\ge\frac\pi4.\tag 5.75$$
So we have deduced that in the inequality (5.64) the upper limit
is constantly positive and is contained in the interval $(-\frac\pi2, \frac\pi2)$.
Also $-\frac14\;r+f_0$ is contained in $(-\frac\pi2,\frac\pi2)$.
Thus by making use of the above substitutions with respect to $X$ and $r$ and simple trigonometrical
identities, after simple transformations we find the following:
$$c_0\sqrt{\frac2{\kp-1}}\sin\beta\sin\delta_2\le e^G\le c_0\sqrt{\frac2{\kp-1}}
\sin\beta\sin\delta_1\tag 5.76$$
where
$$\aligned
&\delta_1=-\sqrt{\rho_1}\arctan\Bigl(\sqrt{\frac{\rho_1\sin^2\beta-1}
{\rho_1(1-\sin^2\beta)}}\Bigr)+\arctan\Bigl(
\sqrt{\frac{\rho_1\sin^2\beta-1}{1-\sin^2\beta}}\Bigr)+f_0
\\\\
&\delta_2=\sqrt{\rho_1}\arctan\Bigl(\sqrt{\frac{\rho_1\sin^2\beta-1}{\rho_1(1-\sin^2\beta)}}
\Bigr)-\arctan\Bigl(\sqrt{\frac{\rho_1\sin^2\beta-1}{1-\sin^2\beta}}\Bigr)+f_0.
\endaligned\tag 5.77$$
On the other hand we have $0\le e^G\le\chi_1$ and so
$$0\le e^G\le\sqrt{\frac2{\kp-1}}\;c_0\sin\beta.\tag 5.78$$
Thus summing up we have
$$\align
&c_0\sqrt{\frac2{\kp-1}}\sin\beta\max[0,\sin\delta_2]\le c_0
\sqrt{\frac2{\kp-1}}\sin\beta\sin\delta_1\tag 5.79
\\\\
&0\le e^{G(\beta_0)}\le c_0\sqrt{\frac2{\kp-1}}\endalign$$
where $\sin^2\beta_0=\dfrac1\rho$, $\beta=\sqrt{\frac{\kp-1}2}(R+c_1)$
and $\beta$ obeys the inequality (5.51).
This is a set of all restrictions for an arbitrary function $G$.

Finally we derive the following expressions which are an exact solution of the
system of equations (5.44).
%\bye 3
$$\aligned
&R=\Psi(\lm_1x+\lm_2y+\lm_3z)\\
&\vphi_1=\eta_1e^G\sin K(R)\\
&\vphi_2=-\eta_1e^G\cos K(R)\\
&\vphi_3=\eta_2\Bigl(\frac
{2c_0^2}{\kp-1}\sin^2\bigl(\sqrt{\frac{\kp-1}2}(R+c_1)\bigr)
-e^{2G}\Bigr)^{1/2}
\\\\
&K(R)=\ep\int_{R_0}^R\frac
{\Bigl\{\Bigr.\dfrac{4c_0^4}{(\kp-1)^2}\sin^4\bigl(\sqrt{\frac{\kp-1}2}
(R+c_1)\bigr)+}
{e^G\Bigl(\dfrac{2c_0^2}{\kp-1}\sin^2\bigl(\sqrt{\frac{\kp-1}2}(R+c_1)\bigr)-e^{2G}
\Bigr)^{1/2}}
\\\\
&\frac{-\Bigl[c_0^2\cos^2\bigl(\sqrt{\frac{\kp-1}2}(R+c_1)\bigr)+e^{2G}\bigl(1+\bigl(
\dfrac {dG}{dR}\bigr)
^2\bigr)\Bigr]\dfrac{2c_0^2}{(\kp-1)}\sin^2\bigl(\sqrt{\frac{\kp-1}2}
(R+c_1)\bigr)+}{}
\\\\
&\frac{-\ep e^{2G}\dfrac{d G}{dR}
\;\dfrac{2c_0^2}{(\kp-1)}\sin\bigl(\sqrt{2(\kp-1)}
(R+c_1)\bigr)\Bigl.\Bigr\}^{1/2}}{}\;dR'+c_2
\endaligned \tag 5.80 $$
$\lm_1,\lm_2,\lm_3$ are given by the right--hand side of
the expressions (5.44) where (5.47) and (5.48) should
be substituted instead of $\sin\al$ and $\cos\al$. The function $\Psi\in C^\infty_0$ and it takes
values only from one arbitrary interval
given by the inequalities (5.51). The arbitrary function $G$
has the implied conditions (5.73).

Let us also discuse the case $\dd GR=0$, $G=G_0=$const. Here we have the following
condition
$$\Biggl|\frac{e^{G_0}\sin\beta\sqrt{\frac2{\kp-1}}}
{\Bigl(\frac{\kp+1}{\kp-1}\sin^2\beta-1\Bigr)^{1/2}\Bigl(\dfrac{2c_0^2}{
(\kp-1)}\sin^2\beta e^{2G_0}\Bigr)^{1/2}}\Biggr|\le 1
\tag 5.81$$
From which we get:
$$e^{2G_0}\Bigl(\frac{\kp+3}{\kp-1}\sin^2\beta-1\Bigr)\le\frac{2c_0^2
(\kp+1)}{(\kp-1)^2}\sin^4\beta\tag 5.82$$
where
$\dfrac1{\rho_1}\le\sin\beta\le1$.

$\dfrac1{\rho_1}=\dfrac{\kp-1}{\kp+1}>\dfrac{\kp-1}{\kp+3}$ \ so we have
$$e^{2G_0}\le\frac{2c_0^2(\kp+1)}{(\kp-1)^2}\;\frac
{\sin^4\beta}{\Bigl(\frac{\kp+3}{\kp-1}\sin^2\beta-1\Bigr)}.\tag 5.83$$
Thus
$$\align
e^{2G_0}&\le
\frac{2c_0^2(\kp+1)}{(\kp-1)^2}\min_{(\tfrac1\rho\le\sin^2\beta\le1)}
\Bigl[\frac{\sin^4\beta}{\frac{\kp+3}{\kp-1}\sin^2\beta-1}\Bigr]=
\tag 5.84
\\\\
&=\frac{2c_0^2(\kp+1)}{(\kp-1)^2}\min\Bigl[g\Bigr(\frac{2(\kp-1)}{\kp+3}\Bigr),g(1)
\Bigr]=
\frac{2c_0^2(\kp+1)}{(\kp-1)^2}\min\Bigl[\frac{4(\kp-1)^2}{(\kp+3)^2},
\frac{\kp-1}2\Bigr]
\endalign$$
where
$$g(\xi)=\frac{\xi^2}{\frac{\kp+3}{\kp-1}\;\xi-1}.$$
Finally we have
$$e^{2G_0}\le\frac{8c_0^2(\kp+1)}{(\kp+3)^2}.\tag 5.85$$
On the other hand we have
$$0\le e^G\le\chi_1\tag 5.86$$
so we have
$$\align
&0\le e^{G_0}\le\min\chi_1=c_0\sqrt{\frac2{\kp+1}}\tag 5.87
\\\\
&0\le e^{G_0}\le
2c_0^2\min\Bigl[\frac1{\kp-1},\frac{4(\kp+1)}{(\kp+3)^2}\Bigr].
\tag 5.88\endalign$$
For $1<\kp\le1+\sqrt{\frac{16}{3}}$ we have
$$\frac1{\kp-1}\ge\frac{4(\kp+1)}{(\kp+3)^2}\tag 5.89$$
Thus finally
$$0\le e^{G_0}\le\frac{8c_0^2(\kp+1)}{(\kp+3)^2}\tag 5.90$$
Therefore it is
the only restriction for constant $G_0$. Inserting
$\dd GR=0$ and $G=G_0=$const obeying (5.90) into expressions (5.80), (5.14) and (5.48)
we derive the solution for constant $G$.

\vskip10pt
Now let us consider the cases for $K=1$. They are similar to those discussed above,
but certain differrences occur, which are pointed out here. Substituting
into equation (2.3) the simple integral element $\ov{(1)}{\lm}$,
$K=1$, i.e. $F_3$ (Table~1 Appendix~A) we get
the following system of equations
$$\align
\dd{\vphi_1}R&=\frac{(\chi_1^2-c^2)^{1/2}}{\chi_2}\Bigl(
-\bigl(\vphi_2\sin\al+\frac{\vphi_1\vphi_3}{\chi_1}\cos\al\bigr)+
\\\\
&+
\bigl(-\vphi_2\cos\al+\frac{\vphi_1\vphi_3}{\chi_1}\sin\al\bigr)\sh\rho\Bigr)+
\frac{c\vphi_1}{\chi_1}\ch\rho
\tag 5.91a
\\\\\\
\dd {\vphi_2}R&=\frac{(\chi_1^2-c^2)^{1/2}}{\chi_2}\Bigl(\bigl(
\vphi_1\sin\al-\frac{\vphi_2\vphi_3}{\chi_1}\cos\al\bigr)+
\\\\
&+(\vphi_1\cos\al+\frac{\vphi_2\vphi_3}{\chi_1}\sin\al\bigr)
\sh\rho\Bigr)+
\frac{c\vphi_2}{\chi_1}\ch\rho
\tag 5.91b
\\\\\\
\dd{\vphi_3}R&=\frac{\chi_2(\chi_1^2-c^2)^{1/2}}{\chi_1}(\cos\al-\sin\al
\sh\rho)+\frac{c\vphi_3}{\chi_1}\ch\rho
\tag 5.91c
\endalign$$
where
$\chi_1^2=\vphi^2_1+\vphi_2^2+\vphi_3^2$, $\chi_2^2=\vphi_1^2+\vphi_2^2$
and
$\al$ and $\rho$ are functions of $\vphi_1, \vphi_2, \vphi_3$ i.e.
$R$.
We write down equations for $\chi_1,\chi_2$ and $\mu_0=\dfrac{\vphi_1}{\vphi_2}$.
$$\aligned
&\dd{\chi_1}R=c\ch\rho\ge c
\\\\
&\dd{\chi_2}R=\ep\frac{(\chi_1^2-c^2)^{1/2}}{\chi_1}(\chi_1^2-\chi_2^2)^{1/2}
(\sin\al\sh\rho-\cos\al)+
\frac{c\chi_2}{\chi_1}\ch\rho
\\\\
&\dd{}R\arctan\mu_0=-\frac{(\chi_1^2-c^2)^{1/2}}{\chi_2}(\sin\al+\cos\al\sh\rho)
\endaligned\tag 5.92$$
where $\ep_1^2=1$ and $c^2=c_0^2-\frac{\kp-1}2\chi_1^2\ge0$

We shift the  arbitrariness from $\al$ and $\rho$ to $\chi_1$
and $\chi_2$ assuming that $\chi_1^2=e^{2H}$, $\chi_2^2=e^{2G}$
and then we search for the restrictions for those
functions. So we have
$$\ch\rho=\frac1c\;e^H\dd HR\quad,\quad
\sh\rho=\ep_2\Bigl(\frac1{c^2}\;e^{2H}\bigl(\dd HR\bigr)^2-1\Bigr)^{1/2}.
\tag 5.93$$
Regarding equations (5.92) as an algebraic equation for a function
$\sin\al$ and using equations (5.93) we achieve the following
$$A=ZB-\ep_3\sqrt{1-Z^2}\tag 5.94$$
where $Z=\sin\al$, $\ep_3^2=1$ and
$$\aligned
&A=\frac{\ep_1e^{(H+G)}\bigl(\dd GR-\dd HR\bigr)}{(e^{2H}-c^2)^{1/2}
(e^{2H}-e^{2G})^{1/2}}
\\\\
&B=\ep_2\Bigl(\frac1{c^2}\;e^{2H}\bigl(\dd  HR\bigr)^2-1\Bigr)^{1/2}\endaligned
\tag 5.95$$
requiring that $|Z|\le1$ and that $\tr$ of the quadratic equation
$$Z^2(B^2+1)-2ABZ+(A^2-1)=0\tag 5.96$$
is non--negative, using the equations (5.96) and (5.7) we
achieve the same restrictions for $H$ and $G$ i.e. (5.16),
(5.31), (5.32) and (5.33). $\sin\al$, $\cos\al$, are given
by formula (5.38) whereas $\ch\rho$ and $\sh\rho$ are given by (5.39) or (5.93),
in the case when $G\ne$const. In the case when $G=G_0=$const
we have the restriction (5.43) instead of (5.16). Also the restriction for
the function $\Psi$  is identical. The only thing that changes is the
form of solution and it is the following:
$$\aligned
&\vphi_1=\eta_2e^G\sin K(R)\qquad \eta_1^2=\eta_2^2=1\\
&\vphi_2=\eta_2 e^G\cos K(R)\\
&\vphi_3=\eta_1\sqrt{e^{2H}-e^{2G}}
\endaligned\tag 5.97$$
where
$$\align
&K(R)=-\int_{R_0}^R\frac{(e^{2H}-c^2)^{1/2}}{e^G}
\;(\sin\al+\cos\al\sh\rho)\,dR'+c
\\
&c=\text{const.}\endalign$$
and $R=\Psi(\lm_1x+\lm_2y+\lm_3z)$.

$\lm_1,\lm_2,\lm_3$ are given by the expressions for $K=1$ from Appendix A (Table 1),
whereas $\sin\al,\cos\al,\ch\rho$ and $\sh\rho$ are given by the expressions
written above. The only difference between (5.97) and (5.37) is the form
of the function $K(R)$. All the arguments concerning the interpretation of
the  solution are still valid.

\vskip8pt

Now let us turn to the analysis of the simple element $\ov{(2)}\lm$,
$K=1$ (i.e. $F_4$, see Table~1, Appendix~A).
Then from equation (2.3) we get the following system of equations
\vskip3pt
$$\align
&\dd {\vphi_1}R=\frac{\ep(\chi_1^2-c^2)^{1/2}}{\chi_2}
\Bigl(-\vphi_2\cos\al+\frac{\vphi_1\vphi_3}{\chi_1}\sin\al\Bigr)+\frac{c\vphi_1}{\chi_1}
\\\\
&\dd {\vphi_2}R=\frac{\ep(\chi_1^2-c^2)^{1/2}}{\chi_2}\Bigl(\vphi_1\cos\al+
\frac{\vphi_2\vphi_3}{\chi_1}\sin\al\Bigr)+\frac{c\vphi_2}{\chi_1}
\tag 5.98
\\\\
&\dd {\vphi_3}R=-\ep\frac{(\chi_1^2-c^2)^{1/2}}{\chi_1}\chi_2\sin\al+\frac{c\vphi_3}{\chi_1}\quad,
\quad\ep^2=1.
\endalign$$
%bye 4
By introducing (as before) variables $\chi_1,\chi_2,\mu_0$ we get the
following equations
$$\align
&\dd{\chi_1}R=c\\\\
&\dd{\chi_2}R=\ep\frac{(\chi_1^2-c^2)^{1/2}}{\chi_1}(\chi_1^2-\chi_2^2)^{1/2}
\sin\al+\frac{c\chi_2}{\chi_1}\tag 5.99
\\\\
&\dd{}R\arctan\mu_0=-\frac{\ep(\chi_1^2-c^2)^{1/2}}{\chi_2}\cos\al\endalign$$
which are identical to (5.45). We shall not derive from that any other
solutions different from (5.80).

According to Appendix A we can write down equations (2.3) for the element
$\ov{(1')}\lm$ for $K=0$ i.e. $F_{1'}$ (see Table 2, Appendix A)
$$\align
&\dd{\vphi_1}R=\frac{\sqrt{\chi_1^2-c^2}}{c\chi_2}\Bigl(\vphi_2\cos\beta-\frac{\vphi_1\vphi_3}{\chi_1}\sin\beta\Bigr)+\frac{\vphi_1}{\chi_1}
\\\\
&\dd{\vphi_2}R=-\frac{\sqrt{\chi_1^2-c^2}}{c\chi_2}\Bigl(\vphi_1\cos\beta+\frac{\vphi_2\vphi_3}{\chi_1}\sin\beta\Bigr)
+\frac{\vphi_2}{\chi_1}\tag 5.100
\\\\
&\dd{\vphi_3}R=\frac{\sqrt{\chi_1^2-c^2}}{c\chi_1}\sin\beta+\frac{\vphi_3}{\chi_1}.
\endalign$$
Introducing variables $\chi_1,\chi_2,\mu_0$ we get:
$$\align
&\dd{\chi_1}R=1\\\\
&\dd{\chi_2}R=-\ep_1\frac{\sqrt{\chi_1^2-c^2}}{\chi_1}(\chi_1^2-\chi_2^2)^{1/2}\sin\beta+\frac{\chi_2}{\chi_1}\tag 5.101
\\\\
&\dd{}R\arctan\mu_0=\frac{\sqrt{\chi_1^2-c^2}}{c\chi_2}\cos\beta
\endalign$$
where of course
$$\vphi_3=\ep_1(\chi_1^2-\chi_2^2)^{1/2}\quad,\quad
\ep_1^2=1. $$
The first equation (5.101) is immediately integrated and we get
$$\chi_1=R+c_1\tag 5.102$$
where $c_1=$const.
Because of the fact that
$$c_0^2\frac2{\kp+1}\le\chi_1^2\le c_0^2\bigl(\frac2{\kp-1}\bigr)\tag 5.103$$
we have a restriction for the variable $R$
$$c_0^2\bigl(\frac2{\kp-1}\bigr)\le (R+c_1)^2\le c_0^2\bigl(\frac2{\kp-1}\bigr)
\tag 5.104$$
Now we deal with the second equation of (5.101) and according to the
procedure proposed here we shift the arbitrariness
from $\beta$ to $\chi_2$, assuming that $\chi_2=e^G$. Then we find the restriction for the
arbitrary function $G$. Thus, treating the second of equations
(5.101) as an equation for a function $\sin\beta$ we get:
$$\sin\beta=\ep_1\frac
{1-(R+c_1)\dd GR}{\bigl((R+c_1)^2-e^{2G}\bigr)^{1/2}
\bigl(\frac{\kp+1}2(R+c_1)^2-c_0^2\bigr)^{1/2}}
\tag 5.105$$
where of course
$$e^{2G}\le (R+c_1)^2.\tag 5.106$$
Obviously $|\sin\beta|\le1$, and this leads to the following
differential inequality:

$$\Biggl|
\frac{1-\dd GR\;(R+c_1)}
{\bigl((R+c_1)^2-e^{2G}\bigr)^{1/2}
\bigl(\frac{\kp+1}2(R+c_1)^2-c_0^2\bigr)^{1/2}}\Biggr|\le1.\tag 5.107$$

Taking $X=\dfrac{R+c_1}{e^G}$ and $R+c_1=Z$ we reduce (5.107) to the following form
$$\Biggl|\frac{X^2\dd XZ}{\bigl(\frac{\kp+1}2Z^2-c_0^2\bigr)^{1/2}\bigl(X^2-1\bigr)^{1/2}}
\Biggr|\le 1.\tag 5.108$$
Then by taking other  independent and dependent variables
$$\eta=\frac{c_0^2}8\sqrt{\frac2{\kp+1}}\Bigl[
(\xi+\sqrt{\xi^2-1})^2-\frac1{(\xi+\sqrt{\xi^2-1})^2}-
4\ln(\xi+\sqrt{\xi^2-1})\Bigr]
\tag 5.109$$
$$
t=\frac18\Bigl[(X+\sqrt{X^2-1})^2-\frac1{(X+\sqrt{X^2-1})^2}\Bigr]
+\frac12\ln(X+\sqrt{X^2-1})\tag 5.110$$
where
$$\xi=\frac1{c_0}\sqrt{\frac{\kp+1}2}Z=\frac1{c_0}\sqrt{\frac{\kp+1}2}(R+
c_1)$$
we reduce (5.108) to the following convenient form
$$\Bigl|\dd t\eta\Bigr|\le 1.\tag 5.111$$
Obviously we have
$$\align
&\dd\eta Z=\Bigl(\bigl(\frac{\kp+1}2\bigr)Z^2-c_0^2\Bigr)^{1/2}
\tag 5.112\\
&\dd tX=\frac{X^2}{(X^2-1)^{1/2}}.\tag 5.113\endalign$$
Let $t(\eta_0)=t_0$. Then from (5.111) we get
$$-(\eta-\eta_0)+t_0\le t\le(\eta-\eta_0)+t_0\quad,\quad
\eta\ge\eta_0\tag 5.114$$
From which we obtain
$$-(\eta-\eta_0)+t_0\le F(X)\le(\eta-\eta_0)+t_0\tag 5.115$$
where
$$\align
F(X)={}&\frac18\Bigl[(X+\sqrt{X^2-1})^2-\frac1{(X+\sqrt{X^2-1})^2}\Bigr]\\
&+\frac12\ln(X+\sqrt{X^2-1}) \tag 5.116\\
X\ge{}& 1\\
\dd FX={}&\frac{X^2}{(X^2-1)^{1/2}}\ge0\tag 5.117
\endalign
$$
So the function $F$ is an increasing one and consequently it possesses the inverse
function $Q$ such as:
$$
Q(F(X))=X\quad\text{for } X\ge1\tag 5.118$$
$$\aligned
&\overline{\Cal D}F=[0,+\infty)\quad,\quad
\overline{\Cal D}Q=[1,+\infty)\\
&\Cal DF=[1,+\infty)\quad,\quad\Cal DQ=[0,+\infty). \endaligned
\tag 5.119$$
We also have
$$\aligned
F(1)=0\quad,&\quad\lim_{X=1^+}\dd FX=+\infty\\\\
\lim_{X\to+\infty}F(X)=+\infty\quad,&\quad\lim_{X\to+\infty}\dd FX=+\infty
\endaligned\tag 5.120$$
and
$$\dd{{}^2F}{X^2}=\frac{X(2X^2-\frac12X-2)}{(X^2-1)^{3/2}}.\tag 5.121$$
In the interval $[1,+\infty)$ \
$\dfrac{d^2F}{dX^2}=0$, for $X_0=\frac18(1+\sqrt{65})>1$ and there
the function $F$ has its point of
inflection.
Thus, from (5.115) we get
$$Q(-(\eta-\eta_0)+t_0)\le\frac{R+c_1}{e^G}\le Q((\eta-\eta_0)+t_0)\quad,\quad
\eta\ge\eta_0\tag 5.122$$
and $t_0\ge(\eta-\eta_0)$
(The function $Q$ is defined only for a non--negative argument $t$).
$$\eta_0\le\eta\le t_0+\eta_0\quad,\quad
t_0\ge0\tag 5.123$$
Now let us examine the following function
$$\eta(\xi)=\frac{c_0^2}8\sqrt{\frac2{\kp+1}}
\Bigl[(\xi-\sqrt{\xi^2-1})^2-\frac1
{(\xi+\sqrt{\xi^2-1})^2}-4\ln(\xi+\sqrt{\xi^2-1})\Bigr]
\tag 5.124$$
for $1\le\xi\le\sqrt{\dfrac{\kp+1}{\kp-1}}=\sqrt{\rho_1}$.
\vskip12pt
\flushpar
$\dsize\dd\eta\xi>0$
for $\xi>1$ and $\dfrac{d\eta}{d\xi}=0$ for $\xi=1$. Thus
it is an increasing function
$$\lim_{\xi=1}\eta(\xi)=0\quad,\quad
\lim_{\xi\to+\infty}\eta(\xi)=+\infty. \tag 5.125$$
We conclude that in the interval $[1,\sqrt{\rho_1}]$ we have:
$$0\le\eta(\xi)\le\frac
{c_0^2}8\sqrt{\frac2{\kp+1}}\Bigl[(\sqrt{\rho_1}+\sqrt{\rho_1-1})^2-
\frac1{(\sqrt{\rho_1}+\sqrt{\rho_1-1})^2}-4\ln(\sqrt{\rho_1}+\sqrt{\rho_1-1})
\Bigr].\tag 5.126$$
It is apparent, that it is convenient to assume $\eta_0=0$ and
$$t_0=\frac{c_0^2}8
\sqrt{\frac2{\kp+1}}\Bigl[(\sqrt{\rho_1}+\sqrt{\rho_1-1})^2-\frac1
{(\sqrt{\rho_1}+\sqrt{\rho_1-1})^2}-4\ln(\sqrt{\rho_1}+\sqrt{\rho_1-1})\Bigr].
\tag 5.127$$
In such a situation we have a maximal interval for $\eta$ and consequently for $R$
as well.

So finally we have
$$t_0=F(X(0))\tag 5.128$$
which leads to
$$e^{2G}\Bigl|_{ (R+c_1)=c_0\sqrt{\tfrac2{\kp-1}}}\Bigr.
=c_0^2\bigl(\frac2{\kp+1}\bigr)\frac1{Q^2(t_0)}.\tag 5.129$$
On the other hand
$$e^{2G}\le(R+c_1)^2 \tag 5.130 $$
so finally we get:
$$\frac{(R+c_1)^2}{Q^2(t_0-\eta)}\le e^{2G}\le\min\Bigl(1,\frac1{Q^2(t_0+\eta)}\Bigr)
(R+c_1)^2.\tag 5.131$$
Now we turn back to the system of equations (5.100) and write down the solution
$$\aligned
&\vphi_1=e^G\sin K(R)\\
&\vphi_2=e^G\cos K(R)\\
&\vphi_3=\ep_1\sqrt{(R+c_1)^2-e^{2G}}\\
&R=\Psi(\lm_1x+\lm_2y+\lm_3z)\endaligned\tag 5.132$$
where
$$\align
&K(R)=\ep_2\int_{R_0}^R\frac
{\sqrt{
((R'+c_1)^2-e^{2G})\bigl(\frac{\kp+1}2(R'+c_1)^2-c_0^2\bigr)-
\bigl(1-\dd GR(R'+c_1)^2\bigr)}}
{e^G\bigl(c_0^2-\frac{\kp-1}2(R'+c_1^2)^2\bigr)^{1/2}
\bigl((R'+c_1)^2-e^{2G}\bigr)^{1/2}}\;dR'+c_2
\\\\
&(R_0+c_1)^2=c_0^2\bigl(\frac2{\kp+1}\bigr),\qquad
\ep_2^2=1
\endalign$$
and
$$\align
&\cos\beta=\ep_2(1-\sin^2\beta)^{1/2}=
\tag 5.133\\
&=\ep_2\sqrt{
\frac
{((R+c_1)^2-e^{2G})\bigl(\frac{\kp+1}2(R+c_1)^2-c_0^2\bigr)
-\bigl(1-\dd GR(R+c_1)^2\bigr)}
{((R+c_1)^2-e^{2G})\bigl(\frac{\kp+1}2(R+c_1)^2-c_0^2\bigr)}}.
\endalign$$
Restrictions (5.131) are imposed on the function $G$ , whereas $\Psi$ is a smooth function taking its value from
the following intervals
$$\align
&\left[c_0\sqrt{\frac2{\kp+1}}-c_1,c_0\sqrt{\frac2{\kp-1}}-c_1\right]\\
&\text{or}\tag 5.134\\
&\left[-c_0\sqrt{\frac2{\kp-1}}-c_1,-c_0\sqrt{\frac2{\kp+1}}-c_1\right]\endalign$$
$\lm_i$ are components of the covector $\ov{(2')}\lm$ for $K=0$ and according to
Appendix~A (5.133), (5.132) and (5.105) should be substituted to them.

In the case $\lm'$, \ $K=1$,
i.e. $F_{2'}$,  and according to Table~2 of Appendix~A we get equations (2.3) in the following form:
$$\align
&\dd{\vphi_1}R=\frac{\sqrt{\chi_1^2-c^2}}{c\chi_2}\Bigl(-\vphi_2\cos\om-\frac{\vphi_1\vphi_3}
{\chi_1}\sin\om\Bigr)+\frac{\vphi_1}{\chi_1}
\\\\
&\dd {\vphi_2}R=\frac{\sqrt{\chi_1^2-c^2}}{c\chi_2}\Bigl(\vphi_1\cos\om-\frac{
\vphi_2\vphi_3}{\chi_1}\sin\om\Bigr)+\frac{\vphi_2}{\chi_1}\tag 5.135
\\\\
&\dd {\vphi_3}R=\frac{\sqrt{\chi_1^2-c^2}}{c\chi_2}+\frac{\vphi_3}{\chi_1}
\endalign$$
%\bye 4
\flushpar
For $\chi_1,\chi_2$ and $\mu_0$ we get
$$\align
&\dd{\chi_1}R=1\\\\
&\dd{\chi_2}R=-\ep_1\frac{(\chi_1^2-c^2)^{1/2}(\chi_1^2)^{1/2}(\chi_1^2-\chi
_2^2)^{1/2}}{c\chi_1}\sin\om+\frac{\chi_2}{\chi_1}
\tag 5.136\\\\
&\dd{}R\arctan\mu_0=-\cos\om
\frac{\sqrt{\chi_1^2-c^2}}{c\chi_2}.\endalign$$
These equations are identical to (5.101) except for the
sign in the third equation. Thus they lead to almost
identical solutions with an identical restriction as before.

Finally we consider the last simple integral elements,
i.e. $F_{1''}$, $F_{2''}$, $F_{3''}$
(compare Table~3, Appendix~A). However these elements do not lead us to a new solutions.

Let us also note that our equation (3.6) is
invariant under permutation of $\vphi_1$, $\vphi_2$, $\vphi_3$. Thus we may obtain exact solutions by permutating
$\vphi_1,\vphi_2,\vphi_3$ in expressions (5.37), (5.80), (5.97) etc.
The solutions break the permutation symmetry $P_3$. This is spontaneous symmetry breaking and it results from
the acceptance of the form of the matrix
$B(\al,l)$ which does not posses this symmetry.

\proclaim Theorem.
There are simple waves for all simple elements collected in Appendix~A.
All details are given above

Proof is also given above.

All described solutions have a gradient catastrophe on a certain
hypersurface $S$. On this surface some shock waves can appear.

%\bye 5

%############################
%    SEC 6
%############################
\newpage

\vskip2.5\baselineskip
\centerline{\bf 6. Gauge and B\"acklund transformation}
\vskip\baselineskip

In this section we define a gauge transformation and an analogue for simple waves.

Now we construct some geometrical structures for simple (Riemann) waves of (2.5).
These structures establish relations between the Riemann waves of (2.5) and allow us to introduce
nonlinear transformations connecting two Riemann waves (exact solutions of
(2.5)).
The nonlinear transformations are of gauge type, and may be treated as B\"acklund
transformation [9, 10] for (2.5).

Specifically, let the matrix $A=(a_{ij})$ have $K$ eigenvalues $\om_i$, $i=1,2,\dots,K$, each of order
$l_i$~, \ $\sum\limits_{i=1}^Kl_i=n$.

In this case we have a natural group acting on simple elements $\lm$, i.e.
$\prod\limits_{i=1}^K\otimes O(l_i)$. Each of the groups $O(l_i)$ acts on the coordinates $V_j$~,
\ $j=\sum\limits_{r=1}^{i-1}l_r$~, \ $\sum\limits_{r=1}^{i-1}l_r+1$~, \ $\dots, \
\sum\limits_{r=1}^il_r$ without destroying the diagonalization of matrix $A$.

It is easy to see that $\al_1,\al_2,\dots,\al_m$ are parameters of the group $\prod\limits_{i=1}^K\otimes O(l_i)$~,
\ $m=\sum\limits_{i=1}^K\frac12\;l_i(l_i-1)$.

Simultaneously we can associate the group $O(p,q)$ with the cone
$$
\sum_{j=1}^n\zeta_jV_j^2=0\ ,\qquad(\zeta_j \ \text{is equal to one of
$\om_i)$}. \tag 6.1 $$
Let
$$\zeta_j=\ep_j|\zeta_j|\qquad\text{where}\quad \ep_j=\text{sgn}\;\zeta_j
. \tag 6.2 $$
By transforming $V_j$ to $V_j'$, $j=1,2,\dots,n$
$$V_j'=\sqrt{|\zeta_j|}\;V_j \tag 6.3 $$
we transform (6.1) into
$$\sum_{j=1}^n\ep_jV_j'{}^2=0 \tag 6.4$$
i.e. into a canonical cone.

The group $O(p,q)$ preserves a quadratic form
$$Q(V',V')=\sum_{i=1}^n\ep_iV_i'{}^2 \tag 6.5 $$
where $p={}$ number of integers $\ep_i$ equal to $1$, $q={}$ number of integers $\ep_i$
equal to $(-1)$ in the sum (6.5). \newline
Obviously $p+q=n$ (we assume that there exist no zero eigenvalues).

Notice that
classes of simple elements, and, in consequence, simple waves, which are
constructed according to section 2, are related to the choice of a concrete chain of subgroups $O(p,q)$.
This chain ends on the two--element group $\{e,-e\}$ or the trivial group $\{e\}$
hence
$$
O(p,q)\supset O(p_1,q_1)\supset O(p_2,q_2)\supset\cdots\supset\{e,-e\}\supset\{e\}
\tag 6.6 $$
where for $p_i,q_1,p_{i+1},q_{i+1}$ we have the following relations either
$$\aligned &p_i=p_{i+1}\\&q_i=q_{i+1}+1\endaligned\qquad\text{or}\qquad
\aligned &p_i=p_{i+1}+1\\&q_i=q_{i+1}\endaligned \tag 6.7 $$
$p_0=p$, $q_0=q$.

In this way the dimension of the space in which the group operates diminishes to 1
according section 2. The choice of the sequence of subscripts $j_0,j_1,\dots,j_{K-2}$ corresponds to one
of the possible chains of subgroups (6.6). Thus with each simple element we can
associate in a natural way the following group
$$L_i=\Bigl[\otimes\prod_{r=1}^KO(l_i)\Bigr]\otimes O(p_i,q_i). \tag 6.8 $$
The origin of each factor of the simple product is, of course, different. In general, we
associate with equation
(2.5)  the group
$$L=\Bigl[\otimes\prod_{r=1}^KO(l_i)\Bigr]\otimes O(p,q). \tag 6.9 $$
The group $L_i$ acts on a submanifold $F_i\subset \Cal E^*$ (the manifold of simple elements
of a chosen class according to the classification from section 2). Since a simple element is a function of a point of
the hodograph space $\Cal H$ (the space of values of the solutions of the equation) we may
construct some
natural fibre bundles associated with the equation. For every class of simple elements we have
a fibre bundle $P_i$ over the base space $\Cal H$ with structural group $L_i$,
typical fibre $F_i$ and projection $\pi_i:\/P_i\to\Cal H$.

It is easy to see that $\text{dim}\;(L_i)=\text{dim}\;(F_i)$ and for every simple element,
$\lm\in F_i^*$ we have
$$\align
\lm=g\cdot\lm_0\ ,\quad\text{where}\ g\in L_i\ \text{and}\ &\lm_0\in F_i \ \tag
6.10\\ &\lm_0=\text{const.} \endalign $$
Taking a local section of $P_i$ we get
$$\lm(u)=g(\al(u))\;\lm_0\ ,\qquad u\in\Cal D\subset\Cal H \tag 6.11 $$
where $\al$ is the set of all parameters of the group $L_i$. But in the case of simple waves
we have the Riemann invariant $R$ (parametrization in the hodograph space $\Cal H$).
Thus we obtain a special structure, a bundle $\Pi_i$ over the base space $\Cal R$ (a straight line
of the Riemann invariant) with structural group $L_i$, typical fibre $F_i$ and with a projection
$\overline {\Pi}_i:\/\Pi_i\to\Cal R$.

For every local section of $\Pi_i$ we get
$$\lm(R)=g(\al(R))\;\lm_0\quad,\quad R\in V\subset\Cal R . \tag 6.12 $$
Every local section $f$ gives us a simple wave belonging to a chosen class of simple waves (simple
elements). If we have two local sections $f$ and $g$ we have two different simple waves of the
same type. If we change the section from $f$ to $g$, we change functions $\al(R)$ to $\beta(R)$
and we get
$$
g(\al(R))=h(R)g(\beta(R))\quad,\quad
\lm(\al(R))=h(R)\lm(\beta(R))\quad,\quad
h(R)\in L\; .
\tag 6.13 $$
Thus we see that the action of the gauge group of $L_i$ over the straight line $\Cal R$
(Riemann invariant) on a simple wave creates a new simple wave of the same type ("gauge
group" means that the parameters of $L_i$ depend on $R$). In section 5 we solved equation
(5.1) using arbitrary functions $\al(R)$. We shift the freedom from the $\al$'s to new,
more convenient functions and we should do this for the functions $\al$ and $\beta$
independently. For $\al$ and $\beta$ one gets algebraic (or transcendental)
equations, which will express $\al$ (or $\beta$)
in terms of new functions and their first derivatives with respect to $R$. The condition of solvability of the algebraic (or
transcendental) equations provide us with restrictions for the new functions. Varying
$\al(R)$ to $\beta(R)$ we change these new functions and their first  derivatives.
Thus we get the gauge transformation connecting two exact solutions (simple waves of the same type).
This transformation is very similar to the classical B\"acklund transformation [9].
For the equation (3.1) we have the following situation
$$L=O(2)\otimes O(1,2) \tag 6.14 $$
and we used the following chains of subgroups of $O(1,2)$
$$\aligned
&O(1,2)\supset\ovj O(1,1)\supset\{e,-e\}\\
&O(1,2)\supset O(2)\supset\{e,-e\}\\
&O(1,2)\supset\ovd O(1,1)\supset\{e,-e\}.\endaligned \tag 6.15 $$
\vskip4pt
\noindent All these chains correspond to the simple elements and simple waves
that we examined and we obtain the following gauge groups
$$\aligned
&L_1=O(2)\otimes \ovj O(1,1)\\&L_2=O(2)\otimes\{e,-e\}\endaligned
\
\aligned&,\ L_{1'}=O(2)\otimes O(2)\\ &,\ L_{1''}=O(2)\otimes\ovd O(1,1)
\endaligned
\
\aligned&,\ L_{2'}=O(2)\otimes\{e,-e\}\ .\\&,\ L_{2''}=O(2)\otimes\{e,-e\}\ .
\endaligned \tag 6.16 $$
The case with $L_1=O(2)\otimes O (2)$ is very interesting because we have simple elements corresponding to that group
with only one arbitrary parameter. Two parameters of $O(2)\otimes O(2)$ become one parameter of the
diagonal group $O(2)$.

Now let us write down the explicit form of the action of the gauge group
of functions $H$ and $G$ for the case (4.1). Let us  suppose that
there exists one exact solution --- a simple wave with parameters $\al$ and
$\rho$ and corresponding arbitrary functions $H$ and $G$.
In this case we have $L_i=O(2)\otimes O(1,1)=L_1$, $\al$ is a parameter
of $O(2)$ and $\rho$ of $O(1,1)$.

We change functions $\rho$ and $\al$ into $\rho+\Delta\rho$ and $\al+\Delta\al$. It is a change of gauge by means of functions $\Delta\rho$ and $\Delta\al$.
And we look at how $G$ and $H$  change. In this way we obtain the explicit
action of the gauge group on the manifold of functions $H$ and $G$ and their
first derivatives with respect to $R$.

We have
$$\text{ch}\;\rho=\frac{e^H}{c(e^H)}\;\frac{dH}{dR}\quad,\quad
\text{ch}\;(\rho+\Delta\rho)=
\frac{e^{H_1}}{c(e^{H_1})}\;\frac{dH_1}{dR}\ , \tag 6.17 $$
but
$$\text{ch}(\rho+\Delta\rho)=\text{ch}\rho \;\text{ch}\Delta\rho+
\text{sh}\rho  \;\text{sh}\Delta\rho. \tag 6.18 $$
Inserting (5.17) and (4.39) into (5.18) we get:
$$
\frac{e^{H_1}}{c(e^{H_1})}\;\frac{dH_1}{dR}=\frac{e^H}{c(e^H)}\;\frac{dH}{dR}
\;\text{ch}(\Delta\rho)+
\ep_2\Bigl(\frac{e^{2H}}{c^2(e^{2H})}\;\Bigl(\frac{dH}{dR}\Bigr)^2-1\Bigr)^{1/2}
\text{sh}(\Delta\rho). \tag 6.19 $$
(6.19) is the nonlinear representation of the gauge group originating from
$O(1,1)$ on the manifold of functions $H$ and $G$ and their first derivatives with respect
to $R$ \ ($H,H_1,G,G_1,\al,\rho,\Delta\al,\Delta\rho$ are functions
of $R$).

Similarly we get for $O(2)$ from (6.20)
$$\sin(\al+\Delta\al)=\sin\al\cos\Delta\al+\sin\Delta\al\cos\al \tag 6.20 $$

%\newpage

$$\aligned
&\ep_1\Biggl\{\frac
{
e^{G_1}\dfrac d{dR}(H_1\text{--}G_1)\Bigl(\dfrac{e^{2H_1}}{c_1^2}
\Bigl(\dfrac{dH_1}{dR}\Bigr)^2\text{--}1\Bigr)^{\frac12}\text{--}
\Bigl[\Bigl(\dfrac{dH_1}{dR}\Bigr)^2(e^{2H_1}\text{--}e^{2G_1})\dfrac{(e^{2H_1}\text{--}c^2)}
{c^2}
\text{+}}
{\dfrac1{c_1^2}\;e^{H_1}\Bigl(\dfrac{dH_1}{dR}\Bigr)^2(e^{2H_1}\text{--}e^{2G_1})^{\frac12}
\cdot(e^{2H_1}\text{--}c_1^2)^{\frac12}}
\\\\
&\quad \frac{\text{--}e^{2G_1}\Bigl(\dfrac d{dR}(H_1\text{--}G_1)\Bigr)^2\Bigr]^
{\frac12}}
{\ }
\Biggr\}=
\\\\
&=\cos(\Delta\alpha)\ep_2\Biggl\{\frac
{e^G\dfrac d{dR}(H\text{--}G)\Bigl(\dfrac{e^{2H}}{c^2}\Bigl(\dfrac{dH}{dR}\Bigr)^2\text{--}1\Bigr)^{\frac12}
\text{--}\Bigl[\Bigl(\dfrac{dH}{dR}\Bigr)^2(e^{2H}\text{--}e^{2G})\dfrac{(e^{2H}\text{--}c^2)}{c^2}
\text{+}}
{\dfrac1{c^2}\;e^H\Bigl(\dfrac{dH}{dR}\Bigr)^2(e^{2H}\text{--}e^{2G})^{\frac12}\cdot
(e^{2H}\text{--}c^2)^{\frac12}}
\\\\
&\quad \frac{\text{--}e^{2G}\Bigl(\dfrac d{dR}\;(H\text{--}G)\Bigr)^2\Bigr]
^{\frac12}}{\ }
\Biggr\}\text{+}
\\\\\\
&\sin(\Delta\alpha)\ep_3\Biggl\{
1\text{-}\frac{
e^{2G}\Bigl(\dfrac d{dR}(H\text{--}G)\Bigr)^2\Bigl(\dfrac{e^{2H}}{c^2}
\Bigl(\dfrac{dH}{dR}\Bigr)^2\text{--}1\Bigr)\text{+}\Bigl(\dfrac{dH}{dR}\Bigr)^2
(e^{2H}\text{--}e^{2G})(e^{2H}\text{--}c^2)\dfrac1{c^2}}
{
\dfrac1{c^4}\;e^{2H}\Bigl(\dfrac{dH}{dR}\Bigr)^4(e^{2H}\text{--}e^{2G})\cdot
(e^{2H}\text{--}c^2)}\text{+}
\\\\\\
&\text{+}\frac{
e^{2G}\Bigl(\dfrac d{dR}(H\text{--}G)\Bigr)^2\text{+}2e^G\dfrac d{dR}(H\text{--}G)\Bigl(
\dfrac{e^{2H}}{c^2}\Bigl(\dfrac{dH}{dR}\Bigr)^2\text{--}1\Bigr)^{\frac12}\cdot
\Bigl[
\Bigl(\dfrac{dH}{dR}\Bigr)^2(e^{2H}\text{--}e^{2G})(e^{2H}\text{--}c^2)
\dfrac1{c^2}\text{+}}
{
\dfrac1{c^4}\;e^{2H}\Bigl(\dfrac{dH}{dR}\Bigr)^4(e^{2H}\text{--}e^{2G})\cdot
(e^{2H}\text{--}c^2)}
\\\\\\
&\quad\frac{\text{+}
e^{2G}\Bigl(\dfrac d{dR}(H\text{--}G)\Bigr)^2\Bigr]^{\frac12}}{ \ }
\Biggr\}^{\frac12}.
\endaligned \tag 6.21 $$
Restrictions on the functions $H,H_1,G,G_1$ and the range of the parameter $R$
given in section~5 have been already imposed (see section 5).

Relations (6.19) and (6.21) are analogues of the B\"acklund transformation for
equation (6.19). At the same time this action is a certain representation of the gauge group (a local one)
on the manifold of arbitrary functions and their first derivatives parametrizing the solution. In this way
the B\"acklund transformation for the  equation (6.19) is
a nonlinear representation of the gauge group
(a local one) which has  originated from the group $L_1$.

From the geometrical point of view the solution which corresponds
to the group $L_2$ is very interesting. The restrictions have forced the
Riemann invariant $R$ to belong to an interval.
But the function $e^{2H}$, which is the  length of the velocity vector,
has the same values on the edges of the interval. So it is
possible to identify these edges. In this way we obtain a fibre bundle over a circle.

%########################
%   SEC 7
%########################
\vskip2\baselineskip
\centerline{\bf 7. Equation for potential nonstationary flow of a compressible
perfect gas}
\centerline{\bf (the second example of an application of the method)}
\vskip\baselineskip

This section is devoted to the equation for potential nonstationary
flow of compressible perfect gas.
We discuss the flat nonstationary flow of a compressible gas described by
the potential of velocity and density
$$\ln\rho=-\Phi_{t}\quad,\quad \vec v=\vec \nabla\Phi\ .\tag 7.1$$
This assumption allows us to describe discontinuities of velocity and density as a change of the gauge of this potential on both sides
of the surface of discontinuity. Eliminating the density $\rho$ from the mass conservation law,
by means of the Euler equation
$$\Bigl(\frac\partial{\partial t}+\vec v\vec \nabla\Bigr)\vec v=-\frac{
\vec \nabla p}\rho=-\frac {c^2}\rho\;
\vec \nabla\rho\ , $$
we get
$$\frac{\partial\ln\rho}{\partial t}+c^2\;\text{div}\;
\vec v-\vec v\Bigl(\frac{\partial\vec  v}{\partial t}+(\vec v
\;\vec \nabla)\vec v\Bigr)=0\ .
$$
Introducing here the potential according to equation (7.1) we find
(References [1--3])
$$
\Phi_{,tt}+2(\Phi_{,x}\Phi_{,xt}+\Phi_{,y}\Phi_{,yt})+2\Phi_{,x}
\Phi_{,y}\Phi_{,xy}+
(\Phi_{,x}^2-c^2)\Phi_{,xx}+(\Phi_{,y}^2-c^2)\Phi_{,yy}=0\ ,
\tag 7.2 $$
where the lowest indices are for partial  derivatives, $\Phi \colon D
\subset R^3 \to R$, $D$ an open set.
The velocity of sound $c^2$
is given by a variant of the (compressible) Bernoulli equation:
$$\align
&\frac{\partial\Phi}{\partial t}+\frac12\;(\vec
\nabla\Phi)^2+\int\frac{dP}\rho=
\text{const}\ ,\quad  modulo\ \text{equation (7.1)},\\
&(1-c^2)\frac{\partial\Phi}{\partial t}+\frac12\;\vec v^2=\text{const.}
\endalign $$
We consider the special situation when the velocity of sound is constant $c^2\approx
c_0^2$ (see References 11 and 12).

The above treatment of the potential, nonsteady, compressible flow is not very
well known for the class of flows considered. Moreover it transforms to the correct equation
(see References 11 and 12) without redefinition of the potential as in Thompson's book
[13]. Thus all hydrodynamical quantities $(\vec v,\rho)$ are treated equally in this
particular potential approach.

It is easy to notice that equation (7.2) is a hyperbolic equation
of the second order. We are interested in finding solutions that can be described
by means of the Riemann invariant method. According to the requirements of the method
we transform equation (7.2) by introducing some new dependent variables into the quasilinear system
of equations of the first order,
$$\aligned
&\phi_{0,t}+2    (\phi_1\phi_{0,x}+\phi_2\phi_{0,y})+2\phi_1\phi_2\phi_{1,y}+
(\phi_1^2-c^2)\phi_{1,x}+(\phi_2^2-c^2)\phi_{2,y}=0\ ,\\
&\phi_{0,x}-\phi_{1,t}=0\quad,\quad \phi_{1,y}-\phi_{2,x}=0\ ,\\
&\phi_{0,y}-\phi_{2,t}=0\ ,\endaligned \tag 7.3 $$
where we introduce the notation
$$\aligned
&\Phi_{,t}=\phi_0\ ,\quad\Phi_{,x}=\phi_1\ ,\quad\Phi_{,y}=\phi_2\ ,\\
&\vec v=(\phi_1(t,x,y),\phi(t,x,y))=\\
&\rho=\exp(-\phi_0(t,x,y))\ .\endaligned
\tag 7.4 $$
Thus equation (7.2) is reduced to an undetermined system of four equations for three functions
$\phi_0,\phi_1,\phi_2$.

It is convenient to do the following transformations:
$$t\to t'=c_0t\ ,\quad\phi_0\to\phi_0'=\phi_0/c_0\ . \tag 7.5 $$
Then we can write (7.3) in the folowing form:
$$\aligned
&\phi_{0,t}'+2(\phi_1'\phi_{0,x}'+\phi_2'\phi_{0,y}')+2\phi_1'\phi_2'\phi_{1,y}'+
(\phi_1'{}^2-1)\phi_{1,x}'+(\phi_2'{}^2-1)\phi_{2,y}'=0,\\
&\phi_{0,x}'-\phi_{1,t}'=0\ ,\quad\phi_{1,y}'-\phi_{2,x}=0,\\
&\phi_{0,y}'-\phi_{2,t}'=0.
\endaligned \tag 7.6 $$
The field of velocity of flow $v$ and the density $\rho$ are then in the following form:
$$\align
&\vec v=c_0(\phi_1'(c_0t,x,y),\phi_2'(c_0t,x,y))\ ,\\
&\rho=\exp(-c_0\phi_0'(c_0t,x,y))\ .\endalign $$

%########################
%   SEC 8
%########################
\vskip2\baselineskip
\centerline{\bf 8. Simple integral elements}
\centerline{\bf (the second example of an application of the method)}
\vskip\baselineskip

In this section we calculate simple integral elements for the
equation for potential nonstationary flow of a compressible perfect gas.

Let us write equation (7.2) using simple integral elements. We get
$$\align
&\gm^0\lm_0\text{+}2(\phi_1\gm^0\lm_1\text{+}\phi_2\gm^0\lm_2)\text{+}2\phi_1\phi_2\gm^1\lm_2\text{+}
(\phi_2^2-c^2)\gm^2\lm_2\text{+}(\phi_1^2-c^2)\gm^1\lm_1\text{=}0,\ \
\tag 8.1a\\
&\gm^\mu\lm_\nu-\gm^\nu\lm_\mu   =0\ ,\qquad \mu,\nu=0,1,2.\tag 8.1b
\endalign $$
From equation (8.1b) we find that the vector $\gm$ is  proportional to the vector $\lm$. Thus, by
inserting $\gm\sim\lm$ into equation (8.1a) we get
a quadratic form with respect to $\lm_0,\lm_1,\lm_2$:
$$\aligned
Q(\lm_0,\lm_1,\lm_2)&=\lm_0^2+2\lm_0(\phi_1\lm_1+\phi_2\lm_2)+
2\phi_1\phi_2\lm_1\lm_2+
\\&+(\phi_2^2-c^2)\lm_1^2+(\phi_2^2-c^2)\lm_2^2=0\ .
\endaligned \tag 8.2 $$

Now, we follow Reference [7] and section~4.
We transform the quadratic form (8.2) to a canonical form and
search for eigenvalues
of matrix $A$ (the matrix of the quadratic form $Q$),
$$\det (A_{ij}-\mu\delta_{ij})=0\ ,\tag 8.3 $$
where
$$A=\left(\matrix 1&\phi_1&\phi_2\\\phi_1&(\phi_1^2-c^2)&\phi_1\phi_2\\
\phi_2&\phi_1\phi_2&(\phi_1^2-c^2)\endmatrix\right)\ .$$
We get an algebraic equation of third order with respect to the quantity $\mu$, i.e.
$$
(c^2+\mu)[\;-\mu^2+\mu(\phi_1^2+\phi_2^2-c^2+1)+c^2\;]=0\ . $$
The eigenvalues $\mu$ are real,
$$\aligned
&\mu_1=-c^2\ ,\\
&\mu_{2,3}=\tfrac12(\phi_1^2+\phi_2^2-c^2+1\pm\sqrt{\Delta})\ ,
\endaligned \tag 8.4 $$
where $\Delta=(\phi_1^2+\phi_2^2-c^2+1)^2+4c^2>0$. Thus the  quadratic
form (8.2)
transforms to a canonical form:
$$
Q(y,y)=-c^2y^2_1+\tfrac12(\phi_1^2+\phi_2^2-c^2+1+\sqrt{\Delta})y^2_2+
\tfrac12(\phi_1^2+\phi_2^2-c^2+1-\sqrt{\Delta})y_3^2\ .
\tag 8.5
$$
According to References [7] and [8] and section~4 we parametrize the covector $\lm$.
To do this we search for a parametric equation of (8.5).

Let us suppose that $y_1\ne0$. Then equation (8.5) may be written in the form
$$
\frac{X^2}{a^2}-\frac{Y^2}{b^2}=2c^2\ ,
\quad\text{where}\quad X=\frac{y_2}{y_1}\ ,\quad Y=\frac
{y_3}{y_1}\ ,
\tag 8.6 $$
and $$
\aligned
&a^2=(\phi_1^2+\phi_2^2-c^2+1+\sqrt\Delta)^{-1}\ ,\\
&b^2=(\sqrt\Delta-\phi_1^2-\phi_2^2+c^2-1)^{-1}\ .\endaligned \tag 8.7 $$
If $y_1=0$, $y_2\ne0$ and $y_3\ne0$, then equation (8.5) is
$$
(v^2-c^2+1+\sqrt\Delta)y_2^2+(v^2-c^2+1-\sqrt\Delta)y_3^2=0\ ,\tag 8.8 $$
where $v^2=\vphi_1^2+\vphi^2=\vec v{}^2$.
Equation (8.6) is the equation of a hyperbola.

Thus we write in a parametric form
$$X=\sqrt2\;ac\;\text{ch}\;\tau\ ,\qquad Y=\sqrt2\;bc\;\text{sh}\;\tau\ ,
\tag 8.9 $$
where $\tau$ is an arbitrary function of $\phi_i$, $i=0,1,2$. Thus the covector $\lm$ (and consequently $\gm$)
become
$$\gm_1\sim\lm^1=B\left(\matrix 1\\x\\y\endmatrix\right)=B\left(
\matrix1\\\sqrt2\;ac\;\text{ch}\;\tau\\\sqrt2\;bc\;\text{sh}\;\tau\endmatrix\right)\ ,
\tag 8.10$$
where $B$ is an orthogonal matrix, which diagonalizes the matrix $A$, i.e.,
$$B^TAB=\left(\matrix\mu_1&0&0\\0&\mu_2&0\\0&0&\mu_3\endmatrix\right)\ ,\quad B^T=B^{-1}\ .
\tag 8.11$$
Matrix $B$ is built from the eigen vectors of matrix $A$ and takes the form
$$
B=h\left(\matrix
0&c^2+1-v^2+\sqrt\Delta&c^2+1-v^2-\sqrt\Delta\\
-\phi_2&2\phi_1&2\phi_1\\\phi_1&2\phi_2&2\phi_2\endmatrix\right)\ ,
\tag 8.12 $$
where
$$h=[v^2\{(c^2+1-v^2+\sqrt\Delta)^2+4v^2\}
\times\{(c^2+1-v^2-\sqrt\Delta)^2+4v^2\}]^{1/2}.
$$
Inserting (8.8) and (8.12) into (8.10) we get a covector $\lm$.

In the case with $y_1=0$ we get
$$X=y_3/y_2=\ep b/a\ ,\quad \ep^2=1\ . \tag 8.13 $$
and finally we obtain
$$\gm_2\sim\lm^2=B\left(\matrix0\\1\\x\endmatrix\right)=B\left(\matrix 0\\1\\
\ep b/a\endmatrix\right)\ . \tag 8.14 $$
Inserting (8.8) and (8.12) into (8.14) we derive a simple element. According to
References [7] and [8] we consider first
the case with $y_2\ne0$ and then with $y_3\ne0$. We may introduce the following coordinate systems:
$$X=\frac{y_1}{y_2}\ ,\quad
\frac{y_3}{y_2}\ ,\qquad\text{or}\qquad
X=\frac{y_1}{y_3}\ ,\quad Y=\frac{y_2}{y_3}\ .
$$
Proceeding as before we get the following simple elements:
$$\gm_3\sim\lm^3=B\left(\matrix x\\1\\y\endmatrix\right)=B\left(\matrix(1/\sqrt2ac
)\cos\tau\\1\\(b/a)\sin\tau\endmatrix\right) \tag 8.15 $$
and $$
\gm_4\sim\lm^4=B\left(\matrix x\\y\\1\endmatrix\right)=B\left(\matrix (1/\sqrt2bc)
\text{sh}\;\tau\\(a/b)\text{sh}\;\tau\\1\endmatrix\right)\ .\tag 8.16$$
Let us consider the cases $y_2=0$ or $y_3=0$. We get
$$X=y_3/y_1=\ep bc\sqrt2\qquad\text{or}\qquad X=y_2/y_1=\ep ac\sqrt2\ . $$
This time we obtain the following simple elements:
$$\gm_5\sim\lm^5=B\left(\matrix1\\0\\x\endmatrix\right)=B\left(\matrix 1\\0\\\ep bc\sqrt2
\endmatrix\right) \tag 8.17 $$
and
$$\gm_6\sim\lm^6=B\left(\matrix1\\x\\0\endmatrix\right)=B\left(\matrix
1\\\ep ac\sqrt2\\0\endmatrix\right)\ . \tag 8.18 $$
Thus we get six kinds of simple elements that will be used for the
construction of solutions, i.e. simple waves and their interactions, the so--called
double and multiple waves. All of those simple elements are presented in
Appendix~B.

\proclaim Theorem.
All simple elements of the equation (7.2) are $\lm^1$, $\lm^2$, $\lm^3$,
$\lm^4$, $\lm^5$.

Proof has been given above.

%##########################
%    SEC 9
%##########################
\vskip2\baselineskip
\centerline{\bf 9. Simple waves}
\centerline{\bf (the second example of an application of the method)}
\vskip\baselineskip

This section is devoted to simple waves for the equation of potential
nonstationary flow for a compressible perfect gas.

We have the following cases (see [14]).

%\newpage

{\bf A. Case I ($\bold{\gm_1\sim\lm^1}$ --- see Appendix B)}

The simple wave, according to the considerations of section~2, is reduced here to fulfilling conditions
(2.3) and (2.4). Inserting the simple integral element (8.10) into
equation (2.3), we get
$$\align
&\frac{d\phi_0}{dR}=c\sqrt2\Biggl[\frac{(c^2+1-v^2+\sqrt\Delta)}{(v^2-c^2+1+\sqrt\Delta)
^{1/2}}\cos\tau+
\frac{(c^2+1-v^2-\sqrt\Delta)}{(c^2-1-v^2+\sqrt\Delta)^{1/2}}\sin\tau\Biggr],
\quad\tag 9.1a\\\\\\
&\frac{d\phi_1}{dR}=-\phi_2+2c\sqrt2\phi_1\Biggl[
\frac{\text{ch}\;\tau}{(v^2-c^2+1+\sqrt\Delta)^{1/2}}+
\frac{\text{sh}\;\tau}{(c^2-1-v^2+\sqrt\Delta)^{1/2}}\Biggr],
\quad\tag 9.1b\\\\\\
&\frac{d\phi_2}{dR}=\phi_1+2c\sqrt2\phi_2\Biggl[
\frac{\text{ch}\;\tau}{(v^2-c^2+1+\sqrt\Delta)^{1/2}}+
\frac{\text{sh}\;\tau}{(c^2-1-v^2+\sqrt\Delta)^{1/2}}\Biggr],\quad
\tag 9.1c
\endalign $$
where $v^2=\phi_1^2+\phi_2^2$, $\Delta=(v^2-c^2+1)^2+4c^2$. We are interested here
in solving the system (7.3) with respect to the potential of the velocity field
$v=(\phi_1,\phi_2)$ and density $\rho=\exp(-\phi_0)$. Now let us assume that the expression
in the square
brackets in equation (9.1b) and (9.1c) is a smooth function of $v^2$, i.e.
$$
\frac{\text{ch}\;\tau}{(v^2-c^2+1+\sqrt\Delta)^{1/2}}+
\frac{\text{sh}\;\tau}{(c^2-1-v^2+\sqrt\Delta)^{1/2}}=f(v^2)\ .
\tag 9.2 $$
That means $\tau=\tau(v^2)=\tau(R)$.

In this case we are able to find a solution in closed form. We introduce the quantity
$\mu_0=\phi_1/\phi_2$ and divide equation (9.1b) [resp. equation (9.1c)] by $\phi_1$ (resp.
$\phi_2$). Then we subtract both sides and finally we integrate and obtain
$$\mu_0=\tan(c_1-R)\ ,\quad c_1=\text{const.} \tag 9.3 $$
Then we introduce the quantity $v^2=\phi_1^2+\phi_2^2$. Thus, by multiplying
equation (9.1b)
by $\phi_1$ and (9.1c) by $\phi_2$ summing both sides, and then integrating
we have $$
F(v^2)=\int_a^{v^2}\frac{dr}{cr\;f(r)}=8\sqrt2R+c_2\ , \tag 9.4 $$
where $a,c_2=\text{const}.$

Now we assume there exists an inverse function $G$ of $F$ such that
$$G(F(r))=r\ ,\quad r>0\ , $$
and we get $$v^2=G(s)\ ,\qquad\text{where}\quad
s=8\sqrt2R+c_2\ .\tag 9.5 $$
The function $G$ is an arbitrary nonnegative function and it obeys the equation
$$\frac{dG}{ds}=\Bigl(\frac{dF}{dv^2}\Bigr)^{-1}=f(G(s))\ . \tag 9.6 $$
From equations (9.3) and (9.5) we can calculate
$$\aligned
&\phi_1=\ep G^{1/2}(8\sqrt2R+c_2)\sin(c_1-R)\ ,\\
&\phi_2=\ep G^{1/2}(8\sqrt2R+c_2)\cos(c_1-R)\ ,\quad \ep^2=1\ .
\endaligned \tag 9.7 $$
Inserting (9.7) into (9.2) and simultaneously using relations (9.4) and (9.6) and then
introducing $G=e^{2H}$ we obtain
$$\align
&\frac{dH}{ds}\Bigl|_{s=8\sqrt2R+c_2}=\frac1{2c}\Bigl[\frac{\text{ch}\;\tau}
{(1+e^{2H}-c^2+\sqrt\Delta)^{1/2}}+
\frac{\text{sh}\;\tau}{(c^2-1-e^{2H}+\sqrt\Delta)^{1/2}}\Bigr]\ ,\quad
\tag 9.8\\\\
&\Delta=(e^{2H}-c^2+1)^2+4c^2\ .  \endalign $$
where $H$ is an arbitrary function of $s$. So, it is obvious that if
the function $\tau$ is given, then the function $H$
is given as well and vice versa. But the function $H$
is more convenient for parametrizing
the simple element. Since we regard the quantity $\tau$ as given, we are led to
solving an ordinary differential equation with respect to $H$. Thus, from
equation (9.8) we get
$$
e^\tau=\al/2\sqrt\Delta\ ,$$
where
$$\al=4\frac{dH}{ds}
+2\Bigl(8\Bigl(\frac{dH}{ds}\Bigr)^2+e^{2H}+1-c^2\Bigr)^{1/2} \tag 9.9 $$
and $$
\text{ch}\;\tau=\frac{\al^2+4\Delta}{4\al\sqrt\Delta}\ ,\quad \text{sh}\;\tau
=\frac{\al^2-4\Delta}{4\al\sqrt\Delta}\ .\tag 9.10 $$
As in [7] we search for a restriction of the function $H$. It is
obvious that it must be
$$\aligned
&(1)\qquad 8\Bigl(\frac{dH}{ds}\Bigr)^2+e^{2H}+1-c^2\ge0\ ,\\
&(2)\qquad \al>0\ .\endaligned \tag 9.11 $$
Condition (2) is easily satisfied by assuming that $dH/ds\ge0$. If $c=\text{const}$,
we may substitute $c=1$ and both conditions are always satisfied [cf.
section~7, equations (7.5) and (7.6)]. Hence, the simple wave corresponding to the simple element (8.12)
has the form
$$\aligned
&\vec v=\ep\exp[H(8\sqrt2+c_2)](\sin(c_1-R),\cos(c_1-R))\ ,\\
&\ep^2=1\ .\endaligned \tag 9.12 $$
The function $R=R(t,x,y)$ is to be understood in the context of
expression (2.3) which means that the three--dimensional vector
$\vec \nabla R(t,x,y)$ is proportional to the vector $\lm^1$. Then

%\newpage
$$
\aligned
R&=
\Psi\Bigl(\frac{\sqrt2c}{4\al\sqrt\Delta}\Bigl[\frac{2-e^{2H}+\sqrt\Delta}
{(e^{2H}+\sqrt\Delta)^{\frac12}}
(\al^2+4\Delta)+\frac{2-e^{2H}-\sqrt\Delta}{(-e^{2H}+\sqrt\Delta)^{\frac12}}
(\al^2-4\Delta)\Bigr]t+
\\\\
&+\ep \exp H\Bigl\{\Bigl[ 4\sqrt2\frac{dH}{ds}\sin(c_1-R)-\cos(c_1-R)\Bigr]x
+
\\\\
&+4\sqrt2\Bigl[\frac{dH}{ds}\cos(c_1-R)+\sin(c_1-R)\Bigr]y\Bigr\}\Bigr)\ .
\endaligned \tag 9.13 $$
The density $\rho$ is given by
$$\rho=\rho_0\exp(-\phi_0)\ ,
\qquad \rho_0=\text{const}, \tag 9.14 $$
where $\phi_0$ is
$$\align
\phi_0&\text{=}
\int_{R_0}^R\frac{c\;dR'}
{\sqrt2(4dH/ds\text{+}[2(8(dH/ds)^2\text{+}e^{2H}\text{--}1\text{--}c^2)]^{\frac12})((e^{2H}\text{--}c^2\text{+}1)^2\text{+}4c^2)^{\frac12}
}\times\\
\vphantom{M}
\\
&\times
\Biggl\{\frac{
c^2\text{+}1\text{--}e^{2H}\text{+}((e^{2H}\text{--}c^2\text{+}1)^2\text{+}4c^2)^{\frac12}}{[e^{2H}\text{--}c^2\text{+}1\text{+}
((e^{2H}\text{--}c^2\text{+}12\text{+}4c^2)]^{\frac12}}
\Bigl(16\Bigl(\frac{dH}{ds}\Bigr)^2\text{+}4\frac{dH}{ds}\Bigl[
2\Bigl(8\Bigl(\frac{dH}{ds}\Bigr)^2\text{+}e^{2H}\text{--}1\text{--}c^2
\Bigr)\Bigr]^{\frac12}\text{+}
\\ \vphantom{M} \\
&\text{+}2e^{4H}\text{+}2c^4\text{+}1\text{--}4c^2e^{2H}\text{+}5e^{2H}\text{+}8c^2\Bigr)\text{+}\frac
{c^2\text{--}1\text{--}e^{2H}\text{--}((e^{2H}\text{--}c^2\text{+}12\text{+}4c^2)^{\frac12}}
{[c^2\text{--}1\text{--}e^{2H}\text{+}((e^{2H}\text{--}c^2\text{+}1)^2\text{+}4c^2)^{\frac12}]}
\times\\ \vphantom{M} \\
&\times\Bigl(16\Bigl(\frac{dH}{ds}\Bigr)^2
\text{+}4\frac{dH}{ds}\Bigl[2\Bigl(8\Bigl(\frac{dH}
{ds}\Bigr)^2\text{+}e^{2H}\text{--}1\text{--}c^2\Bigr)\Bigr]^{\frac12}
\text{--}2e^{4H}\text{--}2c^4\text{--}3\text{+}4c^2e^{2H}\text{--}3e^{2H}
\text{--}5c^2\Bigr)\Biggr\}.
\\ &\hphantom{aaa} \tag 9.15 \\ \endalign $$
The function $H$ is a function of $s=8\sqrt2R+c_2$ and is arbitrary
($c_2=\text{const}$)
and $\dfrac{dH}{ds}\ge0$.

Now we introduce the quantity
$$\delta=\lm_0+\vec v\cdot\vec\lm\ ,\tag 9.16$$
which has a physical interpretation as the velocity of a moving wave with respect
to the medium, whereas $\lm_0$ is the
local velocity of the wave. In our case $\delta$ and $\lm_0$ take the following form:
$$
\align
\lm_0=&\frac{\sqrt2}{4\al\sqrt\Delta }\Bigl[\frac{c^2-e^{2H}+\sqrt\Delta }{(e^{2H}+\sqrt\Delta
)^{1/2}}(\al^2+4\Delta )+\frac{2-e^{2H}-\sqrt\Delta }{(-e^{2H}+\sqrt\Delta )^{1/2}}(\al^2-4\Delta )
\Bigr]\ ,\ \
\\\\
\delta=&\lm_0+\exp(2H)\Biggl\{\Bigl[
4\sqrt2\frac{dH}{ds}\sin(c_1-R)-\cos(c_1-R)\Bigr]\sin(c_1-R)+\ \
\tag 9.17
\\
&+\Bigl[4\sqrt2\frac{dH}{ds}\cos(c_1-R)+\sin(c_1-R)\Bigr]\cos(c_1-R)\Biggr\}\ .
\ \
\endalign $$

\vskip6pt
{\bf B. Case II ($\bold{\gm_2\sim\lm^2}$ --- see Appendix B)}

A simple wave corresponding to the simple integral element (8.14) may be found by integrating the following
system of equations:
$$\align
&\frac{d\phi_0'}{dR}=(2-v^2+\sqrt\Delta )(\sqrt\Delta -v^2)^{1/2}+(2-v^2-\sqrt\Delta )(v^2+\sqrt\Delta )^{1/2}\ ,
\tag 9.17a\\\\
&\frac{d\phi_1'}{dR}=2\phi_1'[2(\sqrt\Delta +2)]^{1/2}\ ,\tag 9.17b\\\\
&\frac{d\phi_2'}{dR}=2\phi_2'[2(\sqrt\Delta +2)]^{1/2}\ ,\tag 9.17c
\endalign $$
where $v^2=\phi_1'{}^2+\phi_2'{}^2$, $\Delta =v^4+4$. We assume here that the velocity of sound $c=1$. By dividing both sides
of equations (9.1b) and (9.1c) and integrating we get
$$\phi_1'=c_1\phi_2'\ ,\qquad c_1=\text{const}.\tag 9.18$$
Now we calculate the quantity $v=|\vec v|$:
$$v=[(1/(c_2-2\sqrt2R)^2-2)^2-4]^{1/4}\ ,\qquad c_2=\text{const}.\tag 9.19 $$
Hence we obtain
$$\aligned
&\phi_1'=\frac{\ep c_1}{\sqrt{c_1^2+1}}\Bigl[\Bigl(\frac1{(c_2-2\sqrt2R)^2}-2\Bigr)
^2-4\Bigr]^{1/4}\ ,\\\\
&\phi_2'=\frac\ep{\sqrt{c_1^2+1}}\Bigl[\Bigl(\frac1{(c_2-2\sqrt2R)^2}-2\Bigr)^2-4\Bigr]^{1/4}
\ ,\\\\
&\ep^2=1\ .\endaligned \tag 9.20 $$
Substituting (9.19) into (9.1a) and then integrating, we have
$$
\phi_0'=-6\Bigl[\sqrt2(c_2-2\sqrt2R)+\arctan\frac{1-(1-2(c_2-2\sqrt2R)^2)^{1/2}}
{\sqrt2(c_2-2\sqrt2R)}\Bigr]\ .
\tag 9.21 $$
Thus a simple wave corresponding to the simple element
(8.14) has the following form:
$$\aligned
&\vec v=\frac{\ep}{\sqrt{c_1^2+1}}\Bigl[\Bigl(\frac1{(c_2-2\sqrt2R)^2}-2\Bigr)^2-4\Bigr]^{1/4}
(c_1,1)\ ,\\\\
&\rho=\rho_0\exp\Bigl[6\sqrt2R)-\arctan\frac{1-(1-2(c_2-2\sqrt2R)^2)^{1/2}}{\sqrt2
(c_2-2\sqrt2R)}\Bigr]\ .
\endaligned \tag 9.22 $$
The dependent variable $R$, i.e., the Riemann invariant, is given explicity
by the formula
$$
\align
R
&\text{=}
\Psi\Bigl(\Bigl[\Bigl(
\frac1{(c_2\text{--}2\sqrt2R)^2}\text{--}\Bigl[\Bigl(
\frac1{(c_2\text{--}2\sqrt2R)^2}
\text{--}2\Bigr)^2\text{--}4
\Bigr]^{\frac12}\Bigr)
%...
\Bigl(
\frac1{(c_2\text{--}2\sqrt2R)^2}\text{--}2\text{--}\Bigl[\Bigl(
\frac1{(c_2\text{--}2\sqrt2R)^2}\text{--}2\Bigr)^2\text{--}4\Bigr]
^{\frac12}\Bigr)^{\frac12}\text{+}
\\\\
&\text{+}\Bigl(4\text{--}\Bigl[\Bigl(\frac1{(c_2\text{--}2\sqrt2R)^2}\text{--}2\Bigr)^2\text{--}4\Bigr]^{\frac12}\text{--}
\frac1{(c_2\text{--}2\sqrt2R)^2}\Bigr)
\Bigl(\Bigl[\Bigl(\frac1{(c_2\text{--}2\sqrt2R)^2}\text{--}2\Bigr)^2\text{--}
4\Bigr]^{\frac12}\text{+}
\endalign$$

$$\align
&\text{+}\frac1{(c_2\text{--}2\sqrt2R)^2}\text{--}2\Bigr)^{\frac12}\Bigr]t
 \text{+}\frac{2\ep}{(c_1^2\text{+}1)^{\frac12}}
\Bigl[\Bigl(\frac1{(c_2\text{--}2\sqrt2R)^2}\text{--}2\Bigr)^2\text{--}4\Bigr]
^{\frac14}\times
\\\\
&\times\Bigl\{\Bigl(\frac1{(c_2\text{--}2\sqrt2R)^2}\text{--}2\text{--}
\Bigl[\Bigl(\frac1{(c_2\text{--}2\sqrt2R)^2}
\text{--}2\Bigr)^2\text{--}4\Bigr]^{\frac12}\Bigr)^{\frac12}\text{+}
\\\\
&\text{+}\Bigl(\frac1{(c_2\text{--}2\sqrt2R)^2}\text{--}2\text{+}\Bigl[\Bigl(\frac1{(c_2\text{--}2
\sqrt2R)^2}\text{--}2\Bigr)^2\text{--}4\Bigr]
^{\frac12}\Bigr)^{\frac12}\Bigr\}(c_1x\text{+}y)\Bigr),
\\ & \ \tag 9.23
\endalign $$
where $\Psi$ is an arbitrary smooth (at least of class $C^2$)
function of one variable. The solution (9.22 ) with condition
(9.23 ) describes one--dimensional, nonstationary flow that goes in the direction
$(c_1,1)$. It is worth mentioning that the solution is defined everywhere
except of $R_0=c_2/2\sqrt2$. The quantities $\lm_0$ and $\delta$ (the total velocity
of the wave and the velocity of the wave with respect to the medium) are
as follows:
$$\align
&\lm_0=
\Bigl[\Bigl(\frac1{(c_2\text{--}2\sqrt2R)^2}\text{--}\Bigl[\Bigl(\frac1{(c_2\text{--}2\sqrt2R)^2}\text{--}2\Bigr)^2\text{--}4\Bigr]^{\frac12}\Bigr)
\Bigl(\frac1{(c_2\text{--}2\sqrt2R)^2}\text{--}2\text{--}\Bigl[\Bigl(\frac1{(c_2\text{--}2\sqrt2R)^2}\text{--}2\Bigr)^2
\text{--}4\Bigr]^{\frac12}\Bigr)^{\frac12}\text{+}
\\ \vphantom{M} \\
&\text{+}
\Bigl(4\text{--}\Bigl[\Bigl(\frac1{(c_2\text{--}2\sqrt2R)^2}\text{--}2\Bigr)^2\text{--}4\Bigr]^{\frac12}\text{--}\frac1{(c_2\text{--}2\sqrt2R)^2}\Bigr)
\Bigl(\Bigl[\Bigl(\frac1{(c_2\text{--}2\sqrt2R)^2}\text{--}2\Bigr)^2\text{--}4\Bigr]^{\frac12}\text{+}
\frac1{(c_2\text{--}2\sqrt2R)^2}\text{--}2\Bigr)\Bigr],
\endalign$$

$$\align
&\delta=
\lm_0\text{+}2\ep(c_1^2\text{+}1)^{\frac12}\Bigl[\Bigl(\frac1{(c_2\text{--}2\sqrt2R)^2}\text{--}2\Bigr)^2\text{--}4\Bigr]^{\frac12}
\Bigl\{\Bigl(\frac1{(c_2\text{--}2\sqrt2R)^2}\text{--}2\text{--}\Bigl[\Bigl(\frac1{(c_2\text{--}2\sqrt2R)^2}\text{--}2\Bigr)^2\text{--}4
\Bigr]^{\frac12}\Bigr)^{\frac12}\text{+}
\\\\\
&\text{+}\Bigl(\frac1{(c_2\text{--}2\sqrt2R)^2}\text{--}2\text{+}\Bigl[\Bigl(\frac1{(c_2\text{--}2\sqrt2R)^2}\text{--}2\Bigr)^2\text{--}4\Bigr]^{\frac12}\Bigr)^
{\frac12}\Bigr\}.
\\& \ \tag 9.24 \endalign $$

Moreover we have a restriction:
$$\bigl(1/{(c_2-2\sqrt2R)^2}-2\bigr)^2-4\ge0\ . \tag 9.25 $$
This inequality may be easily solved and we obtain
$$R\ge(c_2-2)/2\sqrt2\quad\text{or}\quad R\le(2+c_2)/2\sqrt2\ .$$

\vskip10pt

{\bf C. Case III ($\bold{\gm_3\sim\lm^3}$ --- see Appendix B)}

A simple wave corresponding to simple integral elements (8.18) may be found by integrating
the following system of ordinary differential equations:~

$$\align
&\frac{d\phi_0}{dR}=(c^2+1-v^2+\sqrt\tr)+(c^2+1-v^2-\sqrt\tr)\times
\frac{(v^2-c^2+1+\sqrt\tr)^{1/2}}
{(c^2-1-v^2+\sqrt\tr{1/2}}\sin\tau,
\\\\\\
&\frac{d\phi_1}{dR}=\frac{-\phi_2}{c\sqrt2}(v^2-c^2+1+\sqrt\tr)^{1/2}\cos\tau+
2\phi_1\Bigl[1+\Bigl(\frac{v^2-c^2+1+\sqrt\tr}{c^2-1-v^2+\sqrt\tr}\Bigr)^{1/2}\sin\tau\Bigr]
, \\\\\\
&\frac{d\phi_2}{dR}=\frac{\phi_1}{c\sqrt2}(v^2-c^2+1+\sqrt\tr)^{1/2}\cos\tau+
2\phi_2\Bigl[1+\Bigl(\frac{v^2-c^2+1+\sqrt\tr}{c^2-1-v^2-\sqrt\tr}\Bigr)^{1/2}
\sin\tau\Bigr],
\\ &\ \tag 9.26 \endalign$$
where $v^2=\phi_1^2+\phi_2^2$, $\tr=(v^2-c^2+1)^2+4c^2$.

To solve the equations it is convenient to introduce a new variable $\mu_0=
\phi_1/\phi_2$. For quantities $\mu_0$ and $v^2$ we obtain the following system of equations:
$$\align
&\frac{dv^2}{dR}=\Bigl[1+\Bigl(\frac{v^2-c^2+1+\sqrt\tr}{c^2-1-v^2+\sqrt\tr}
\Bigr)^{1/2}\sin\tau
\Bigr], \tag 9.27 \\\\
&\frac d{dR}\arctan\mu_0=\frac{(\sqrt\tr-v^2+c^2-1)^{1/2}}{c\sqrt2}\cos\tau,
\tag 9.28 \endalign $$
where $\tau$ is an arbitrary function of $R$. It is convenient to substitute
$$v^2=e^{2H} \tag 9.29 $$
and $H$ parametrizes the simple element (8.15).

We find the restrictions for $H$. Substituting (9.29 ) into (9.27 ) we have
$$
\frac12\Biggl|\Bigl(\frac{dH}{dR}-2\Bigr)\Bigl(\frac{c^2-1-e^{2H}+\sqrt\tr}
{e^{2H}-c^2+1+\sqrt\tr}\Bigr)\Biggr|=|\sin\tau|\le1 \ . \tag 9.30 $$
Using the relations between trigonometrical functions we calculate $\cos\tau$ and substitute
into (9.28 )
$$\aligned
&\frac d{dR}\arctan\mu_0=
\ep_1\frac{(\sqrt\tr-e^{2H}+c^2-1)^{1/2}}{c\sqrt2}\times
\Bigl[\pm\frac14\Bigl(\frac{dH}{dR}-2\Bigr)^2\frac{(c^2-1-e^{2H}
+\sqrt\tr)}{(e^{2H}-c^2+1+\sqrt\tr)}\Bigr]^{1/2},
\\\\
&\ep_1^2=1.  \endaligned \tag 9.31 $$
The differential  inequality (9.30 ) leads us to
the solution for which the length of the vector $\vec v$ is constant, i.e.
$H=H_0=\text{const}$. Then by inserting the quantity $H=H_0$ into
(9.27 ) and (9.28 ) and integrating we finally get
$$\aligned
&\vec v=ce^{H_0}(\ep_2\sin(K(R)+c_1),\cos(K(R)+c_1))\ ,\\
&\rho=\rho_0\exp[-2c(R-R_0)(2-e^{2H_0})]\ ,\\\\
&K(R)=\ep_1\Bigl[\frac{e^{2H_0}[(e^{4H_0}+4)^{1/2}-e^{2H_0}]}
{(e^{4H_0}+4)^{1/2}+e^{2H_0}}\Bigr](R-R_0)\ ,  \\\\
&\ep_1^2=\ep_2^2=1\ ,\\\\
&R=\Psi(2c(2-e^{2H_0})t+e^{H_0}\bigl\{[4\ep_2\sin(K(R)+c_1)\\
&\qquad-\ep_1e^{H_0}\cos(K(R)+c_1)]x+
[4\cos(K(R)+c_1)+\ep_1\ep_2 e^{H_0}\sin(K(R)+c_1)]y\bigr\}.
\endaligned \tag 9.32 $$
The quantities $\lm_0$ and $\delta$ are
$$\aligned &\lm_0=2c(2-e^{2H_0})\ ,\\
&\delta=2c(2-e^{2H_0})+4e^{2H_0}\ ,\endaligned \tag 9.33 $$
and $c$ is the velocity of sound.

\vskip20pt
{\bf D. Case IV ($\bold{\gm_5\sim\lm^5}$ --- see Appendix B)}

A simple wave corresponding to the simple integral element (8.17) may be found by integrating
the following system of ordinary differential equations:
$$\align
&\frac{d\phi_0'}{dR}=\frac{\ep_1\sqrt2(2-v^2+\sqrt\tr)}
{(\sqrt\tr-v^2)^{1/2}}\ ,
\tag 9.34a\\\\
&\frac{d\phi_1'}{dR}=-\phi_2'+2\phi_1'\frac{\ep_1\sqrt2}
{(\sqrt\tr-v^2)^{1/2}}\ ,
 \tag 9.34b \\\\
&\frac{d\phi_2'}{dR}=+\phi_1'+2\phi_2'\frac{\ep_1\sqrt2}
{(\sqrt\tr-v^2)^{1/2}}\ ,
\tag 9.34c\endalign $$
where $\tr=v^4+4$, $v^2=\phi_1'{}^2+\phi_2'{}^2$, and we have assumed that the velocity
of sound equals 1. By dividing equation (9.34b) by $\phi_1'$ and equation (9.34c)
by $\phi_2'$ and subtracting both sides of them and integrating we obtain
$$\phi_1'/\phi_2'=\tan(c_1-R)\ ,\quad c_1=\text{const}. \tag 9.35 $$
Then we calculate $v^2$. We get
$$\align
F_1(v^2)&=\frac23((v^4+4)^{1/2} -v^2)^{3/2}+4\sqrt2\arctan\Bigl(\frac
{(v^4+4)^{1/2}
-v^2}2\Bigr)^{1/2}+\\\\
&+2\sqrt2\ln\frac
{((v^4+4)^{1/2}-v^2)^{1/2}+\sqrt2}{((v^4+4)^{1/2}-v^2)^{1/2}-\sqrt2}
=4\sqrt2R+c_2\ ,\tag 9.36\\\\
&c_2=\text{const}.
\endalign  $$
The function $F_1$ is monotone and consequently it has an inverse function $G_1$.
 Thus we have
$$v^2=G_1(4\sqrt2R+c_2)\ .\tag 9.37 $$
The domain of $G_1$ is $(-\infty,+\infty)$ and its range is $(0,+\infty)$.

From equations (9.35), (9.37), and (9.34a) we get
$$\aligned
&\phi_1'=\ep_2G_1^{1/2}(4\sqrt2R+c_2)\sin(c_1-R)\ ,\\\\
&\phi_2'=\ep_2G_1^{1/2}(4\sqrt2R+c_2)\cos(c_1-R)\ ,\\\\
&\phi_0'=\ep_2\sqrt2\int_{R_0}^R\frac
{(2-G_1+(G_1^2+4)^{1/2})}{((G_1^2+4)^{1/2}-G_1)^{1/2}}\;dR'+c_3\ ,\\\\
&c_3=\text{const}\ , \endaligned \tag 9.38 $$
where $G_1(F_1(x))=x$ and $G_1$ is a function of $4\sqrt2R'+c_2$ (see
Fig.~5).

Thus the  simple wave corresponding to the simple element (8.17) is
 $$\aligned
&\vec v=\ep_2 cG_1^{1/2}(4\sqrt2R+c_2)(\sin(c_1-R),\cos(c_1-R))\ ,\\\\
&\rho=\rho_0\exp\Bigl(-\ep_2\sqrt2\int_{R_0}^R\frac
{(2-G_1+(G_1^2+4)^{1/2})}{((G_1^2+4)^{1/2}-G_1)^{1/2}}\;dR'\Bigr)
\\\\ &\rho_0=\text{const}\ .
\endaligned
\tag 9.39 $$
Moreover one can perform the integration in the second formula of
(9.39) getting
$$\align
&\rho=\rho_0'\exp\Bigl(-4\ep_2\bigl(2\ln G_1-G_1+(4+G_1^2)^{1/2}\cdot
(1-\tfrac2G_1)\bigr)\Bigl|_{s=4\sqrt2R+c_2}\Bigr) \tag 9.39'
\\&\rho_0'=\text{const}\ .\endalign
$$
The dependent variable $R$, (i.e., Riemann invariant) is given in an implicit form
$$\aligned
R=&
\Psi\Bigl(\bigl[\frac{\ep_1\sqrt2(2-G_1+(G_1^2+4)^{1/2})}
{((G_1^2+4)^{1/2}-G_1)^{1/2}}\Bigr]ct+
\\\\
&+\ep_2G_1^{1/2}\Bigl\{\Bigl[2\sin(c_1-R)\frac{\ep_1\sqrt2}{((G_1^2+4)^{1/2}-G_1)^{1/2}}
-\cos(c_1-R)\Bigr]x+
\\\\
&+\Bigl[2\cos(c_1-R)\frac{\ep_1\sqrt2}{((G_1^2+4)^{1/2}-G_1)^{1/2}}+\sin(c_1-R)
\Bigr]
y\Bigr\}\Bigr)\ ,
\endaligned \tag 9.40 $$
where $\Psi$ is an arbitrary smooth (at least of $C^2$ class)
function of one variable. The quantities $\lm_0,\delta$, i.e.,
respectively a local wave velocity and velocity of a moving wave with respect to
the medium, equal
$$\aligned
&\lm_0=\frac{c\ep_1\sqrt2(2-G_1+(G_1^2+4)^{1/2})}{((G_1^2+4)^{1/2}-G_1)^{1/2}}\ ,\\\\
&\delta=\frac{\ep_1\sqrt2(2+G_1+(G_1^2+4)^{1/2})}{((G_1^2+4)^{1/2}-G_1)^{1/2}}\ .
\endaligned \tag 9.41 $$

\vskip10pt
{\bf E. Case V ($\bold{\gm_4\sim\lm^4}$ --- see Appendix B)}

A simple wave corresponding to a simple element (8.16) can be found by
integration of the following system of equations:
$$\aligned
&\frac{d\phi_0}{dR}=(c^2+1-v^2-\sqrt\tr)\Bigl(\frac
{\sqrt\tr+v^2+c^2-1}{\sqrt\tr+v^2-c^2+1}\Bigr)^{1/2}
\text{ch}\;\tau+(c^2+1-v^2-\sqrt\tr),
\\ &\vphantom{M}\\
&\frac{d\phi_1}{dR}=\frac{-\phi_2}{c\sqrt2}(\sqrt\tr-v^2+c^2-1)^{1/2}\text{sh}
\;\tau+2\phi_1\Bigl[1+\Bigl(\frac
{\sqrt\tr-v^2+c^2-1}{\sqrt\tr+v^2-c^2+1}\Bigr)^{1/2}\text{ch}\;\tau\Bigr],
\\&\vphantom{M}\\
&\frac{d\phi_2}{dR}=\frac{\phi_1}{c\sqrt2}(\sqrt\tr-v^2+c^2-1)^{1/2}\text{sh}\;\tau+2\phi_2\Bigl[
1+\Bigl(\frac{\sqrt\tr-v^2+c^2-1}{\sqrt\tr+v^2-c^2+1}\Bigr)^{1/2}
\text{ch}\;\tau\Bigr],
\endaligned \tag 9.42 $$
where $v^2=\phi_1^2+\phi_2^2$, $\tr=(v^2-c^2+1)^{1/2}+4c^2$.

We introduce new dependent variables $v^2$ and $\mu_0=\phi_1/\phi_2$ and we get
$$\align
&\frac12\;\frac{dv^2}{dR}=2v^2\Bigl[1+\Bigl(
\frac{v^2-c^2+1+\sqrt\tr}{c^2-1-v^2+\sqrt\tr}\Bigr)^{1/2}\text{ch}\;\tau\Bigr],
\tag 9.43 \\\\
&\frac d{dR}\arctan\mu_0=\frac{(\sqrt\tr-v^2+c^2-1)^{1/2}}{c\sqrt2}\;\text{sh}\;\tau,
\tag 9.44\endalign $$
where $\tau$ is an arbitrary function of $R$. It is convenient to substitute
$$v^2=e^{2H}\ .\tag 9.45 $$
Thus, the simple element (8.16) may be parametrized by the function $H$ instead of
$\tau$
and we find a restriction for $H$. By inserting (9.45)
into (9.43) we find
$$
1\le\text{ch}\;\tau=\frac12\Bigl(\frac{dH}{dR}-1\Bigr)\Bigl(\frac
{c^2-1-e^{2H}+\sqrt\tr}{e^{2H}-c^2+1+\sqrt\tr}\Bigr)^{1/2}\ . \tag 9.46 $$
Using relations between hyperbolic functions we calculate the quantity $\text{sh}\;\tau$
and then substitute it into equation (9.44) to get
$$\aligned
&\frac d{dR}\arctan\mu_0=
\frac{\ep_1(\sqrt\tr-e^{2H}+c^2-1)^{1/2}}{c\sqrt2}\Bigl[1-\frac14\Bigl(
\frac{dH}{dR}-2\Bigr)^2\frac{(\sqrt\tr-e^{2H}+c^2+1)}
{(\sqrt\tr+e^{2H}-c^2+1)}\Bigr]^{1/2},
\\\\
&\ep_1^2=1\ . \endaligned \tag 9.47 $$
The differential inequality (9.46) leads to a solution for which the function $H$ is constant,
$H=H_0=\text{const}$. It means that the length of the vector $\vec v$ is constant. Thus by
substituting $H=H_0$ into equations (9.42) and (9.47) and then integrating, we get
$$\aligned
&\phi_1'=\ep_2e^{H_0}\sin(K(R)+c_1)\ , \ep_2^2=1\ ,\\
&\phi_2'=e^{H_0}\cos(K(R)+c_1)\ ,\endaligned \tag 9.48 $$
where
$$\aligned
K(R)&=\ep_1\Bigl[\frac {e^{2H_0}(\sqrt\tr-e^{2H_0}+2)}{(\sqrt\tr+e^{2H_0})}
\Bigr]^{1/2}
(R-R_0)\ ,
\\\\
\tr&=e^{4H_0}+4\ ,
\\
c_1&=\text{const}\ ,\endaligned \tag 9.49 $$
and
$$
\phi_0'=c_2+[-(2-e^{2H_0}+\sqrt\tr)(\sqrt\tr-e^{2H_0})/(\sqrt\tr+e^{2H_0})+
(2-e^{2H_0}-\sqrt\tr)](R-R_0),
\tag 9.50 $$
where $c_2=\text{const}$.

Finally we have
$$\aligned
&\vec v=c\;e^{H_0}(\ep2\sin(K(R)+c_1),\cos(K(R)+c_1))\ ,
\\\\
&\rho=\rho_0\exp\{c[(2-e^{H_0}+\sqrt\tr)(\sqrt\tr-e^{2H_0})/(\sqrt\tr+e^{2H_0})
-(2-e^{2H_0}-\sqrt\tr)](R-R_0)\},
\\\\
&\rho_0=\text{const}. \endaligned
\tag 9.51 $$
The Riemann invariant $R$ is given in an implicit form as
$$\aligned
R&=
\Psi\Bigl(\Bigl[(2-e^{2H_0}-\sqrt\tr)-(2-e^{2H_0}\sqrt\tr)
\frac{\sqrt\tr-e^{2H_0}}{\sqrt\tr+e^{2H_0}}\Bigr]ct+
\\&\vphantom{M}\\
&+
e^{H_0}\Bigl\{\Bigl[4\ep_2\frac{e^{2H_0}}{\sqrt\tr+e^{2H_0}}\sin(K(R)+c_1)-
\frac{\ep_1(\sqrt\tr-e^{2H_0})^{1/2}e^{H_0}}{(\sqrt\tr+e^{2H_0})^{1/2}}\cos(K(R)+c_1)
\Bigr]x+
\\&\vphantom{M}\\
&+\Bigl[4\frac{e^{2H_0}}{\sqrt\tr+e^{2H_0}}\cos(K(R)+c_1)+\frac{\ep_1\ep_2
(\sqrt\tr
-e^{2H_0})^{1/2}}{(\sqrt\tr+e^{2H_0})^{1/2}}\sin(K(R)+c_1)\Bigr]y
\Bigr\}\Bigr),
\endaligned \tag 9.52 $$
where $\Psi$ is an arbitrary smooth function of one variable. The local velocity of
the wave and the velocity with respect to the medium are constant and given by
$$\aligned
&\lm_0=c\Bigl[(2-e^{2H_0}-\sqrt\tr)-(2-e^{2H_0}+\sqrt\tr)\frac{(\sqrt\tr-e^
{2H_0})}
{(\sqrt\tr+e^{2H_0})}\Bigr],
\\
&\delta=\lm_0+\frac{4e^{4H_0}}{\sqrt\tr+e^{2H_0}},
\endaligned \tag 9.53 $$
where $c=\text{const}$ and is the velocity of sound.

\newpage

{\bf F. Case VI ($\bold{\gm_6\sim\lm^6}$ --- see Appendix B)}

A simple wave corresponding to the simple
element (8.18) may be found by integration of the following system of equations~:
$$\aligned
&\frac{d\phi_0'}{dR}=\frac{\ep_1\sqrt2(2-v^2+\sqrt\tr)}{(v^2+\sqrt\tr)^{1/2}}\ ,
\quad \ep_1^2=1\ ,
\\\\
&\frac{d\phi_1'}{dR}=-\phi_2'+2\phi_1'\frac{\ep_1\sqrt2}{(v^2+\sqrt\tr)^{1/2}}\ ,
\\\\
&\frac{d\phi_2'}{dR}=\phi_1'+2\phi_2'\frac{\ep_1\sqrt2}{(v^2+\sqrt\tr)^{1/2}}\ ,
\endaligned \tag 9.54 $$
where $\tr=v^4+4$, $v^2=\phi_1'{}^2+\phi_2'{}^2$.

We assume that the velocity of sound equals 1 and introduce new variables $\mu_0=\phi_1'/\phi_2'$ and $v^2$.
As before we integrate equations for these variables and we get
$$\align
&\mu_0=\phi_1'/\phi_2'=\tan(c_1-R)\ , \quad c_1=\text{const}\ ,
\tag 9.55 \\\\
&F_2(v^2)=-2\arctan\Bigl(\frac{((v^4+4)^{1/2}-v^2)^{1/2}-\sqrt2}2\Bigr)^{1/2}
+2\bigl((v^4+4)^{1/2}-v^2)\bigr)^{1/2}+
\\\\
&\qquad+\ln\Bigl[\frac{((v^4+4)^{1/2}-v^2)^{1/2}-\sqrt2}
{((v^4+4)^{1/2}-v^2)^{1/2}+\sqrt2}\Bigr]=2\ep_1\sqrt2R+c_2\ , \quad
c_2=\text{const}.
\tag 9.56 \endalign $$
The function $F_2$ is monotone. So, in the interval $(0,+\infty)$ it posesses an inverse function
$G_2$ such as
$$G_2(F_2(x))=x \ . \tag 9.57 $$
We get
$$v^2=G_2(2\ep_1\sqrt2R+c_2)\ . $$
From (9.55), (9.57), and (9.54) we have
$$\aligned
&\phi_1'=\ep_2G_2^{1/2}(2\ep_1\sqrt2R+c_1)\sin(c_1-R)\ ,\\\\
&\phi_2'=\ep_2G_2^{1/2}(2\ep_1\sqrt2R+c_1)\cos(c_1-R)\ ,\\\\
&\phi_0'=\ep_1\sqrt2\int_{R_0}^R\frac
{(2-G_2+(G_2^2+4)^{1/2})}
{((G_2^2+4)^{1/2}+G_2)^{1/2}}\;dR'+c_3\ ,\quad c_3=\text{const.}
 \endaligned \tag 9.58 $$
where the function $G_2$ is given by a transcendental equation
\vskip8pt
$$\aligned
&\bigl(
2\bigl((c_1^2+4)^{1/2}-G_2\bigr)\bigr)^{1/2}+\ln\Bigl[\frac
{((G_2^2+4)^{1/2}-G_2)^{1/2}-\sqrt2}{((G_2^2+4)^{1/2}-G_2)^{1/2}+\sqrt2}\Bigr]
+
\\\\
&-2\arctan\Bigl(\frac{(G_2^2+4)^{1/2}-G_2}2\Bigr)^{1/2}=
2\ep_1\sqrt2R+c_2\ .
\endaligned \tag 9.59 $$
Thus a simple wave corresponding to the simple element (8.18) is
$$\aligned
&\vec v=\ep_2G_2^{1/2}(2\ep_1\sqrt2R+c_2)(\sin(c_1-R), \cos(c_1-R))\ ,\\\\
&\rho=\rho_0\exp\Bigl(-\ep_1\sqrt2\int_{R_0}^R\frac
{(2-G_2+(G_2^2+4)^{1/2})}{((G_2^2+4)^{1/2}+G_2)^{1/2}}\;dR'\Bigr)\\\\
&\rho_0=\text{const}\ .
\endaligned \tag 9.60 $$
Moreover one can performe an integration in the second formula of equation (9.60) getting
$$\align
&\rho=\rho_0'\exp\Bigl(-\tfrac14\bigl(2\ln G_2-G_2+
(4+G_2^2)^{1/2}(1-\tfrac2{G_2})\bigr)\Bigr|_{s=2\ep_1\sqrt2R+c_2}
\Bigr)\tag 9.60'\\
&\rho_0'=\text{const}\ . \endalign $$
The dependent variable $R$, the Riemann invariant, is given in an
implicit form as
$$\aligned
R&=
\Psi\Bigl(\Bigl[\frac{\ep_1\sqrt2(2-G_2+(G_2^2+4)^{1/2})}{(G_2+(G_2^2+4)^{1/2}
)^{1/2}}\Bigr]ct+
\\\\
&+\ep_2G_2^{1/2}\Bigl\{\Bigl[\frac{2\ep_1\sqrt2}{(G_2+(G_2^2+4)^{1/2})^{1/2}}\sin(c_1-R)-\cos(c_1-R)
\Bigr]x+
\\\\
&+\Bigl[\frac{2\ep_1\sqrt2}{(G_2+(G_2^2+4)^{1/2})^{1/2}}\cos(c_1-R)+\sin(c_1-R)\Bigr]y\Bigr\}\Bigr)\ ,
\endaligned \tag 9.61 $$
where $\Psi$ is an arbitrary function of one variable.
The quantities $\lm_0$, $\delta$, which are, respectively, the local wave velocity and
the velocity of the wave with respect to the medium, are given by
$$\aligned
&\lm_0=\frac{c\ep_1\sqrt2(2-G_2+(G_2^2+4)^{1/2})}{(c_1+(G_2^2+4)^{1/2})^{1/2}}\ ,
\\\\
&\delta=\frac{\ep_1\sqrt2(2+G_2+(G_2^2+4)^{1/2})}{G_2+(G_2^2+4)^{1/2})^{1/2}}\ .
\endaligned \tag 9.62 $$
Thus these simple waves are the basis for searching for a wider class of
solution, the so called double waves and
multiple waves. The superposition of this type may be very interesting from the physical point of view
and they will be considered in future papers.

It is interesting to notice that our calculations can be extended to the
three--dimensio\-nal
case, but this will cause very tedious and laborous algebra. It seems that
the assumption of the constancy of the velocity of sound can be abandoned.
However, we cannot use some mathematival tricks in the above calculations and
we probably cannot get compact results.

\proclaim Theorem.
There are simple waves for simple elements described in Appendix~B. All
details are given above.

The proof is also given above.

All described solutions have a gradient catastrophe on a certain
hypersurface $S$. On this hypersurface some shock waves can appear.

\newpage
\vskip2\baselineskip
\centerline{\bf Conclusions}
\vskip\baselineskip

In this paper we formulate an improved Riemann invariant method and apply it for two
important equations of gas dynamics. Using the method, we find several wide
classes of exact solutions with interesting properties which have physical
inter\-pre\-ta\-tions
and some applications in gas dynamics.

\newpage
%**************************
\vskip2\baselineskip
\centerline{\bf Appendix A --- simple elements (the first example)}
\vskip\baselineskip
 
{\bf Table 1.} Covectors $\lm$ ,\quad $(F_1,F_2)$
$$
\underset{(K=0)}\to{\ovj\lm}\text{=}
\pmatrix
\dfrac{(\chi_1^2\text{--}c^2)^{\frac12}}{\chi_2}
\Bigl[\text{--}\Bigl(\vphi_2\sin\al\text{+}\dfrac
{\vphi_1\vphi_3}
{\chi_1}\cos\al\Bigr)\text{+}\Bigl(\vphi_2\cos\al\text{--}\dfrac
{\vphi_1\vphi_3}{\chi_1}\sin\al\Bigr)
\text{sh}\rho
\Bigr]\text{+}\dfrac{c\vphi_1}{\chi_1}\text{ch}\rho
\\\\
\dfrac{(\chi_1^2\text{--}c^2)^{\frac12}}{\chi_2}\Bigl[\Bigl(
\vphi_1\sin\al\text{--}\dfrac{\vphi_2\vphi_3}{\chi_1}\cos\al
\Bigr)\text{--}\Bigl(\vphi_1\cos\al\text{+}\dfrac{\vphi_2\vphi_3}{\chi_1}
\sin\al\Bigr)\text{sh}\rho\Bigr]\text{+}\dfrac
{c\vphi_2}{\chi_1}\text{ch}\rho
\\\\
\dfrac{\chi_2(\chi_1^2\text{--}c^2)^{\frac12}}{\chi_1}(\cos\al\text{+}\sin\al
\;\text{sh}\rho)\text{+}\dfrac{c\vphi_3}
{\chi_1}\text{ch}\rho
\endpmatrix $$
\vskip10pt
$$
\underset{(K=0)}\to{\ovd\lm}=
\pmatrix
\dfrac{\ep(\chi_1^2-c^2)^{\frac12}}{\chi_2}\Bigl(\vphi_2\cos\al-\dfrac{\vphi_1\vphi_3}{\chi_1}\sin\al\Bigr)
+\dfrac{c\vphi_1}{\chi_1}
\\\\
\dfrac{-\ep(\chi_1^2-c^2)^{\frac12}}{\chi_2}\Bigl(\vphi_2\cos\al+\dfrac
{\vphi_2\vphi_3}{\chi_1}\sin\al\Bigr)+\dfrac{c\vphi_2}{\chi_1}
\\\\
\dfrac{\ep(\chi_1^2-c^2)^{\frac12}}{\chi_1}\chi_2\sin\al+
\dfrac{c\vphi_3}{\chi_1}
\endpmatrix
$$
\vskip10pt
Covectors $\lm$, \quad $K=1$, \quad $(F_3,F_4)$
\vskip10pt
$$
\underset{(K=1)}\to{\ovj\lm}\text{=}\pmatrix
\dfrac{(\chi_1^2\text{--}c^2)^{\frac12}}{\chi_2}\Bigl[\text{--}\Bigl(
\vphi_2\sin\al\text{+}\dfrac{\vphi_1\vphi_3}{\chi_1}
\cos\al)\text{+}\Bigl(\text{--}\vphi_2\cos\al\text{+}\dfrac{\vphi_1\vphi_3}
{\chi_1}\sin\al\Bigr)\text{sh}\rho\Bigr]\text{+}\dfrac{c\vphi_1}{\chi_1}
\text{ch}\rho
\\\\
\dfrac{(\chi_1^2\text{--}c^2)^{\frac12}}{\chi_2}
\Bigl[\Bigl(\vphi_1\sin\al\text{--}
\dfrac{\vphi_2\vphi_3}{\chi_1}\cos\al\Bigr)\text{+}\Bigl(\vphi_1\cos\al\text{+}
\dfrac{\vphi_2\vphi_3}{\chi_1}\sin\al\Bigr)
\text{sh}\rho\Bigr]\text{+}\dfrac{c\vphi_2}
{\chi_1}\text{ch}\rho
\\\\
\dfrac{\chi_2(\chi_1\text{--}c^2)^{\frac12}}{\chi_1}(\cos\al\text{--}
\sin\al\;\text{sh}\rho)\text{+}\dfrac{c\vphi_3}
{\chi_1}\text{ch}\rho
\endpmatrix
$$
\vskip10pt
$$
\underset{(K=1)}\to{\ovd\lm}
=\pmatrix
\dfrac{\ep(\chi_1^2-c^2)^{\frac12}}{\chi_2}\Bigl(-\vphi_2\cos\al+\frac{\vphi_1\vphi_3}{\chi_1}
\sin\al\Bigr)+\dfrac{c\vphi_1}{\chi_1}
\\\\
\dfrac{\ep(\chi_1-c^2)^{\frac12}}{\chi_2}\Bigl(\vphi_1\cos\al+\dfrac{\vphi_2\vphi_3}{\chi_1}
\sin\al\Bigr)+\dfrac{c\vphi_2}{\chi_1}
\\\\
\dfrac{-\ep(\chi_1^2-c^2)^{\frac12}}{\chi_1}\chi_2\sin\al+\dfrac{c\vphi_3}{\chi_1}
\endpmatrix
$$

\vskip20pt
{\bf Table 2.} Covectors $\lm', K=0$ \quad( $F_{1'},F_{2'}$)
$$
\underset{(K=0)}\to{\ov{(1')}\lm}=
\pmatrix
\dfrac{(\chi_1^2-c^2)^{\frac12}}{c\chi_2}(\vphi_2\cos\beta-\dfrac{\vphi_1\vphi_3}{\chi_1}
\sin\beta)+\dfrac{\vphi_1}{\chi_1}
\\\\
\dfrac{-(\chi_1^2-c^2)^{\frac12}}{c\chi_1}(\vphi_1\cos\beta+\dfrac{\vphi_2\vphi_3}{\chi_1}\sin\beta
)+\dfrac{\vphi_2}{\chi_1}
\\\\
\dfrac{(\chi_1^2-c^2)^{\frac12}}{c\chi_1}\sin\beta+\dfrac{\vphi_3}{\chi_1}
\endpmatrix
$$
\vskip10pt
Covectors
$\lm'$, $K=1$
\vskip10pt
$$\underset{(K=1)}\to{\ov{(1')}\lm}=
\pmatrix
\dfrac{(\chi_1-c^2)^{\frac12}}{c\chi_2}(-\vphi_2\cos\om-\dfrac{\vphi_1}{\chi_1}\sin\om
\Bigr)+\dfrac{\vphi_1}{\chi_1}
\\\\
\dfrac{(\chi_1^2-c^2)^{\frac12}}{c\chi_2}\Bigl(\vphi_1\cos\om-\dfrac{\vphi_2\vphi_3}{\chi_1}\sin\om
\Bigr)+\dfrac{\vphi_2}{\chi_1}
\\\\
\dfrac{(\chi_1^2-c^2)^{\frac12}}{c\chi_1}\sin\om+\dfrac{\vphi_3}{\chi_1}
\endpmatrix $$

\vskip20pt
{\bf Table 3.} Covectors $\lm''$, \quad $(F_{1''},F_{2''}, F_{3''})$
\vskip10pt
$$
\underset{(K=0)}\to{\ov{(1'')}\lm}=\pmatrix
\dfrac{(\chi_1^2-c^2)^{\frac12}}{\chi_2}\Bigl[-\dfrac{\vphi_1\vphi_3}{\chi_1}
(\text{sh}\rho\cos\al 
+\sin\al)-\vphi_2(\text{sh}\rho\sin\al+\cos\al)\Bigr]+\dfrac{c\vphi_1}{\chi_1}\text{ch}\rho
\\\\
\dfrac{(\chi_1^2-c^2)^{\frac12}}{\chi_2}\Bigl[-\dfrac{\vphi_2\vphi_3}{\chi_1}(\text{sh}\rho\cos\al+\sin\al)+
\vphi_1(\text{sh}\rho\sin\al+\cos\al)\Bigr]+\dfrac{c\vphi_2}{\chi_1}\text{ch}\rho
\\\\
\dfrac{(\chi_1-c^2)^{\frac12}}{\chi_1}\chi_2(\text{sh}\rho\cos\al+\sin\al)+\dfrac
{c\vphi_3}{\chi_1}\text{ch}\rho
\endpmatrix $$
\vskip10pt
$$
\underset{(K=1)}\to{\ov{(1'')}\lm}=\pmatrix
\dfrac{(\chi_1^2-c^2)^{\frac12}}{\chi_2}\Bigl[-\dfrac{\vphi_1\vphi_3}{\chi_1}
(\text{sh}\rho\cos\al-\sin\al)-\vphi_2(\text{sh}\rho\sin\al-\cos\al)\Bigr]+
\dfrac{c\vphi_1}{\chi_1}\text{ch}\rho
\\\\
\dfrac{(\chi_1^2-c^2)^{\frac12}}{\chi_2}\Bigl[\dfrac{\vphi_2\vphi_3}{\chi_1}
(-\text{sh}\rho+\vphi_1(\cos\al+\text{sh}\rho\sin\al)\Bigr]
+\dfrac{c\vphi_2}{\chi_1}\text{ch}\rho
\\\\
\dfrac{(\chi_1^2-c^2)^{\frac12}}{\chi_1}\chi_2(\text{sh}\rho\cos\al-\sin\al)-\dfrac{c\vphi_3}
{\chi_1}\text{ch}\rho
\endpmatrix
$$
\vskip10pt
$$\ov{(2'')}\lm=
\pmatrix
\dfrac{-\ep(\chi_1^2-c^2)^{\frac12}}{c\chi_2}\Bigl(\vphi_2\sin\al+\dfrac{\vphi_1\vphi_3}{\chi_1}
\cos\al\Bigr)+\dfrac{\vphi_1}{\chi_1}
\\\\
\dfrac{\ep(\chi_1^2-c^2)^{\frac12}}{c\chi_2}\Bigl(\vphi_1\sin\al-\dfrac{\vphi_2\vphi_3}{\chi_1}\cos\al\Bigr)
+\dfrac{\vphi_2}{\chi_1}
\\\\
\dfrac{\ep(\chi_1^2-c^2)^{\frac12}}{c\chi_1}\chi_2\cos\al+\dfrac{\vphi_3}{\chi_1}
\endpmatrix $$
\vskip10pt\flushpar
$K=0,1$
\vskip10pt\flushpar
where
$\al,\rho,\beta,\om$ are arbitrary smooth (at least of class $C^2$)
functions of $\vphi_i$, $i=1,2,3$ and
$\chi_1^2=\vphi_1^2+\vphi_2^2+\vphi_3^2$, $\chi_2^2=\vphi_1^2+\vphi_2^2$, $\ep^2=1$.

\newpage
\vskip2\baselineskip
\centerline{\bf Appendix B --- simple elements (the second example)}
\vskip\baselineskip

The covectors $\lm$ are
\vskip15pt
$$
\lm^1=
\pmatrix
c\sqrt2\Bigl[\dfrac
{(c^2+1-v^2+\sqrt\tr)}{(v^2-c^2+1+\sqrt\tr)^{1/2}}\text{ch}\tau+
\dfrac{(c^2+1-v^2-\sqrt\tr)}{(c^2-1-v^2+\sqrt\tr)^{1/2}}\text
{sh}\tau\Bigr]
\\\\
-\phi_2+2c\sqrt2\phi_1\Bigl[\dfrac{\text{ch}\tau}{(v^2-c^2+1+\sqrt\tr)^{1/2}}
+\dfrac{\text{sh}\tau}{(c^2-1-v^2+\sqrt\tr)^{1/2}}\Bigr]
\\\\
\phi_1+2c\sqrt2\phi_2\Bigl[\dfrac{\text{ch}\tau}{(v^2-c^2+1+\sqrt\tr)^{1/2}}
+\dfrac{\text{sh}\tau}{(c^2-1-v^2+\sqrt\tr)^{1/2}}\Bigr]
\endpmatrix $$
\vskip15pt
$$\lm^2=\pmatrix
(c^2\text{+}1\text{--}v^2\text{+}\sqrt\tr)
(\sqrt\tr\text{--}v^2\text{+}c^2\text{--}1)^{1/2}\text{+}
(c^2\text{+}1\text{--}v^2\text{--}\sqrt\tr)(v^2\text{--}c^2\text{+}1\text{+}\sqrt\tr)^{1/2}
\\\\
2\phi_1[\sqrt\tr\text{--}v^2\text{+}c^2\text{--}1)^{1/2}\text{+}
(v^2\text{--}c^2\text{+}1\text{+}\sqrt\tr)^{1/2}]
\\\\
2\phi_2[\sqrt\tr\text{--}v^2\text{+}c^2\text{--}1)^{1/2}\text{+}
(v^2\text{--}c^2\text{+}1\text{+}\sqrt\tr)^{1/2}]
\endpmatrix
$$
\vskip15pt
$$\lm^3=\pmatrix
(c^2+1-v^2+\sqrt\tr)+(c^2+1-v^2-\sqrt\tr)\dfrac
{(v^2-c^2+1+\sqrt\tr)^{1/2}}{(c^2-1-v^2+\sqrt\tr)^{1/2}}\sin\tau
\\\\
\dfrac{-\phi_2}{c\sqrt2}(v^2-c^2+1+\sqrt\tr)^{1/2}\cos\tau+2\phi_1\Bigl[
1+\Bigl(\dfrac
{v^2-c^2+1+\sqrt\tr}{c^2-1-v^2+\sqrt\tr}\Bigr)^{1/2}\sin\tau\Bigr]
\\\\
\dfrac{\phi_2}{c\sqrt2}(v^2-c^2+1+\sqrt\tr)^{1/2}\cos\tau+2\phi_2\Bigl[
1+\Bigl(\dfrac{v^2-c^2+1+\sqrt\tr}{c^2-1-v^2+\sqrt\tr}\Bigr)^{1/2}\sin\tau
\Bigr]
\endpmatrix $$
\vskip15pt
$$
\lm^4=\pmatrix
\dfrac{\ep c\sqrt2(c^2+1-v^2+\sqrt\tr)}{(c^2-1-v^2+\sqrt\tr)^{1/2}}
\\\\
-\phi_2+2\phi_1\dfrac{\ep c\sqrt2}{(c^2-1-v^2+\sqrt\tr)^{1/2}}
\\\\
\phi_1+2\phi_2\dfrac{\ep c\sqrt2}{(c^2-1-v^2+\sqrt\tr)^{1/2}}
\endpmatrix $$
\vskip15pt
$$
\lm^5=\pmatrix
(c^2+1-v^2+\sqrt\tr)\Bigl(\dfrac{\sqrt\tr-v^2+c^1-1}{\sqrt\tr+v^2-c^2+1}\Bigr)^{1/2}
\text{ch}\tau+
(c^2+1-v^2-\sqrt\tr)
\\\\
\dfrac{-\phi_2}{c\sqrt2}(\sqrt\tr-v^2+c^2-1)^{1/2}\text{sh}\tau+2\phi_2\Bigl[
1+\Bigl(\dfrac{\sqrt\tr-v^2+c^2-1}{\sqrt\tr+v^2-c^2+1}\Bigr)^{1/2}\text{ch}\tau
\Bigr]\\\\
\dfrac{\vphi_1}{c\sqrt2}(\sqrt\tr-v^2+c^2-1)^{1/2}\text{sh}\tau+
2\vphi_2\Bigl[1+\Bigl(\dfrac{\sqrt\tr-v^2+c^2-1}{\sqrt\tr+v^2-c^2+1}\Bigr)^{1/2}
\text{ch}\tau\Bigr]
\endpmatrix 
$$
\vskip15pt
$$
\lm^6=\pmatrix
\dfrac{\ep c\sqrt2(c^2+1-v^2+\sqrt\tr)}{(v^2-c^2+1+\sqrt\tr)^{1/2}}
\\
-\phi_2+2\phi_1\dfrac{\ep c\sqrt2}{(v^2-c^2+1+\sqrt\tr)^{1/2}}
\\
\phi_1+2\phi_2\dfrac{\ep c\sqrt2}{(v^2-c^2+1+\sqrt\tr)^{1/2}}
\endpmatrix
$$
\vskip10pt\flushpar
where $\tau$ is an arbitrary (at least of class $C^2$) 
function of $\phi_i$, $i\text{=}0,1,2$ and
$v^2\text{=}\phi_1^2+\phi_2^2$, 
$\tr=(v^2-\nomathbreak c^2+\nomathbreak1)^2+\nomathbreak4c^2$.

\newpage
\vskip2\baselineskip
\centerline{\bf Acknowledgements}
\vskip\baselineskip

We thank Professor R. R{\c a}czka for continuous support during
the preparation of the paper.

We thank the referee for critical comments.

The financial support of the National Science Foundation under Contract 
No. I.N.T. 73--2000 2A01 is highly appreciated.

\newpage
\vskip2\baselineskip
\centerline{\bf References}
\vskip\baselineskip

\item{[1]}
G.~E.~B.~~R~i~e~m~a~n~n, {\it \"Uber die Fortpflanzung ebener Luftwellen von endlicher
Schwin\-gungsweite}, Abhandl. K\"onigl. Ges. Wiss., G\"ottingen, 8, (1869).
\item{[2]}
M.~~B~u~r~n~a~t, {\it Theory of simple waves for nonlinear systems of partial differential equations
and applications to gas dynamics}, Arch. Mech. Stos. 18, (1966), p.~4.
\item{[3]}
Z.~~P~e~r~a~d~z~y~{\'n}~s~k~i, {\it Nonlinear plane K--waves and Riemann invariants},
Bull. Acad. Pol. Sec. Sci. Tech. 19, (1971), p.~9
\item{\ }
Z.~~P~e~r~a~d~z~y~{\'n}~s~k~i, {\it Geometry of nonlinear interaction in partial differential
equations}, Report of IPPT, Warsaw, 1981 (in Polish).
\item{[4]}
K.~~B~o~r~s~u~k, {\it Multidimensional analytic geometry}, PWN, Warszawa 1969.
\item{[5]}
L.~~L~a~n~d~a~u, E.~~L~i~f~s~h~i~t~z, {\it Mechanics of continuous media},
Mir, Moskwa 1952 (in Russian).
\item{[6]}
J.~~S~z~a~r~s~k~i, {\it Differential inequalities}, PWN, Warszawa 1967.
\item{[7]}
M.~W.~~K~a~l~i~n~o~w~s~k~i, {\it On the old--new method of solving nonlinear equations}, J. of Math. Physics
25  (1984) p.~2620.
\item{\ }
M.~W.~~K~a~l~i~n~o~w~s~k~i, {\it On certain method of solving nonlinear equations}, Lett. Math. Phys. 
6 (1983) p.~17.
\item{[8]}
M.~W.~~K~a~l~i~n~o~w~s~k~i, {\it On the old--new method of solving nonlinear
equations}, Warsaw University Preprint JFT (13) 80, Warsaw 1980.
\item{[9]}
F.~A.~E.~~P~i~r~a~n~i, D.~C.~~R~o~b~i~n~s~o~n, W.~F.~~S~h~a~d~w~i~c~k, {\it
Local Jet Bundle Formulation of B\"acklund Transformation}, D. Reidel
Publishing Company, Dortrecht, Boston, London 1979. 
\item{[10]}
M.~W.~~K~a~l~i~n~o~w~s~k~i, {\it Gauge transformation for simple waves}, Lett.
Math. Phys.~7 (1983) p.~479.
\item{[11]}
L.~W.~~K~r~o~k~k~o, {\it Foundation of Gas--dynamic}, Naukova Dumka, Kiev 1983
(in Russian). 
\item{[12]}
P.~~P~o~l~o~v~i~n, Sov. J. Exper. Theor. Phys. 41 (1961) p.~394 (in Russian).
\item{[13]}
P.~A.~~T~h~o~m~p~s~o~n, {\it Compressible Fluid  Dynamics}, Mc Graw -- Hill, New York 1972.
\item{[14]}
M.~W.~~K~a~l~i~n~o~w~s~k~i, A.~~G~r~u~n~d~l~a~n~d, {\it Simple waves for
equation of potential, nonstationary flow of compressible gas}, J. Math. Phys.
27 (1986) p.~1906. 

\bye